\documentclass[trsc]{informs4} 

\OneAndAHalfSpacedXI 

\usepackage{natbib}
\bibpunct[, ]{(}{)}{,}{a}{}{,}%
\def\bibfont{\small}%
\usepackage{amssymb,url,revsymb,verbatim,epstopdf}
\usepackage{mathrsfs}
\usepackage{amssymb,amsmath}
\usepackage{lineno}
\usepackage{multirow}
\usepackage[colorlinks,
linkcolor=blue,
anchorcolor=blue,
citecolor=blue]{hyperref}
\usepackage{enumerate}
\usepackage{booktabs}
\usepackage{threeparttable}
\usepackage{float}

\usepackage[labelformat=simple]{subcaption}

\usepackage{color}
\usepackage{bm}

\usepackage{graphicx}
\usepackage{dcolumn}

\usepackage[utf8]{inputenc}
\usepackage[T1]{fontenc}
\usepackage{mathptmx}
\usepackage{epstopdf}

\TheoremsNumberedThrough     

\EquationsNumberedThrough    

\begin{document}
	
	

	\TITLE{Urban traffic resilience control - An ecological resilience perspective}
	\ARTICLEAUTHORS{%

		\AUTHOR{Shengling Gao}
		\AFF{School of Mathematical Sciences, Beihang University, Beijing, China. \EMAIL{shengling\_gao@buaa.edu.cn}}
		
		\AUTHOR{Zhikun She}
        \AFF{Corresponding author. School of Mathematical Sciences, Beihang University, Beijing, China.  \EMAIL{zhikun.she@buaa.edu.cn}}
		
		\AUTHOR{Quanyi Liang}
		\AFF{Co-Corresponding author. School of Mathematical Sciences, Beihang University, Beijing, China. \EMAIL{qyliang@buaa.edu.cn}}

		\AUTHOR{Nan Zheng}
		\AFF{Co-Corresponding author. Institute of Transport Studies, Department of Civil Engineering, Monash University, Clayton, Australia. \EMAIL{Nan.Zheng@monash.edu}}
		
		\AUTHOR{Daqing Li}
		\AFF{School of Reliability and Systems Engineering, Beihang University, Beijing, China. \EMAIL{daqingl@buaa.edu.cn}}
	
	} 
	
	\ABSTRACT{
Urban traffic resilience has gained increased attention, with most studies adopting an engineering perspective that assumes a single optimal equilibrium and prioritizes local recovery. On the other hand, systems may possess multiple metastable states, and ecological resilience is the ability to switch between these states according to perturbations. Control strategies from these two different resilience perspectives yield distinct outcomes.  In fact, ecological resilience oriented control has rarely been viewed in urban traffic, despite the fact that traffic system is a complex system in highly uncertain environment with possible multiple metastable states. This absence highlights the necessity for urban traffic ecological resilience definition. To bridge this gap, this paper defines urban traffic ecological resilience as the ability to absorb uncertain perturbations by shifting to alternative states. The goal is to generate a system with greater adaptability, without necessarily returning to the original equilibrium. Our control framework comprises three aspects: portraying the recoverable scopes; designing alternative steady states; and controlling system to shift to alternative steady states for adapting large disturbances. Among them, the recoverable scopes are portrayed by attraction region; the alternative steady states are set close to the optimal state and outside the attraction region of the original equilibrium; the controller needs to ensure the local stability of the alternative steady states, without changing the trajectories inside the attraction region of the original equilibrium as much as possible. Note that, this paper gives inner and outer estimations of attraction region with explicit algebraic expressions, as the attraction region for nonlinear systems are usually very complex and difficult to obtain. Comparisons with classical engineering resilience oriented urban traffic resilience control schemes show that, proposed ecological resilience oriented control schemes have better adaptability and can generate greater resilience. These results will contribute to the fundamental theory of future resilient intelligent transportation system.
	}%
	
	
	%
	\maketitle
	%
	\section{Introduction}\label{s1}
	Traffic congestion is plaguing most of megacities~(\cite{huang2020transportation,zeng2019switch}).
	Congested transportation systems are highly vulnerable,  and often fall into collapse under extreme events such as poor weather and accidents.  Such vulnerability of transportation network operations is a "persistent urban disease". Therefore, how to  sustain functional  road transportation systems  has been a long-standing research topic, in particular the challenging objective is to improve the resilience of operations towards uncertain perturbations.
	
\begin{figure}[H]
		\centering
		\begin{subfigure}{.3\textwidth}
			\centering
			\includegraphics[height=3cm]{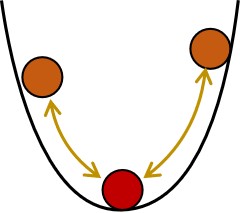}
			\caption{Engineering resilience }
			\label{fig23.1}
		\end{subfigure}
		\begin{subfigure}{.5\textwidth}
			\centering
			\includegraphics[height=3cm]{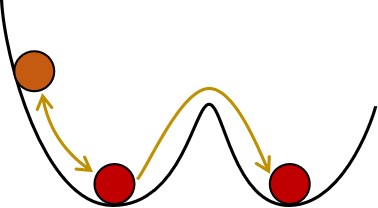}
			\caption{Ecological resilience}
			\label{fig23.2}
		\end{subfigure}
		\caption{Schematic diagram of (a) engineering resilience  and  (b) ecological resilience  (\cite{liao2012theory,rolfer2022resilience})}
		\label{fig23}
	\end{figure}

	Resilience, first proposed by \cite{holling1973resilience} in the ecology field, denotes the system ability to absorb unexpected disturbances and to keep persistence. Over the past few decades, resilience has been greatly developed in  various disciplines including ecology, engineering, and psychology (\cite{martin2011resilience,hosseini2016review}).
	In general, there are two perspectives on resilience: engineering resilience and  ecological resilience~(\cite{holling1996engineering}).  
	Noteworthy,  this categorization of resilience is discipline-independent. One can use ball-and-cup heuristic to illustrate these two perspectives  (\cite{scheffer1993alternative,rolfer2022resilience,liao2012theory,walker2004resilience}), as shown in Fig.~\ref{fig23}: when the ball is at the cup bottom, it symbolizes  a steady state, known as an “attractor”; the shape of the cup portrays the recoverable phase space of the corresponding attractor, usually described as the shape of potential function (landscape) or attraction basin; the yellow arrows represent perturbations and system responses.  Fig.~\ref{fig23.1} illustrates the engineering resilience, which usually assumes there exists a single steady state and characterizes the system ability to absorb disturbances and recover to this single steady state. Studies related to engineering resilience typically aim at  rapid local or global recovery of a single steady-state system (\cite{tilman1994biodiversity,o1986hierarchical,pimm1984complexity}).
	In contrast, Fig.~\ref{fig23.2} shows the ecological resilience, which features multiple metastable states and  the perturbations are absorbed  through a multi-state shifting, without requiring the returning to a single steady state. Clearly, ecological resilience emphasizes persistence with uncertainty~(\cite{holling1973resilience,2006Resilience,2022Urban}). 
	As such, ecological resilience can transform a system under instability from one basin of attraction to another~(\cite{holling1973resilience}), even for cases where they are far from the stable equilibrium as defined otherwise in engineering resilience.
	
	Road network resilience is an emerging research subject. Most of the developed works  are towards  engineering resilience, aiming at a rapid recovery to a non-congested equilibrium or a single optimal equilibrium. For example, the classic perimeter control schemes developed based on the macroscopic fundamental diagram (short for MFD) characterization of congestion dynamics, have received much attention in recent years~
	(\cite{geroliminis2008existence,haddad2012stability,aghamohammadi2020dynamic,laval2017minimal,johari2021macroscopic,2015Robust,2023Optimal,haddad2020adaptive,mariotte2017macroscopic,HADDAD202044,RAMEZANI20151,LI2021103043,9906860,kouvelas2023linear,tsitsokas2023two, YANG2019439,dantsuji2021simulation,ampountolas2017macroscopic,YANG201832,su2020neuro}).
	The perimeter control schemes regulate the proportion of  entering traffic at the perimeter of the networks via traffic signal control.
	In this direction,
	\cite{haddad2012stability} 
	calculated the "region of attraction" (short for RA) and designed a state feedback control to increase the RA for enlarging the recoverable area towards  a single non-congested stable equilibrium.
	One pre-condition though, is that the system is within the states as defined by RA. In face of the so-called “hyper-congestion” which is far beyond the stable equilibrium state (e.g. a congestion density value twice as much as the critical density) and beyond RA, recovering to the desired system equilibrium  cannot be guaranteed, and the system will inevitably fail, e.g. becoming unrecoverable grid lock. On the other hand,
	Zhong et al.~(\cite{zhong2020dynamic,zhong2018robust,zhong2018boundary}) and Huang et al.~(\cite{huang2020dynamic}) developed control schemes where  the control would aim at a designated point, for example the point near the critical density or critical accumulation where the network flow reaches its maximum. However, the sufficient conditions of these controllers require that the disturbance to the systems, i.e. the incoming traffic demand to the road networks, cannot be high and fast-varying over time, which limits the  controllability  towards real world traffic situations. Most recently, 
	\cite{gao2022resilient} proposed a novel control scheme based on RA, denoted as RCS-single, focusing specifically towards the recovery to the single optimal equilibrium from hyper-congested states. This is one of the first exploratory works looking at the “resilience” of system control. Nevertheless, this work was still within the framework of a single equilibrium system, and the recovery speed is very slow.
	For systems where instability is coupled with complex disturbances, in the case of hyper-congestions accompanied by persistently high travel demand,  the existing controls are  not only time-consuming,  but also the objective of returning to single equilibrium is impractical. 
	In literature, the impact of  large changes in travel demand causing the instability of MFD was pointed out in multiple studies, such as  \cite{daganzo2011macroscopic,zhong2020modeling,haddad2020adaptive}.

	Urban traffic is a complex dynamic system, which may exist  multiple metastable states ~(\cite{zeng2020multiple}). 
	Given the above discussion on the limitation of the single-objective control schemes, a more realistic way  appears to be guiding the system to switch between multiple metastable states under perturbations. While the engineering resilience-oriented system controls have its values, there is a need to address the limitation towards system perturbations. 
	Future resilient traffic system, that is capable of withstanding large and rapidly changing environments and contingencies, absorbing environmental uncertainties and developing different adaptive landscapes, would be by nature an ecosystem-like dynamic system. In other words, controlling such system should embrace an ecological perspective, where the ecological operation essentially enables the system to shift  among alternative stable states~(\cite{holling1996engineering,folke2010resilience,walker2012resilience,gunderson2000ecological,dakos2022ecological,scoones2020transformations}). 
	Therefore, this work develops  an ecological resilience control for managing congested urban road systems.  The control aim is to  enhance road traffic flow systems with greater adaptability towards different perturbations, such as congestion, unusual travel demand, sudden infrastructure closure, and accidents. Our work will, for the first time, provide an analytical modeling framework and the proves on  system shifting  between alternative steady states. Our work will demonstrate that such system is superior and much more resilient,  comparing to controlled systems that aiming at returning to  a single state, such as an original equilibrium or a pre-defined optimal state.

	Built upon the MFD-based traffic flow representation system, this paper proposes a methodology for deriving ecological resilience control under large disturbances. Our control methodology comprises three aspects: portraying the recoverable system states, designing the target steady system states, and developing control mechanisms to regulate system towards steady states. Specifically, the RA framework will be adopted but extended, where the recoverable scopes are portrayed by attraction region; the alternative steady states are set close the optimal state and outside the attraction region of the original equilibrium; the controller is devised to guarantee the local stability of the alternative steady states, without changing the dynamics around the original equilibrium as much as possible. We showcase the proposed methodology in two-region road systems (i.e. two congested road networks whose traffic flow and congestion interact with each other, and each road network has its own MFD to describe its traffic dynamics). We first derive the explicit algebraic expressions of the boundaries of inner and outer estimations of the attraction regions. Building upon these theoretical boundary delineations, we then design two distinct resilience control schemes, denoted as RCS-1 and RCS-2, respectively. Considering a four-equilibria system as an illustration, we test two resilient control schemes. The control performance can be seen in Fig.~\ref{fig22} and Fig.~\ref{fig15} (more details and discussions will be provided in section \ref{s5}). CPC (denotes constant perimeter control) and RCS-single (proposed by \cite{gao2022resilient}) are two benchmark methods. As a general remark, the proposed resilience control schemes and RCS-single can prevent the evolution of vehicle density (also called vehicle accumulation) $n_1$  or $n_2$ to jam vehicle density (corresponding to zero completion flow), while CPC is unable to impede such occurrences. Furthermore, the resilience measures summarized in Table~\ref{tab4} indicate that, RCS-1 and RCS-2 outperform  RCS-single as they exhibit greater resilience measures in most cases, corresponding to smaller resilience triangle (resilience loss). This superiority stems from  the ability of RCS-1 and RCS-2 to guide the trajectories towards alternative stable states that are more easily attainable, thereby enhancing adaptability. Moreover, the local landscape (local Lyapunov function (\cite{WANG2021100993,Willems1971,BLANCHINI1995451})) near the target point of controlled system with RCS-2 clearly shows the effect of our control scheme.
	
	The main contributions of this study encompass three key aspects: 
	\begin{enumerate}[(1)]
		\item We establish a theoretical foundation for the approximate depiction of recoverable states for multi-equilibrium systems, by deriving the inner and outer estimations of attraction regions with explicit algebraic expressions for multi-equilibrium MFD dynamics, utilizing invariant sets. Note that the computation of RA for multiple equilibria nonlinear systems is especially intricate and hard to obtain.
		
		\item We innovatively propose a leap from single-steady state control (from an engineering resilience perspective) to multi-steady states control (from an ecological resilience perspective), by leveraging the developed theory of approximate depiction of recoverable states and the stability theory of switch systems. This effort yields explicit control methodological, providing a clear pathway for the implementation of a control system that navigates across different stability landscapes.
		
		\item We innovatively design a resilience measure utilizing the classical resilience triangle and completion flow. And it  intuitively depicts the loss of completed trips compared to the maximum potential number of trips that could have been completed during the recovery period.

	\end{enumerate}

	
	\section{The control system dynamics and its ecological resilience}\label{s2}
	This paper showcases the advancement  and  performance of an ecological resilience-oriented control in a two-region system (known also as two-reservoir system in literature) traffic flow control problem, where the traffic dynamics of the two regions follow an MFD-based representation. Sec. \ref{s2.1} firstly gives the mathematical description of the MFD-based modeling framework, Sec. \ref{s2.2} illustrates the ecological resilience control framework for MFD dynamics, and then Sec. \ref{s2.3} defines ecological resilience measure.
	\subsection{Two-region MFD dynamics}\label{s2.1}
	
\begin{figure}
		\centering
		\includegraphics[width=7cm]{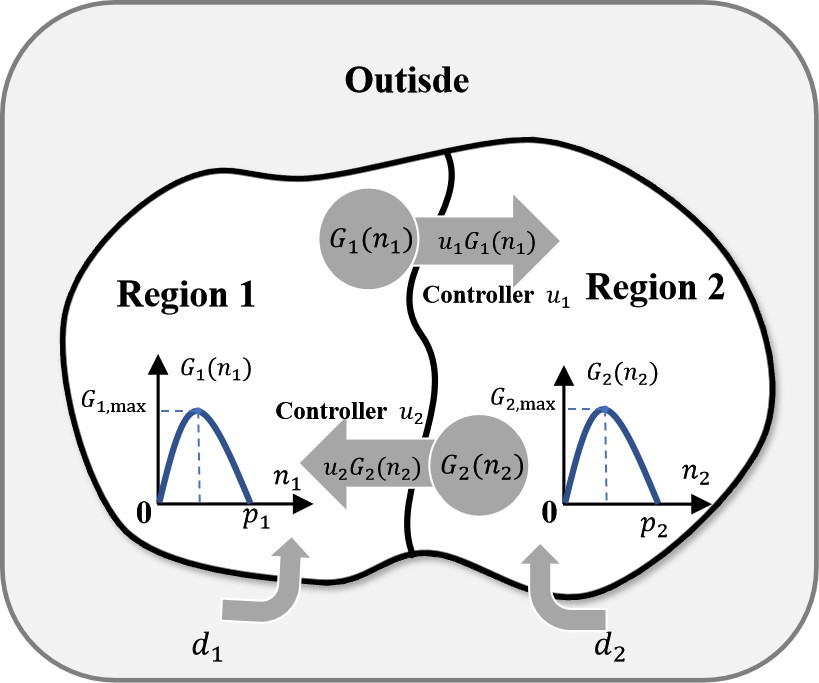}
		\caption{Two region MFD dynamics (\cite{aboudolas2013perimeter,gao2022resilient}).}
		\label{fig30}
	\end{figure}

	To start with, we consider a two-region MFD system, illustrated in Fig. \ref{fig30}. The urban traffic network is partitioned into three parts: Region 1, Region 2, and the Outside. Our primary focus lies on Regions 1 and 2, and each endowed with its own MFD. The MFD delineates the correlation between the travel completion flow $G_i(n_i(t))$ and vehicle accumulation (density) $n_i$ of region $i$ at time $t$, manifesting a single peak, characterized by an initial ascent succeeded by a descent. Classical perimeter control methodologies regulate the traffic flow ratio entering Regions 2 and 1 through signal controls, denoted as allowed pass rates $u_1$ and $u_2$. As illustrated in  Fig. \ref{fig30}, for each region at time $t$, there are two inflows (marked as gray arrows directed toward region i) and one completion flow $G_i(n_i(t))$ (marked as gray circle). The first inflow, ${u_j} {G_j}{(n_j (t))}$ ($j= 1, 2$, $j$ $\neq$ $i$) signifies the allowed traffic inflow transferred from $j$ to i with the pass rate $u_j$. The second inflow, denoted as $d_i$, represents the fixed net inflow from the outside to $i$, encapsulating uncontrollable demand, commonly known as disturbance. 
	
\begin{figure}
		\centering
		\includegraphics[width=16cm]{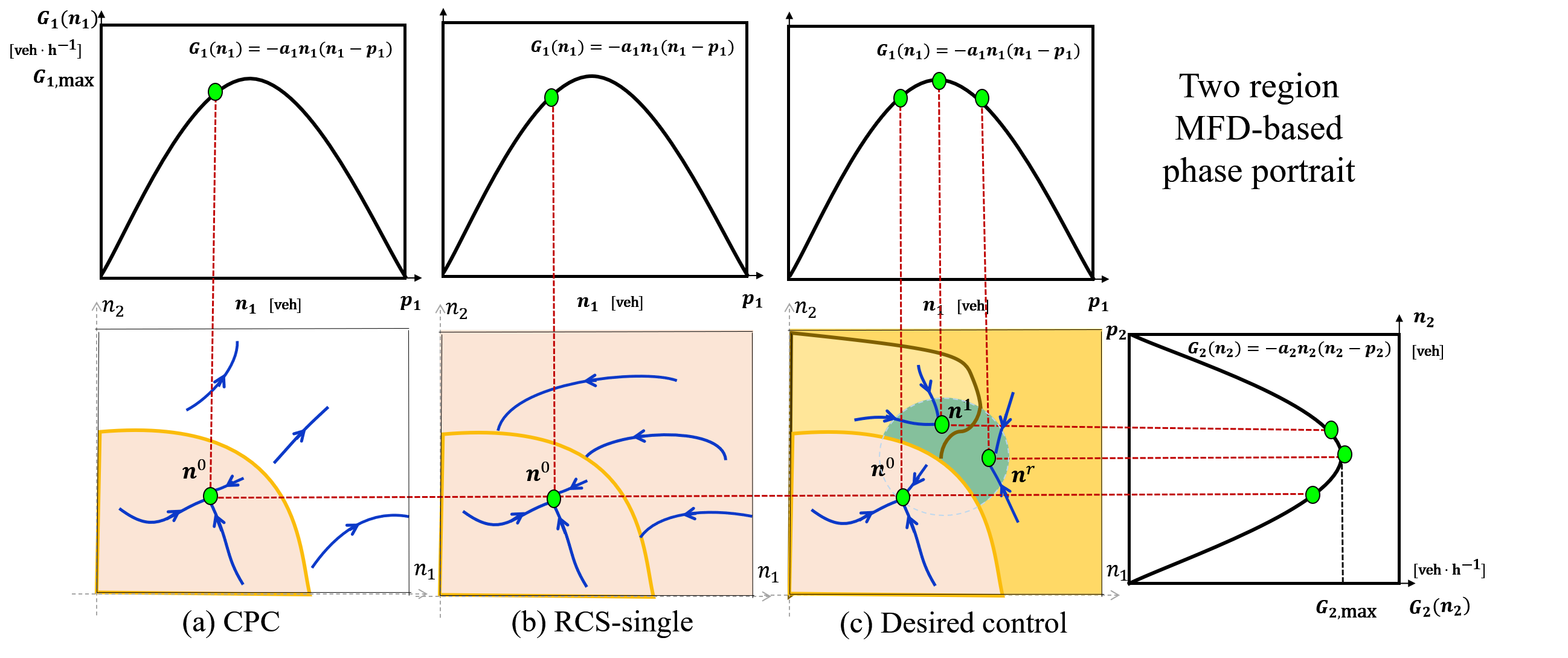}
		\caption{Schematic representation of the phase portraits under (a) CPC, (b) RCS-single, and (c) desired control. Note that desired phase portraits exhibit multiple metastable states. 
			The attraction region of each metastable state is demarcated by different colors. The green area signifies the selectable range of the alternative steady states.
			the green circle represents the steady states; 
			the blue lines with arrows depict state trajectories, indicating the evolution of a point (state) over time. 
			The phase portraits and the Macroscopic Fundamental Diagram (MFD) exhibit a correlation. Each point in the phase portraits corresponds to the vehicle density (accumulation) in two regions, and the corresponding completion flows can be identified on the MFD.
		}
		\label{fig01}
	\end{figure}

	The model can be described by MFD dynamic (\cite{aboudolas2013perimeter,gao2022resilient}):
	\begin{equation}\label{the model}
		\begin{split}
			\frac{\mathrm{d}{n}_1(t)}{\mathrm{d}t}= F_1({n}_1(t),{n}_2(t)):= -{G_1}{(n_1 (t))}+{u_2} {G_2}{(n_2 (t))}+d_1,\\
			\frac{\mathrm{d}{n}_2(t)}{\mathrm{d}t}= F_2({n}_1(t),{n}_2(t)):= -{G_2}{(n_2 (t))}+{u_1} {G_1}{(n_1 (t))}+d_2,
		\end{split}
	\end{equation}
	with the boundary condition:
	\begin{equation}\label{boundary condition}
		\begin{split}
			&0 \leq n_1(t) \leq p_1,\, 0 \leq n_2(t) \leq p_2,\\
			&0 \leq u_1 \leq u_{1,max}, \, u_{1,max}=\min\left\{1,\frac{G_{2,max}}{G_{1,max}}\right\},\\
			&0 \leq u_2 \leq u_{2,max},\, u_{2,max}=\min\left\{1,\frac{G_{1,max}}{G_{2,max}}\right\}.
		\end{split}
	\end{equation}
	Let $\bm{n}=(n_1(t),n_2(t))$, $\bm{F}(\bm{n})=(\left( F_1(n_1(t),n_2(t),F_2(n_1(t),n_2(t))\right)$, and $\dot{\bm{n}}=\frac{\mathrm{d}{\bm{n}}}{\mathrm{d}t}$, system \eqref{the model} can be simplified to $\dot{\bm{n}}= \bm{F}(\bm{n})$.
	Note that perimeter controllers $u_1$ and $u_2$ are assumed to be constants.
	Therefore, they can be denoted as constant perimeter control, short for CPC.
	The fixed net inflow $d_i$ is also assumed to be constants. 
	Moreover, as a realistic scene, 
	we here use the parabolic-MFD: $G_i(n_i)=-{a_i}{n_i}({n_i-{p_i}})$ with opening size $a_i>0$ and jam accumulation $p_i$ of region $i$, as shown in Fig.~\ref{fig30} and Fig.~\ref{fig01}.
	Then we have the maximum capacity $G_{i,max}=\frac{{a_i}{{p_i}^2}}{4}$ at the critical vehicle accumulation $\frac{p_i}{2}$.
	Note that
	$d_j \leq {G_{j,max}}$, $u_i{G_{i,max}} \leq {G_{j,max}}$ and $u_1\times u_2\neq 1$ are satisfied to avoid overflow. 
	Without loss of generality, we can assume $G_{1,max}\leq G_{2,max}$, 
	then the condition of $u_1$ and $u_2$ in Eq.~(\ref{boundary condition}) can be simplified as Condition $(\mathbb{H})$:
	 \begin{equation}\label{u1u2 condition}
0 \leq u_1 \leq 1,\ 0 \leq u_2 \leq \frac{G_{1,max}}{G_{2,max}}, ~\text{but} ~u_1\times u_2\neq 1.
	 \end{equation}
	
	For the two-region MFD dynamic \eqref{the model} with boundary condition \eqref{boundary condition} under traffic demand $(d_1,d_2)$, 
	given the MFD function $(G_1(n_1),G_2(n_2))$, 
	we aim to generate a  transportation system from ecological resilience perspective by combining the stability characteristics of the MFD dynamics. 
	
	To achieve this objective, we first introduce concepts related to stability characteristics. The stability characteristics can be elucidated by phase portrait, where the phase portrait is a geometric representation of the dynamic system trajectories on the phase plane. 
	As shown in Fig.~\ref{fig01}(a), 
	the phase plane is $n_1-n_2$, and each point (state) in $n_1-n_2$ plane denotes the vehicle accumulation in two regions. Progressing to the right or upward along the $n_1-n_2$ plane signifies a corresponding increase in the vehicle accumulation within Region 1 (Region 2). Moreover, the axis $n_2$ in Fig.~\ref{fig01}(a) corresponds to the ordinate of the right MFD and axis $n_1$ corresponds to the abscissa of the upper MFD. The completion flow $G_1(n_1)$($G_2(n_2)$) increases with the vehicle accumulation before the critical vehicle accumulation $\frac{p_1}{2}$($\frac{p_2}{2}$), reaching a peak $G_{1,max}$ ($G_{2,max}$), and subsequently decreases until it diminishes to zero at the right (upper) boundary, signifying a traffic jam. 
	As shown in Fig.~\ref{fig01}(a), the attraction region (light pink region) of the equilibrium point $\bm{n}^0=(n^0_1,n^0_2)$ (green circle) is a set of the initial points. The trajectories start from 	these initial points will converge towards $n^0$, while trajectories start from outside attraction region will move away from $n^0$ and towards either the upper or right boundary. Note that equilibrium point $\bm{n}^0=(n^0_1,n^0_2)$ satisfies $F_1(n^0_1,n^0_2)=0,~F_2(n^0_1,n^0_2)=0$.
	We can introduce the definition of attraction region for the general equilibrium  $\bm{n}^\ast=(n^\ast_1,n^\ast_2)$ as follows:
	\begin{definition}~(\cite{gao2022resilient})
		For system \eqref{the model} with boundary restrictions \eqref{boundary condition}, denoting $n(t)=(n_1(t),n_2(t))$ and $\mathbb{D}=\left\lbrace (n_1(t),n_2(t))|0 \leq n_i(t) \leq p_i,\, i=1,2\right\rbrace,$ 
		define 
			 \begin{equation}\label{attraction region definition}
			\mathcal{R}(n^\ast)=\left\lbrace n_{init} \in \mathbb{D}| n(0)=n_{init}\  and \lim_{t \to +\infty}n(t) = n^\ast\right\rbrace,
		\end{equation}
		as the attraction region for $n^\ast=(n^\ast_1,n^\ast_2)$.	
	\end{definition}
	
	Note that, the classical concept of the "attraction domain" was originally defined for asymptotically stable equilibria (\cite{wang2020inner,zheng2018inner,ratschan2010providing,wang2020estimating}). Especially, for MFD dynamic (\ref{the model}), \cite{gao2022resilient} 
	defined the "attraction region" for the unstable single equilibrium point. Differently, we here in this paper define for the general equilibrium $(n^\ast_1,n^\ast_2)$, including both unstable equilibrium and stable equilibrium. 
	
	\subsection{Ecological resilience control framework for MFD dynamics} \label{s2.2}
	
	In the traffic flow theory and traffic control community, how to recover the traffic system from congestion has been a challenging research question. The classical perimeter control schemes, which utilizes MFD as references for optimal control, target on recovering the traffic system to a specified equilibrium. Such schemes are known to be constrained and uneconomical for hyper-congestion conditions, because these conditions are far from the  single optimal control objective and there exist  complex  dynamics in the system that prevent the system from even getting closer to the targeted equilibrium. For this reason, our approach defines a new control objective, built upon which a corresponding control methodology is developed.
	
	In essence, ecological resilience considers the existence of multiple metastable states in a dynamic system, and it enables the  system  to shift between states as a way to absorb or adapt to rapid external changes. Inspired by this nature, we aim to construct an ecological resilience traffic system resembling Fig.~\ref{fig01}(c), with multiple metastable states (marked as green circles), where multiple metastable states includes the original uncongested equilibrium $\bm{n}^0$ and alternative steady states 
	$\bm{n}^i=({n}^i_1,{n}^i_2)$ ($i=1,\cdots,r$ with  number of  alternative steady states $r$).
	
	To generate the ecological resilience traffic system resembling Fig.~\ref{fig01}(c), our first critical issue is how to select multiple metastable states. Initially, we aim to leverage the inherent landscape of the system as much as possible. Therefore, we preserve the original non-congested equilibrium point of the system, denoted as $\bm{n}^0$. This point coincidentally aligns with the single equilibrium point set by the classical CPC control system (as shown in  Fig.~\ref{fig01}(a)) or the single equilibrium point of the RCS-single control system proposed by \cite{gao2022resilient} (as depicted in  Fig.~\ref{fig01}(b)), satisfying $F_1(\bm{n}^0)=0,~F_2(\bm{n}^0)=0$. 
	After determining $\bm{n}^0$, the next step involves selecting alternative stable states $\bm{n}^i=({n}^i_1,{n}^i_2)$. There are two key criteria for selecting alternative stable states: one to position them as close to the optimal and critical vehicle accumulation $\left( \frac{p_1}{2},\frac{p_2}{2}\right) $ as possible; the other  
		is to position them outside the attraction region $\mathcal{R}(\bm{n}^0)$ (depicted in the light pink region) of the original non-congested equilibrium point $\bm{n}^0$. 
		The first criterion seeks to increase the completion flow, given that such completion flow  tends to decline as the vehicle accumulation gets further from the critical vehicle accumulation (see Fig.~\ref{fig01}(c)). 
		The second criterion is crucial because states within the attraction region of $\bm{n}^0$ will spontaneously recover to $\bm{n}^0$. Here, we provide a quantitative description of the selectable range. For the first criterion, we aim to select $\bm{n}^i=(x,y)$, such that the absolute distance between completion flow $(G_1(x),G_2(y))$ and maximum completion flow  $(G_{1,max},G_{2,max})$ is less than $\Delta_0$, where $\Delta_0$ denotes the absolute distance between completion flow of the original equilibrium $\bm{n}^0$ and maximum completion flow. That is,
		\begin{equation}\label{scope of n1}
			\left| G_1(x)-G_{1,max}\right| +\left| G_2(y)-G_{2,max}\right|\leq \Delta_0,
		\end{equation}
		substituting $G_i(n_i)=-{a_i}{n_i}({n_i-{p_i}})$ and $G_{i,max}=\frac{{a_i}{{p_i}^2}}{4}$, yields
		\begin{equation}\label{new scope of n1}
			\mathcal{E}=\left\lbrace (x,y)|\frac{\left( x-\frac{p_1}{2}\right)^2}{a_2}+\frac{\left( y-\frac{p_2}{2}\right)^2}{a_1}\leq \frac{\Delta_0}{a_1a_2}\right\rbrace, 
		\end{equation}
		$\mathcal{E}$ corresponds to the interior of an ellipse. Combining the second criterion, the selectable range of alternative stable states $\mathcal{S}$ is 
		\begin{equation}\label{lastscope of n1}
			\mathcal{S}=\left\lbrace (x,y)|(x,y) \in \mathcal{E} \bigwedge (x,y)  \notin \mathcal{R} \right\rbrace, 
		\end{equation}
		as the green region shown in Fig.~\ref{fig01}(c).

	Once multiple metastable states have been chosen, the next consideration is how to devise a control scheme that enables the system to recover to appropriate steady states. Controlling a system with multi-steady states poses more challenges compared to a system with a single stable state. The first challenge lies in determining which state the system should recover to, and the second challenge is devising a strategy to facilitate the system recovery to the respective target stable state. To address the first challenge, we need to identify the attraction region of the original equilibrium. It is essential to note that, for nonlinear systems, the attraction region is typically intricate, and obtaining explicit algebraic expressions for it can be challenging. In this direction, \cite{gao2022resilient} derived the attraction region with explicit algebraic expressions for the single-equilibrium MFD dynamics. Different from their work, we consider multi-equilibria (including two-equilibria and four-equilibria) systems in this paper. Given the involvement of multiple equilibrium points and nontrivial boundary shapes, their approach can not be applicable here. To this end, our first effort is the estimation of the attraction region for the multi-equilibria systems (see Sec. \ref{s3} and App. \ref{A}). Upon obtaining an estimation of the attraction region with explicit expressions for the boundaries (including inner and outer estimations) of the original uncongested equilibrium point, a natural boundary between the original uncogested  equilibrium  and the alternative steady state  emerges.
	
	The second challenge, i.e.,  how to devise a strategy to facilitate the system recovery to the respective target steady state is also intricate. To tackle this issue, we make a second effort using a switched controlled system and design two control schemes based on the inner and outer estimations of the attraction region (see Sec. \ref{s4}). 
	It is essential to emphasize that \cite{gao2022resilient} also explored this direction. 
	Their switched controlled system is depicted in  Fig.~\ref{fig01}(b), where trajectories originating outside the attraction region are directed towards its boundary before spontaneously converging towards  the single equilibrium. 
	The proposed control, however, addresses a significantly different and more complex problem. 
	We aim to utilize the escaped trajectories (from attraction region) and guide them to appropriate steady points. 
	Therefore, this novel framework faces new technical barriers, 
	namely how to find a feasible control that can recover system to  $\varepsilon-Neighborhood$ of target point $(\bar{n}_1,\bar{n}_2)$ within recovery time $t_f$ under corresponding perturbations, 
	where $\varepsilon$ denotes an arbitrarily small positive number; 
	and $\varepsilon-Neighborhood$ of a state $(\bar{n}_1,\bar{n}_2)$ 
	is described as  
	the set $\{(n_1,n_2):~(n_1-\bar{n}_1)^2+(n_2-\bar{n}_2)^2\leq \varepsilon\}$. 
	Note that the target point $(\bar{n}_1,\bar{n}_2)$ can be any state in multiple metastable states, depending on the size of the perturbation; and system can switch between multiple metastable states in facing with different perturbations. 
	Moreover, the system can also switch between optimal and suboptimal states, if the optimal state satisfies the conditions of the alternative steady states.

	\subsection{Measuring urban traffic resilience} \label{s2.3}
	
	We have depicted the envisioned ecological traffic resilience system and presented the resilience control framework. However, a quantitative indicator for traffic ecological resilience is still absent. It is imperative to establish a quantitative measure for traffic ecological resilience, thereby enhancing a more intuitive perception of the ecological resilience of traffic and facilitating a more straightforward evaluation and reference for control effectiveness.
	
	The quantitative assessment of resilience can  be categorized into two main types: general resilience measure methods and structure-based model-driven resilience measure (\cite{HOSSEINI201647}). We prioritize general resilience measure in this paper, given that the second type is more reliant on specific models. Among general resilience measure, the “resilience triangle” introduced in  ~\cite{bruneau2003framework} has gained notable attention. Originally designed for assessing a community resilience loss during earthquakes, it calculates the resilience loss through a definite integral:
	
	\begin{figure}
		\centering
		\includegraphics[width=10cm]{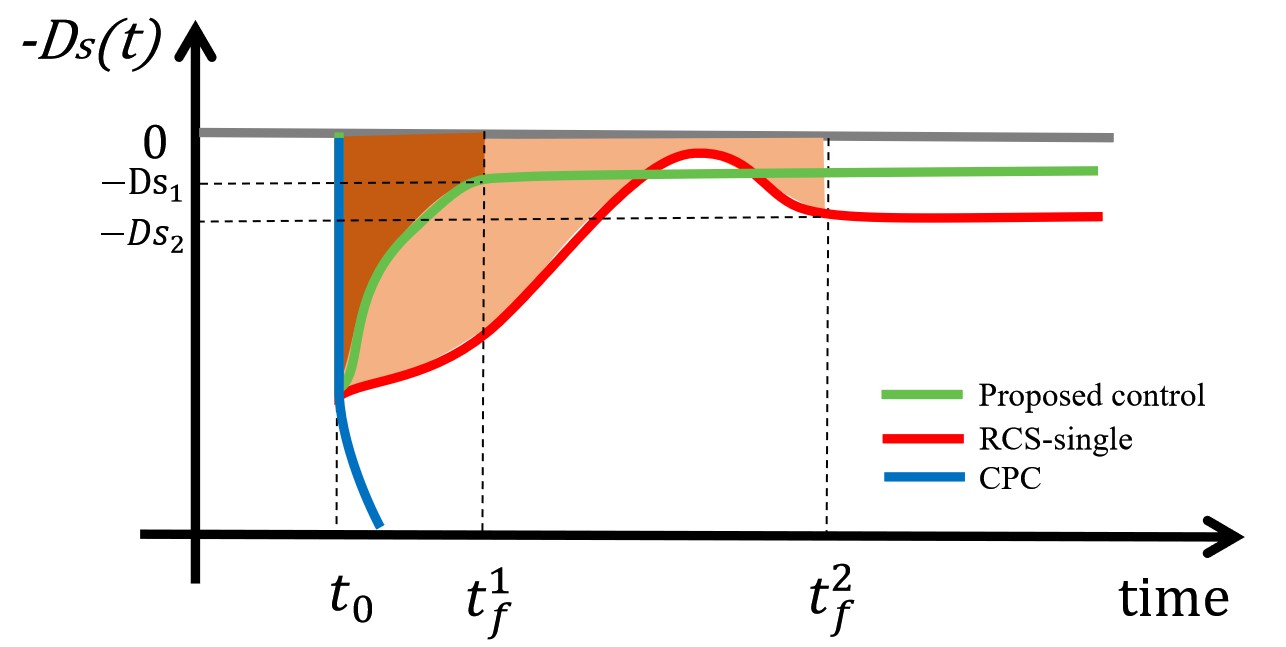}
		\caption{Schematic representation of resilience triangle.  Here we consider hyper-congestion (large $D_s(t_0)$) scenario, and we assume that the original equilibrium is a non-congested equilibrium (not the optimal state). CPC is unable to recover the system, and instead leading to a rapid collapse. Proposed control (RCS-single) recovers the system to a functional level with deviation $-Ds_1$ ($-Ds_2$) from the maximum completion flow at $t_f^1$ ($t_f^2$), having a smaller (larger) resilience loss, depicted as the dark (light) brown shaded region. Note that the resilience loss is the absolute value of the area enclosed by the resilience triangle.
		Proposed control outperforms RCS-single, as proposed control can recover system to a functional state with less deviation ($Ds_1 < Ds_2$) faster ($t_f^1 < t_f^2$).}
		\label{fig31}
	\end{figure}
	
	\begin{equation}\label{cla resilience}
		RL= \int_{t_0}^{t_1}\left( 1-Q(t)\right) \mathrm{d}t,
	\end{equation}
	where $Q(t)$ represents community service quality at time $t$, expressed as a percentage scale from 0 to 100$\%$. $Q(t)$ starts declining at $t_0$ and returns to the normal state at $t_1$. The deviation between the quality of degraded infrastructure and the normal infrastructure quality is quantified through $1-Q(t)$. A higher RL indicates lower resilience, while a lower RL suggests higher resilience.

	In this direction, \cite{gao2022resilient} marks one of the initial attempts to provide a quantitative interpretation of urban traffic resilience. 
		They defined urban traffic resilience as the integral of the absolute deviation between vehicle density (vehicle accumulation) and the optimal vehicle density during the recovery period.
		However, the defined resilience measure lacks an evident physical interpretation. 
		Thus, we revise the definition of deviation $D_s(t)$ as the difference between the completion flow $G_i(n_i) (i=1,2)$ and the maximum completion flow $G_{i,max}$. That is:
		\begin{equation}\label{congestion deviation}
			D_s(t)=\left| G_1(n_1)-G_{1,max}\right| +\left| G_2(n_2)-G_{2,max}\right|.
		\end{equation}
		Note that,  the unit of $D_s(t)$ is $[veh/h]$.
	Based on the defined $D_s(t)$, we can define the ecological traffic resilience measure as follows:
	\begin{equation}\label{tran resilience}
		R=\int_{t_0}^{t_f}\left( 0-D_s(t)\right) \mathrm{d}t= -\int_{t_0}^{t_f}D_s(t)  \,\mathrm{d}t.
	\end{equation}
	Here $t_f$ is the time when system recovers and stabilizes at functional steady states. Formula (\ref{tran resilience}) implies that $R$ is negative. Thus, $-R$ corresponds to the area of the resilience triangle, depicted as the dark brown shaded region in \ref{fig31}.
	Noteworthily, the unit of resilience measure $R$ is $[veh]$, and it denotes the loss of completed trips compared to the maximum potential number of trips that could have been completed during the recovery period.

	We depict a schematic diagram of the resilience triangle under three different controls, as shown in Fig. \ref{fig31}.  
		Notably, under hyper-congestion, CPC fails to recover system and  instead rapidly collapses system; 
		while RCS-single and the proposed control can ensure recoverability. 
		The resilience control proposed in this paper outperforms RCS-single, as it enables the system to recover to an alternative stable states with higher completion flow within a shorter time.
		Thus, the resilience triangle (loss) under our proposed control method (the dark brown area in Fig. \ref{fig31}) is smaller than that of RCS-single (the light brown area).
		Note that the original equilibrium is assumed to be a stable non-congested state, rather than an optimal state.


	\section{Stability analysis and spontaneous attraction region estimation for multi-equilibria system} \label{s3}
	In the previous section, for multi-equilibria system, we defined ecological resilience of urban traffic with an MFD-based representation. In this section, we will derive the global phase portrait and the spontaneous attraction region for the multi-equilibria system under CPC ($u_1$ and $u_2$). Sec. \ref{s3.1} discusses and verifies the local stability of the MFD dynamic; Sec. \ref{s3.2} derives and illustrates the global phase portrait; and Sec. \ref{s3.3} presents the estimation of spontaneous attraction region that are used to design the proposed control in  Sec.~\ref{s4}. Note that multiple equilibria systems include both two-equilibrium systems and four-equilibrium systems. According to \cite{gao2022resilient},	two equilibria exist if and only if  Condition $(\mathbb{K}^{2a})$:    $\frac{{u_2}{{d_2}}+d_1}{1-{u_1}{u_2}}=G_{1,max}$ $\wedge$ $\frac{{u_1}{{d_1}}+d_2}{1-{u_1}{u_2}}<G_{2,max}$
	or Condition $(\mathbb{K}^{2b})$: $\frac{{u_2}{{d_2}}+d_1}{1-{u_1}{u_2}}<G_{1,max}$ $\wedge$ $\frac{{u_1}{{d_1}}+d_2}{1-{u_1}{u_2}}=G_{2,max}$ holds;
	four equilibria exist if and only if Condition $(\mathbb{K}^{4})$: $\frac{{u_2}{{d_2}}+d_1}{1-{u_1}{u_2}}<G_{1,max}$ $\wedge$  $\frac{{u_1}{{d_1}}+d_2}{1-{u_1}{u_2}}<G_{2,max}$ holds. 
	We in this section mainly present the theoretical analysis for two-equilibria system. The theoretical analysis for four-equilibria system can be found in App \ref{A}.
	
	\subsection{Local stability verification}\label{s3.1}
	Consider the two-equilibria case with Condition $(\mathbb{K}^{2a})$. 
	According to Condition $(\mathbb{K}^{2a})$, we have:
	\begin{equation}\label{d1d2 in case2}
		\begin{split}
			&d_1=\left( 1-u_1u_2\right) G_{1,max}-{u_2}d_2,\\
			&d_2< G_{2,max}-{u_1}G_{1,max}.
		\end{split}
	\end{equation}
	Bringing the first equation of \eqref{d1d2 in case2} into system \eqref{the model}, the system dynamics become:
	\begin{eqnarray}\label{dynamics systems 3}
		\frac{\mathrm{d}{n}_1(t)}{\mathrm{d}t}
		&=&-{u_2}{{a_2}{(n_2(t)-\frac{p_2}{2})^2}}+{{a_1}{(n_1(t)-\frac{p_1}{2})^2}}
		+u_2\left(\frac{a_2p^2_2}{4}-u_1\frac{a_1p^2_1}{4}-d_2 \right),\nonumber\\
		\frac{\mathrm{d}{n}_2(t)}{\mathrm{d}t}
		&=&-{u_1}{{a_1}{(n_1(t)-\frac{p_1}{2})^2}}+{{a_2}{(n_2(t)-\frac{p_2}{2})^2}}
		+\left(\frac{u_1a_1p^2_1}{4}-\frac{a_2p^2_2}{4}+d_2\right).
	\end{eqnarray}
	For systems \eqref{dynamics systems 3}, 
	the two equilibria can be denoted as  $P^{2a}_k=(\frac{p_1}{2},p^{2a}_k)$ ($k=1,2$),
	where $p^{2a}_k=\frac{{p_2}+(-1)^k\times M}{2}$, $M=\sqrt{{p_2}^2-\frac{4{d_2} + {u_1}a_1p^2_1}{{a_2}}}$. 
	Upon substituting Formula \eqref{d1d2 in case2} into Formula \eqref{the model}, the system \eqref{the model} reduces by one variable, $d_1$, significantly simplifying our analysis. However, Formula \eqref{dynamics systems 3} remains cumbersome. To further simplify the system for subsequent analysis, we shift equilibrium point $P^{2a}_1$ of the system \eqref{dynamics systems 3} to $(0,0)$. That is, letting $x_1(t)= n_1(t)-\frac{p_1}{2}$ and $x_2(t)= n_2(t)-p^{2a}_1$, we can simplify system \eqref{dynamics systems 3} as:
	\begin{eqnarray}\label{sim dynamics systems 3}
		\frac{\mathrm{d}{x}_1(t)}{\mathrm{d}t}&=&H^{2a}_1({x_1}(t),{x_2}(t))
		:={u_2}{a_2}M{x_2}(t) +\left( {{a_1}{x_1}^2(t)}-{u_2}{{a_2}{x_2}^2(t)}\right),\nonumber\\
		\frac{\mathrm{d}{x}_2(t)}{\mathrm{d}t}&=&H^{2a}_2({x_2}(t),{x_2}(t))
		:=-{a_2}M{x_2}(t)+\left( -{u_1}{{a_1}{x_1}^2(t)}+{{a_2}{x_2}^2(t)}\right),
	\end{eqnarray}
	
	The two equilibria for the new system \eqref{sim dynamics systems 3} are  $\hat{P}^{2a}_1=(0,0)$ and $\hat{P}^{2a}_2=(0,M)$. 
	Noteworthy, system \eqref{sim dynamics systems 3} possesses the same properties as system \eqref{dynamics systems 3} since system \eqref{sim dynamics systems 3} is obtained by shifting system \eqref{dynamics systems 3}. Due to the simpler formulation of system \eqref{sim dynamics systems 3}, we in the following use system \eqref{sim dynamics systems 3} to derive local stability characteristics, global phase portraits, and attraction region estimations. The properties obtained are equally valid for system \eqref{dynamics systems 3}. Therefore, all our propositions and theorems are summarized for system \eqref{dynamics systems 3}.
	
	
	From the aforementioned conversion, we derive the following proposition for its two equilibria.
	
	\begin{proposition}\label{saddle-node point}
		Under Condition $(\mathbb{H})$, the two equilibria $P^{2a}_1$ and $P^{2a}_2$ of system \eqref{dynamics systems 3} are both saddle-node points.
	\end{proposition}
	
	Proof: We have that
	$P^{2a}_1$ and $P^{2a}_2$ of system \eqref{dynamics systems 3} are equivalent to $\hat{P}^{2a}_1$ and $\hat{P}^{2a}_2$ of system \eqref{sim dynamics systems 3}, thus we consider the stability of $\hat{P}^{2a}_1$ and $\hat{P}^{2a}_2$. For $\hat{P}^{2a}_1$, we have 
	$\left(
	\begin{array}{ccccc}
		\frac{\partial H^{2a}_1(x_1,x_2)}{\partial x_1} & \frac{\partial H^{2a}_1(x_1,x_2)}{\partial x_2} \\
		\frac{\partial H^{2a}_2(x_1,x_2)}{\partial x_1} & \frac{\partial H^{2a}_2(x_1,x_2)}{\partial x_2}\\	\end{array}	\right)\Bigg|_{(0,0)}=\left(
	\begin{array}{ccccc}
		0 & {u_2}{a_2}M \\
		0 & -{a_2}M \\
	\end{array}
	\right)$, 
	indicating that:
	$$ Rank \left\lbrace \left(
	\begin{array}{ccccc}
		\frac{\partial H^{2a}_1(x_1,x_2)}{\partial x_1} & \frac{\partial H^{2a}_1(x_1,x_2)}{\partial x_2} \\
		\frac{\partial H^{2a}_2(x_1,x_2)}{\partial x_1} & \frac{\partial H^{2a}_2(x_1,x_2)}{\partial x_2}\\	\end{array}	\right)\right\rbrace\Bigg|_{(0,0)}=1,$$
	and 
	$$div(H^{2a}_1,H^{2a}_2) |(0,0)=\frac{\partial H^{2a}_1(0,0)}{\partial x_1}+\frac{\partial H^{2a}_2(0,0)}{\partial x_2} \ne 0.$$
	
	According to Theorem 7.1 in \cite{zhang2006qualitative}, we can analyze the nature of equilibrium points  $\hat{P}^{2a}_1$ by calculating the bifurcation function $B(x_1(t))$ (defined in \cite{zhang2006qualitative}), yields:
	$$B(x_1(t))=H^{2a}_1(x_1,x_2(x_1))=H^{2a}_1(x_1,\pm\sqrt{\frac{u_1a_1x^2_1}{a_2}+\frac{M^2}{4}}+\frac{M}{2})=(1-u_1u_2)x^2_1(t),$$
	where $x_2(x_1)=\pm\sqrt{\frac{u_1a_1x^2_1}{a_2}+\frac{M^2}{4}}+\frac{M}{2}$ is obtained by  letting $H^{2a}_2(x_1,x_2)=0$.
	Since Condition $(\mathbb{H})$ holds, we have $B(x_1(t))>0$ holds for all $x_1(t) \in \left[-\frac{p_1}{2}~ \frac{p_1}{2}\right]$ except $x_1(t)=0$. Moreover, the power exponent of function $B(x_1(t))$ is 2. Thus, $(0,0)$ is a saddle-node point~(see \cite{zhang2006qualitative} Theorem 7.1(iii)), indicating that $P^{2a}_1$ is a saddle-node point. Similarly, we can prove that $P^{2a}_2$ is a saddle-node point. $\hfill \square$
	
	Proposition \ref{saddle-node point} reveals that, for system \eqref{the model} under $(\mathbb{K}^{2a})$, the two equilibria $P^{2a}_1$ and $P^{2a}_2$ are both saddle-node point. Similarly, for system \eqref{the model} under $(\mathbb{K}^{2b})$, we can also prove that two equilibria $P^{2b}_1$ and $P^{2b}_2$ are both saddle-node point. 	These results will help us to reveal the local stability of the two equilibria, which will assist in subsequent derivations of attraction region, providing  foundation for understanding the global properties of the system.
	
    For the four-equilibria system, we also try to capture the local stability through linearization, and the detailed analysis can be found in App. \ref{A.1}.  In App. \ref{A.1}, we denote the four equilibria as  $P^4_m$ ($m=1,2,3,4$), and we have that $P^4_1$ is a locally stable node, $P^4_2$ and $P^4_3$ are saddle points, $P^4_4$ is an unstable node, which can be summarized as Proposition \ref{stability in 4} in App. \ref{A.1}.

	\subsection{Global phase portrait derivation}\label{s3.2}
	To capture more properties of the system, we aim to delve further into its global nature. Global analysis of dynamical systems is often intricate, lacking a unified analytical approach. For MFD dynamics, we overcome this gap by first dividing the phase space into several subregions using demarcation lines, followed by delineating the trajectory trends within each subregion, and ultimately analyzing and deriving the global phase portrait leveraging the trajectory trends in each subregion. For simplicity in analysis, we continue to consider the more straightforward formulation of Formula \eqref{sim dynamics systems 3}. Note that, the aforementioned demarcation lines are obtained by solving  $\frac{\mathrm{d}{x}_1(t)}{\mathrm{d}t}$=0 or $\frac{\mathrm{d}{x}_2(t)}{\mathrm{d}t}$=0 respectively. The trajectory trends in each region are determined by  $\frac{\mathrm{d}{x}_1(t)}{\mathrm{d}t}$ and $\frac{\mathrm{d}{x}_2(t)}{\mathrm{d}t}$. For instance, if $\frac{\mathrm{d}{x}_1(t)}{\mathrm{d}t}$ > 0 ($\frac{\mathrm{d}{x}_2(t)}{\mathrm{d}t}$> 0), the trajectories move to the right (upward), while if $\frac{\mathrm{d}{x}_1(t)}{\mathrm{d}t}$< 0 ($\frac{\mathrm{d}{x}_2(t)}{\mathrm{d}t}$< 0), the trajectories move to the left (downward). Furthermore, considering that the demarcation lines  $\frac{\mathrm{d}{x}_1(t)}{\mathrm{d}t}$=0 or $\frac{\mathrm{d}{x}_2(t)}{\mathrm{d}t}$=0 take different shapes under four distinct parameter conditions, we will separately examine these under four conditions.

\begin{figure}
		\centering
		\begin{subfigure}{.35\textwidth}
			\centering
			\includegraphics[width=5.5cm]{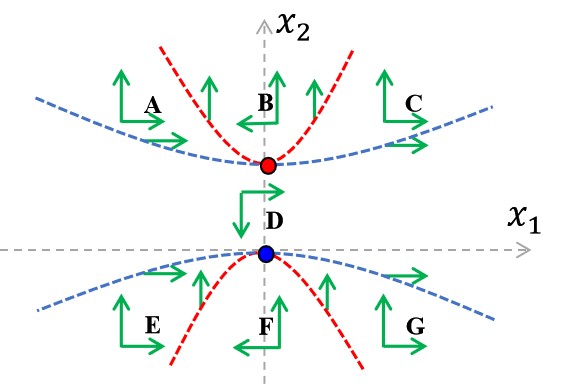}
			\caption{$\mathbb{K}^{2a} \wedge \mathbb{H}_1$}
			\label{fig03.1}
		\end{subfigure}
		\begin{subfigure}{.35\textwidth}
			\centering
			\includegraphics[width=5.5cm]{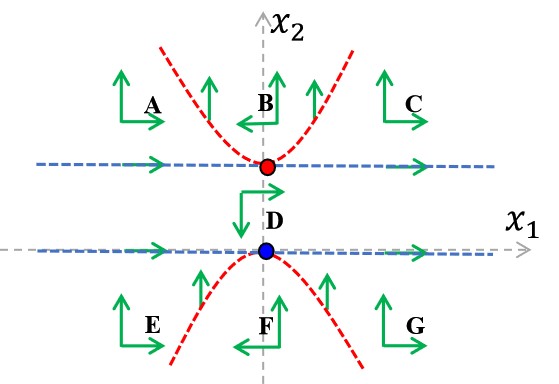}
			\caption{$\mathbb{K}^{2a} \wedge \mathbb{H}_2$}
			\label{fig03.2}
		\end{subfigure}
		\begin{subfigure}{.35\textwidth}
			\centering
			\includegraphics[width=5.5cm]{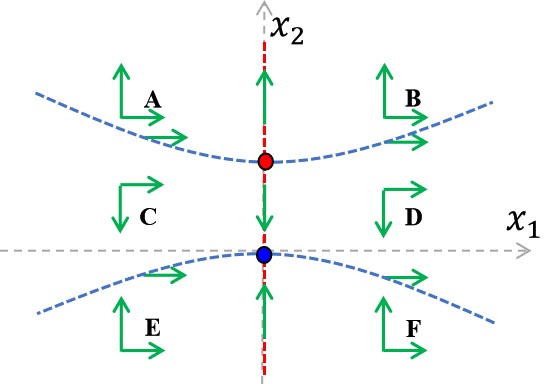}
			\caption{$\mathbb{K}^{2a} \wedge \mathbb{H}_3$}
			\label{fig03.3}
		\end{subfigure}
		\begin{subfigure}{.35\textwidth}
			\centering
			\includegraphics[width=5.5cm]{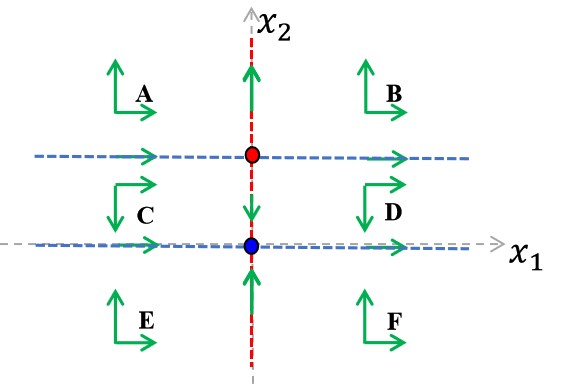}
			\caption{$\mathbb{K}^{2a} \wedge \mathbb{H}_4$}
			\label{fig03.4}
		\end{subfigure}
		\caption{Theoretical schematic phase portraits of the two equilibria for system \eqref{sim dynamics systems 3} under Condition $(\mathbb{K}^{2a} \wedge \mathbb{H})$. Red circle: undesired saddle-node equilibrium. Blue circle: desired saddle-node equilibrium. Red dotted curves: $\dot{x}_1=0$, blue dotted curves: $\dot{x}_2=0$, green arrows: trajectory directions. }
		\label{fig03}
	\end{figure}

	First consider Condition $(\mathbb{H}_1)$: $0< u_1,~u_2 \leq 1 \wedge u_1u_2 \neq 1$. Under this condition, both demarcation lines manifest as hyperbolas, illustrated by the red and blue dotted curves in Fig. \ref{fig03.1}, with the expressions $\frac{\left( x_2-\frac{M}{2}\right)^2}{{\frac{M^2}{4}}}-\frac{x^2_1}{\frac{u_2 a_2 M^2}{4a_1}}=1$
	and $\frac{\left( x_2+\frac{M}{2}\right) ^2}{{\frac{M^2}{4}}}-\frac{x^2_1}{\frac{ a_2 M^2}{4 u_1 a_1}}=1$. 
	In addition, we have $\frac{\mathrm{d}{x}_1(t)}{\mathrm{d}t}=0$ and $\frac{\mathrm{d}{x}_2(t)}{\mathrm{d}t}= \left(1-u_1 u_2\right) {a_2}{x_2(t)}\left( {x_2(t)-M}\right)>0$ on the red dotted curves except for $(0,0)$ and $(0,M)$;
	and we have $\frac{\mathrm{d}{x}_2(t)}{\mathrm{d}t}=0$ and
	$\frac{\mathrm{d}{x}_1(t)}{\mathrm{d}t}=  \frac{1-u_1 u_2}{u_1} {a_2}{x_2(t)}\left( {x_2(t)-M}\right) >0$ on the blue dotted curves except for $(0,0)$ and $(0,M)$.
	It is worth mentioning that, since $\frac{u_2 a_2 M^2}{4a_1}<\frac{a_2 M^2}{4 u_1 a_1}$ holds,  
	the location of these four curves are is uniquely determined, as shown in Fig.~\ref{fig03.1}. 
	Then,  the plane $\mathbb{R}^2$ can be divided into seven unbounded regions by these four curves.
	Moreover, 
	Table \ref{tab2} summarizes the symbols of $\frac{\mathrm{d}{x}_i(t)}{\mathrm{d}t}$ ($i=1,2$) in these seven regions. 
	On the basis of Table \ref{tab2},
	the phase portrait of \eqref{sim dynamics systems 3} under Condition $(\mathbb{H}_1)$ can be obtained, as shown in Fig.~\ref{fig03.1}.
	
	\begin{table}[H]
		\caption{Symbols of the seven regions for system \eqref{sim dynamics systems 3} under Condition $(\mathbb{H}_1)$.}
		\label{tab2}
		\centering
		\fontsize{6.5}{8}\selectfont
		\setlength{\tabcolsep}{4.5mm}{
			\begin{threeparttable}
				\begin{tabular}{cccccccc}
					\toprule
					\multirow{2}{*}{Derivative}&
					\multicolumn{7}{c}{Symbol in region}\cr
					\cmidrule(lr){2-8}
					&A&B&C&D&E&F&G\cr
					\midrule
					$\frac{\mathrm{d}{x}_1(t)}{\mathrm{d}t}$&$+$&$-$&$+$&$+$&$+$&$-$&$+$\cr
					$\frac{\mathrm{d}{x}_2(t)}{\mathrm{d}t}$&$+$&$+$&$+$&$-$&$+$&$+$&$+$\cr
					\bottomrule
				\end{tabular}
		\end{threeparttable}}
	\end{table}
	
	Subsequently, we can verify that regions B and E are both positive invariant sets by utilizing proof by contradiction. Given that the trajectories starting from B and E will not escape, we can further obtain that there exist no close orbits for system \eqref{sim dynamics systems 3} under Condition $(\mathbb{H}_1)$.

	When Condition $(\mathbb{H}_2)$: $u_1=0 \wedge 0< u_2 \leq 1$, $(\mathbb{H}_3)$: $0< u_1\leq 1 \wedge u_2 =0$ or $(\mathbb{H}_4)$: $u_1=0 \wedge u_2 =0$ hold, 
	the two demarcation lines can manifest as one hyperbola and one straight line (as shown in Fig.~\ref{fig03.2} and \ref{fig03.3}) or as two straight lines (as shown in Fig.~\ref{fig03.4}). After obtaining the demarcation lines, we follow similar steps as above. As such, we verify that no close orbits exist under these conditions, and the corresponding phase portraits are illustrated  in Fig.~\ref{fig03.2}, \ref{fig03.3} and \ref{fig03.4}. Similarly, we can also obtain the phase portrait for system \eqref{the model} under Condition $(\mathbb{K}^{2b})$, and there also exists no close orbit. 
	
	For the four-equilibria system, we also try to capture the global phase portraits, and the detailed analysis can be found in App. \ref{A.2}. Note that the steps of global phase portrait derivation differ from that used for two-equilibrium system. Due to the increased number of equilibria, the combination of demarcation lines becomes intricately complex. On one hand, variations in demarcation line types (e.g., two hyperbolas, one hyperbola and one straight line, two straight lines, etc.) lead to different phase portraits. On the other hand, 
    different relative positions of demarcation lines (e.g., whether the foci of two hyperbolas lie on the same axis) add another layer of complexity. The method for two equilibria are insufficient to address the challenges posed by four equilibria. Furthermore, the increased number of equilibrium points also hinders us from directly determining the locations of corresponding separatrices. 
	New methods are required to capture the global phase portrait. We solve these in App. \ref{A.2} by subdividing Condition $(\mathbb{K}^{4})$ into five new distinct conditions. That is,
		Condition $(\hat{\mathbb{K}}^{4}_1)$: $M_1<0\wedge M_2<0$; Condition $(\hat{\mathbb{K}}^{4}_2)$: $M_1=0\wedge M_2<0$; Condition $(\hat{\mathbb{K}}^{4}_3)$: $M_1<0\wedge M_2=0$; Condition $(\hat{\mathbb{K}}^{4}_4)$: $M_1>0 \wedge M_2<\min\left\lbrace {-\frac{M_1}{u_2}}, {-u_1M_1}\right\rbrace$;
		Condition $(\hat{\mathbb{K}}^{4}_5)$: $M_1<\min\left\lbrace {-u_2M_2},{-\frac{M_2}{u_1}} \right\rbrace \wedge M_2>0$, where
		$M_1=d_1 +u_2 \frac{{a_2 p_2}^2}{4}-\frac{{a_1 p_1}^2}{4}$; and $M_2=d_2 +u_1 \frac{{a_1 p_1}^2}{4}-\frac{{a_2 p_2}^2}{4}$. 
		The five conditions mentioned above correspond to different positions of demarcation lines.
		In App. \ref{A.2}, by discussing the shape of critical demarcation lines under each condition, we have summarized  20 distinct phase portraits for four-equilibria system and verifies that no closed orbit exists.

	\subsection{Spontaneous attraction region estimation}\label{s3.3}
	
	In the following, for two-equilibria system \eqref{dynamics systems 3}, 
	the attraction region for the unstable saddle-node point $P^{2a}_1$ will be derived.  Note that, \cite{gao2022resilient} developed a method to find certain special trajectories, e.g., $x_2(t)=qx_1(t)$ for system with single-equilibrium. This form of special solution cannot apply here for multi-equilibrium systems. In this case, the theoretical attraction region boundaries are identified by finding certain separatrices of the equivalent systems \eqref{sim dynamics systems 3}, 
	and the inner and outer attraction regions can be obtained by finding the positive invariant sets, under the aforementioned four conditions respectively.

\begin{figure}
		\centering
		\begin{subfigure}{.33\textwidth}
			\centering
			\includegraphics[width=5cm]{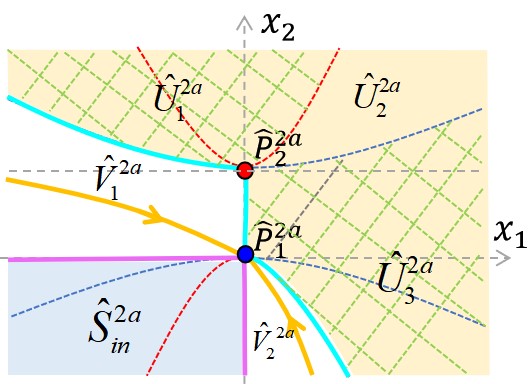}
			\caption{$\mathbb{K}^{2a} \wedge \mathbb{H}_1$}
			\label{fig04.1}
		\end{subfigure}
		\begin{subfigure}{.33\textwidth}
			\centering
			\includegraphics[width=5cm]{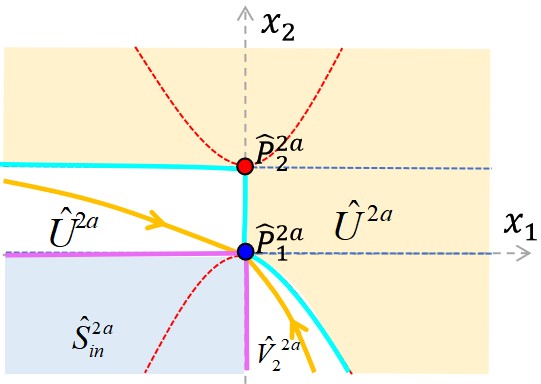}
			\caption{$\mathbb{K}^{2a} \wedge \mathbb{H}_2$}
			\label{fig04.2}
		\end{subfigure}
		\begin{subfigure}{.33\textwidth}
			\centering
			\includegraphics[width=5cm]{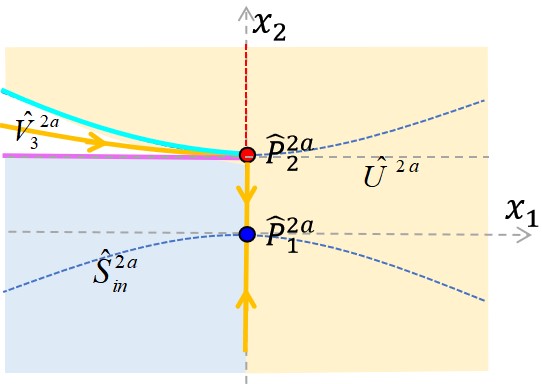}
			\caption{$\mathbb{K}^{2a} \wedge \mathbb{H}_3$}
			\label{fig04.3}
		\end{subfigure}
		\begin{subfigure}{.33\textwidth}
			\centering
			\includegraphics[width=5cm]{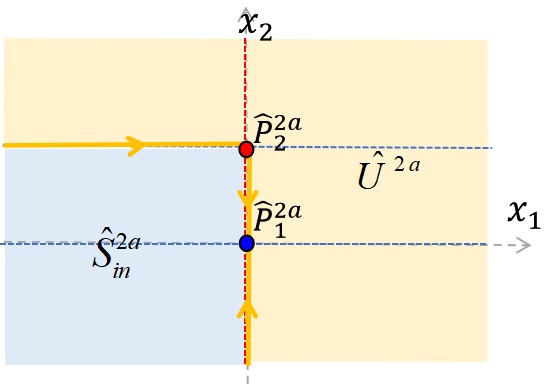}
			\caption{$\mathbb{K}^{2a} \wedge \mathbb{H}_4$}
			\label{fig04.4}
		\end{subfigure}
		\caption{Certain separatrices of $\hat{P}^{2a}_1$=$(0,0)$ or $\hat{P}^{2a}_2$=$(0,M)$, and the inner and outer estimations of attraction regions for system \eqref{sim dynamics systems 3}. Red dotted curves $\dot{x}_1=0$; blue dotted curves: $\dot{x}_2=0$; yellow curves: separatrices, also serving as boundaries of attraction region; purple lines: boundary of the inner estimation of attraction region; bright blue lines: boundary of the outer estimation of attraction region. The green dashed grid is to distinguish $\hat{U}^{2a}_1$, $ \hat{U}^{2a}_2$ and $\hat{U}^{2a}_3$.}
		\label{fig04}
	\end{figure}

	Firstly, we consider Condition $(\mathbb{H}_1)$. As shown in Fig.~\ref{fig04.1}, 
	the two red dotted demarcation curves  
	can be  denoted as $x_2=\hat{r}^{2a}_{i}(x_1)$ ($i=1,2$), where $\hat{r}^{2a}_{i}(x_1)=(-1)^{i}\sqrt{\frac{a_1}{u_2a_2}x_1^2+\frac{M^2}{4}}+\frac{M}{2}$. 
	Similarly,  
	the two blue dotted demarcation lines  
	can be denoted as $x_2=\hat{b}^{2a}_{i}(x_1)$ ($i=1,2$), where $\hat{b}^{2a}_{i}(x_1)=(-1)^{i}\sqrt{\frac{u_1a_1}{a_2}x_1^2+\frac{M^2}{4}}+\frac{M}{2}$. 
	Obviously, the vertexes of the demarcation lines are exactly the two equilibria $\hat{P}^{2a}_1=(0,0)$ and $\hat{P}^{2a}_2=(0,M)$.
	
	Assuming the lower blue dotted line $x_2=\hat{b}^{2a}_{1}(x_1)$ intersects $x_1=-k_1$ with $k_1>0$ at $(-k_1,x^1_D)$ and $(-k_1,x^2_D)$, where $x^1_D< x^2_D$. 
	The trajectory starting from $(-k_1,x^1_D)$ and $(-k_1,0)$ will go into $\hat{P}^{2a}_1$ in E as $t\rightarrow +\infty$; 
	and the trajectory starting from $(-k_1,x^2_D)$ will escape A from the upper boundary of A and stay in B moving away from $\hat{P}^{2a}_1$ as $t\rightarrow +\infty$. 
	Thus, due to the continuous dependence theorem of solutions on initial conditions~(\cite{zhang2006qualitative}),
	there exists a $D^\ast$ with $0<D^\ast<x^2_D$, such that the trajectory starting from any $(-k_1,x_D)$ with $0<x_D\leq D^\ast$ will enter E and then go to $\hat{P}^{2a}_1$ in E as $t\rightarrow +\infty$, 
	and the trajectory starting from any $(-k_1,x_D)$ with $D^\ast<x_D\leq x^2_D$ will enter A, 
	then escape A from the upper boundary of A and stay in B moving away from $\hat{P}^{2a}_1$ as $t\rightarrow +\infty$. 
	Thus, the trajectory starting from  $(-k_1,x_{D^\ast})$, donated as $\hat{\Phi}^1(\hat{P}^{2a}_1)$= $\hat{\Phi}^{2a,1}_t(-k_1,x_{D^\ast})$, is a stable separatrix of $\hat{P}^{2a}_1$, as shown in Fig.~\ref{fig04.1}.
	Similarly,
	assuming the lower red dotted line $x_2=\hat{r}^{2a}_{1}(x_1)$ intersect $x_2=-k_2$ with $k_2>0$ at $(x^1_F,-k_2)$ and $(x^2_F,-k_2)$. 
	By similar analysis, we can obtain that
	there exists an  $F^\ast$ with $0<F^\ast<x^2_F$, 
	such that the trajectory starting from any $(x_F,-k_2)$ with $0<x_F\leq F^\ast$ will enter E and then go to $\hat{P}^{2a}_1$ in E as $t\rightarrow +\infty$, 
	and the trajectory starting from any $(x_F,-k_2)$ with $F^\ast<x_F\leq x^2_F$ will enter G, 
	then escape G from the upper boundary of G and stay in D moving away from $\hat{P}^{2a}_1$ as $t\rightarrow +\infty$. 
	Thus, the trajectory starting from  $(x_{F^\ast},-k_2)$, 
	donated as $\hat{\Phi}^2(\hat{P}^{2a}_1)=\hat{\Phi}^{2a,2}_t(x_{F^\ast},-k_2)$, 
	is a stable separatrix of $\hat{P}^{2a}_1$, as shown in Fig.~\ref{fig04.1}.
	
	Subsequently,
	combining with phase portrait  Fig.~\ref{fig03.1}, we can verify that the region $\hat{S}^{2a}_{in}$  is a positive invariant set utilizing proof by contradiction, where $\hat{S}^{2a}_{in}=\left\{ (x_1,x_2)|x_i \leq 0,\ i=1,2 \right\}$, depicted as the light blue zone  in Fig.~\ref{fig04.1}.
	Or else, there will be at least one trajectory originated from $\hat{S}^{2a}_{in}$, 
	and enter $\hat{V}^{2a}_1$ (or $\hat{V}^{2a}_2$) through the upper boundary (or right boundary) of $\hat{S}^{2a}_{in}$, 
	where $\hat{V}^{2a}_1=\left\{ (x_1,x_2)|x_1 < 0,~ 0 < x_2 < \hat{b}^{2a}_{2}(x_1) \right\}$ 
	and $\hat{V}^{2a}_2=\left\{ (x_1,x_2)|x_1 > 0,~ x_2 < \hat{r}^{2a}_{1}(x_1) \right\}$. 
	Obviously, it is contradictory with the trajectory direction in phase portrait \ref{fig03.1}. 
	Further, joining Fig.~\ref{fig04.1} with  Fig.~\ref{fig03.1}, 
	we have that: 
	the trajectory starting from any point in $D \wedge \hat{S}^{2a}_{in}$ will escape D from the lower boundary of D and entering E,
	and the trajectory starting from any point in $F \wedge \hat{S}^{2a}_{in}$ will escape F from the left boundary of F and entering E.
	Since E is a positive invariant set, and we have $\dot{x}_1>0$ and $\dot{x}_2>0$ in E, the trajectory starting from any point in E will go to $\hat{P}^{2a}_1$ as $t\rightarrow +\infty$. 
	Thus, the trajectory starting from any point in the positive invariant set $\hat{S}^{2a}_{in}$  will go to $\hat{P}^{2a}_1$ as $t\rightarrow +\infty$, implying 
	$\hat{S}^{2a}_{in} \subset \mathcal{R}(\hat{P}^{2a}_1)$,  i.e., $\hat{S}^{2a}_{in}$ is an inner estimation of attraction region~(\cite{zheng2018inner,wang2020inner}) for $\hat{P}^{2a}_1$. 
	
	Similarly, 
	region $\hat{U}^{2a}=\hat{U}^{2a}_1 \vee \hat{U}^{2a}_2 \vee \hat{U}^{2a}_3$ (the light yellow region shown in Fig.~\ref{fig04.1}) is also a positive invariant set, 
	where 
	\begin{eqnarray*}
		\hat{U}^{2a}_{1}&=&\left\{ (x_1,x_2)|\left(x_1 < 0\wedge x_2 \geq \hat{b}^{2a}_{2}(x_1)\right) \vee \left(x_1 > 0\wedge x_2 \geq \hat{r}^{2a}_{2}(x_1) \right)  \right\},\\
		\hat{U}^{2a}_{2}&=&\left\{ (x_1,x_2)| x_1 > 0\wedge \hat{b}^{2a}_{2}(x_1) \leq x_2 \leq \hat{r}^{2a}_{2}(x_1) \right\},\\
		\hat{U}^{2a}_{3}&=&\left\{ (x_1,x_2)| x_1 \geq 0 \wedge \hat{r}^{2a}_{1}(x_1)\leq x_2 <\hat{b}^{2a}_{2}(x_1) \right\}/(0,0).
	\end{eqnarray*}
	Note that  $\hat{U}^{2a}_1$ and $\hat{U}^{2a}_3$ are also positive invariant sets. 
	In addition, the trajectory starting from point in $\hat{U}^{2a}_2$ will escape from the upper boundary of $\hat{U}^{2a}_2$ and enter $\hat{U}^{2a}_1$ moving far from $\hat{P}^{2a}_1$, 
	or escape from the lower boundary of $\hat{U}^{2a}_2$ and enter $\hat{U}^{2a}_3$ moving far from $\hat{P}^{2a}_1$. 
	Thus, we can obtain that $\mathcal{R}(\hat{P}^{2a}_1) \subset \hat{S}^{2a}_{out}=\mathbb{R}^2 \setminus \hat{U}^{2a}$, i.e., the region $\hat{S}^{2a}_{out}$ is an outer estimation of attraction region for $\hat{P}^{2a}_1$.
	Further,
	we can obtain that 
	$\hat{\Phi}^1(\hat{P}^{2a}_1)\in\hat{V}^{2a}_1$ and $\hat{\Phi}^2(\hat{P}^{2a}_1)\in\hat{V}^{2a}_2$, 
	as the yellow lines shown in Fig.~\ref{fig04.1}.
	
	When Condition $(\mathbb{H}_2)$ hold, we can obtain two stable separatrices $\hat{\Phi}^1(\hat{P}^{2a}_1)$ and $\hat{\Phi}^2(\hat{P}^{2a}_1)$ of $\hat{P}^{2a}_1$;
	the inner estimation of attraction region $\hat{S}^{2a}_{in}$;
	and outer estimation of attraction region $\hat{S}^{2a}_{out}$  by similar analysis.
	Furthermore, we have   
	$\hat{\Phi}^1(\hat{P}^{2a}_1)\in\hat{V}^{2a}_1$ and $\hat{\Phi}^2(\hat{P}^{2a}_1)\in\hat{V}^{2a}_2$, 
	as shown in Figs.~\ref{fig04.2}. 
	When $(\mathbb{H}_3)$ or $(\mathbb{H}_4)$ hold, we can first simplify the system \eqref{sim dynamics systems 3} by bringing in  corresponding conditions. Then, by repeating the above analysis, 
	we can obtain corresponding results, as shown in Figs. \ref{fig04.3} and \ref{fig04.4}. 
	
	Consequently, for the system \eqref{sim dynamics systems 3}, the estimation of the attraction region for the saddle-node points $\hat{P}^{2a}_1$ has been obtained.
	Further,
	letting $n_1(t)= x_1(t)+\frac{p_1}{2}$ and $n_2(t)= x_2(t)+\frac{p_2-M}{2}$,
	the estimation of attraction region for $P^{2a}_1$ can be obtained. 
	Denoting 
	$b^{2a}_{i}(n_1)=\hat{b}^{2a}_{i}(n_1-\frac{p_1}{2})+\frac{p_2-M}{2}$ and $r^{2a}_{i}(n_1)=\hat{r}^{2a}_{i}(n_1-\frac{p_1}{2})+\frac{p_2-M}{2}$, where $i=1,2$, 
	we can obtain the theorem as below.
	
	\begin{theorem}\label{attraction region in 3}
		For the unstable saddle-node point $P^{2a}_1$, under Condition $(\mathbb{H})$, we have $S^{2a}_{in} \subseteq S^{2a}_A=\mathcal{R}(P^{2a}_1) \subseteq S^{2a}_{out}$, where the attraction region $S^{2a}_A$, the inner estimation of attraction region $S^{2a}_{in}$ and the outer estimation of attraction region $S^{2a}_{out}$ are defined as follows:
		
		\begin{enumerate}[(1)]
			\item Under Conditions $(\mathbb{H}_1)$ and $(\mathbb{H}_2)$, 
			$S^{2a}_A=\left\{ (n_1,n_2)| n_1 \leq g^{2a}_2(n_2),~ n_2 \leq g^{2a}_1(n_1) \right\}$, 
			where 
			$n_2 = g^{2a}_1(n_1)$ and $n_1 = g^{2a}_2(n_2)$ denotes the two stable separatrices $\Phi^1({P}^{2a}_1)$ and $\Phi^2(P^{2a}_1)$ of $P^{2a}_1$, respectively;
			
			$S^{2a}_{in}=\left\{ (n_1,n_2)| n_1 \leq \frac{p_1}{2},~ n_2 \leq \frac{p_2-M}{2} \right\};$ $S^{2a}_{out}=S^{2a,1}_{out}\vee S^{2a,2}_{out}$, where
			
			$S^{2a,1}_{out}=\left\{ (n_1,n_2)| n_1 \leq \frac{p_1}{2},~ n_2 < b^{2a}_{2}(n_1)\right\}$ and  $S^{2a,2}_{out}=\left\{ (n_1,n_2)| \frac{p_1}{2} \leq n_1 \leq p_1,~ n_2 < r^{2a}_{1}(n_1)\right\}.$		
			
			\item Under Condition $(\mathbb{H}_3)$,		 
			$S^{2a}_A=\left\{ (n_1,n_2)| n_1 \leq \frac{p_1}{2},\   n_2 < g_3^{2a}(n_1) \right\}$, where $n_2=g_3^{2a}(n_1)$ denotes the stable separatrix  $\Phi({P}^{2a}_2)$ of $P^{2a}_2$;
			$S^{2a}_{in}=\left\{ (n_1,n_2)| n_1 \leq \frac{p_1}{2},~ n_2 < \frac{p_2+M}{2} \right\}$;
			$S^{2a}_{out}=S^{2a,1}_{out}$;		
			
			\item Under Condition $(\mathbb{H}_4)$, $S^{2a}_{in}=S^{2a}_A=S^{2a}_{out}=\left\{ (n_1,n_2)| n_1 \leq \frac{p_1}{2},\   n_2 < \frac{{p_2}+M}{2} \right\}$.
		\end{enumerate}
		Note that $n_i\geq 0$ ($i=1,2$) holds all the time.
	\end{theorem}
	
\begin{figure}
		\centering
		\begin{subfigure}{.35\textwidth}
			\centering
			\includegraphics[width=5.5cm]{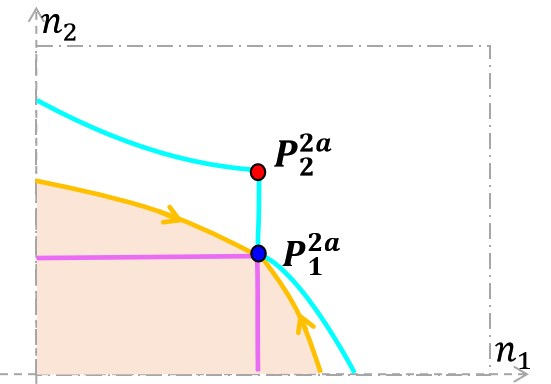}
			\caption{$\mathbb{K}^{2a} \wedge \mathbb{H}_1$}
			\label{fig05.1}
		\end{subfigure}
		\begin{subfigure}{.35\textwidth}
			\centering
			\includegraphics[width=5.5cm]{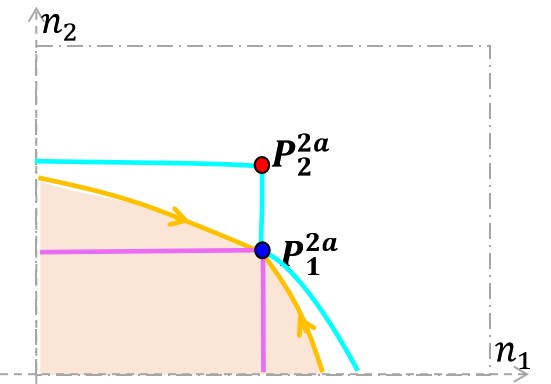}
			\caption{$\mathbb{K}^{2a} \wedge \mathbb{H}_2$}
			\label{fig05.2}
		\end{subfigure}
		\begin{subfigure}{.35\textwidth}
			\centering
			\includegraphics[width=5.5cm]{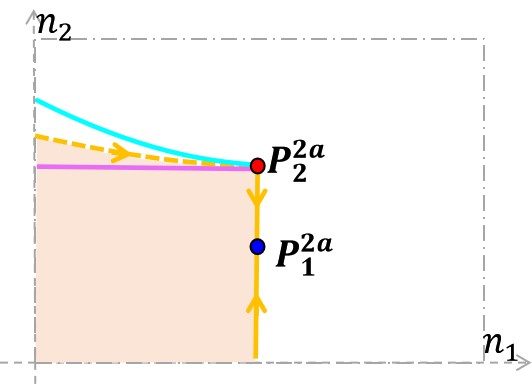}
			\caption{$\mathbb{K}^{2a} \wedge \mathbb{H}_3$}
			\label{fig05.3}
		\end{subfigure}
		\begin{subfigure}{.35\textwidth}
			\centering
			\includegraphics[width=5.5cm]{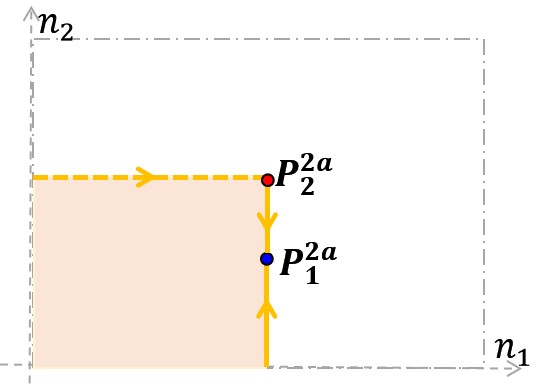}
			\caption{$\mathbb{K}^{2a} \wedge \mathbb{H}_4$}
			\label{fig05.4}
		\end{subfigure}
		\caption{Attraction region of the $P^{2a}_1$ for system \eqref{dynamics systems 3} under Condition $\mathbb{K}^{2a} \wedge \mathbb{H}_i$ ($i=1,2,\cdots,4$). Red circle: undesired saddle-node equilibrium. Blue circle: desired saddle-node equilibrium. The yellow curves: the separatrices, also serving as boundaries of attraction region.  Purple lines and bright blue lines denote the boundary of inner and outer estimations of attraction regions, respectively. }
		\label{fig05}
	\end{figure}

	The geometries of $S^{2a}_A$ in Theorem~\ref{attraction region in 3} are illustrated as the pale pink region shown in Fig.~\ref{fig05}.

	\begin{figure}
	\centering
	\begin{subfigure}{.35\textwidth}
		\centering
		\includegraphics[width=5.5cm]{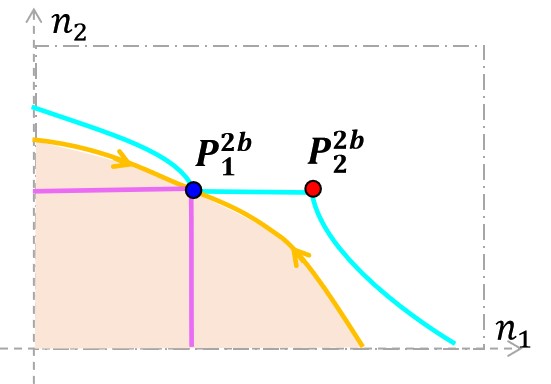}
		\caption{$\mathbb{K}^{2b} \wedge \mathbb{H}_1$}
		\label{fig06.5}
	\end{subfigure}
	\begin{subfigure}{.35\textwidth}
		\centering
		\includegraphics[width=5.5cm]{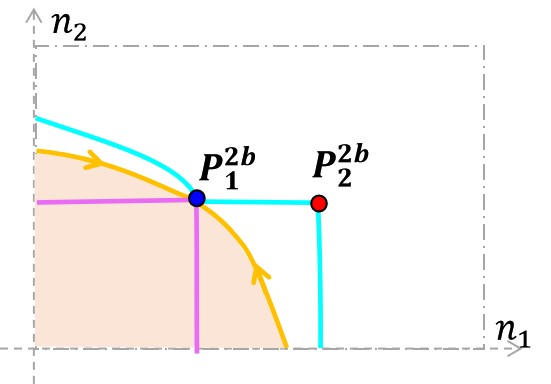}
		\caption{$\mathbb{K}^{2b} \wedge \mathbb{H}_2$}
		\label{fig06.6}
	\end{subfigure}
	\begin{subfigure}{.35\textwidth}
		\centering
		\includegraphics[width=5.5cm]{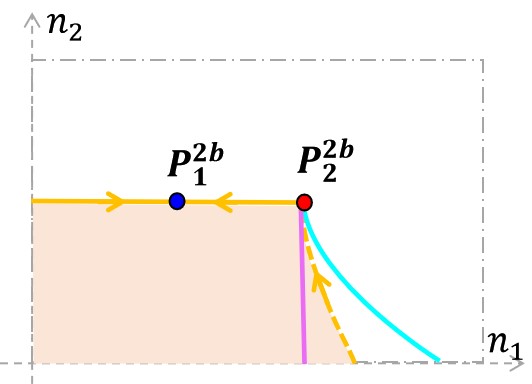}
		\caption{$\mathbb{K}^{2b} \wedge \mathbb{H}_3$}
		\label{fig06.7}
	\end{subfigure}
	\begin{subfigure}{.35\textwidth}
		\centering
		\includegraphics[width=5.5cm]{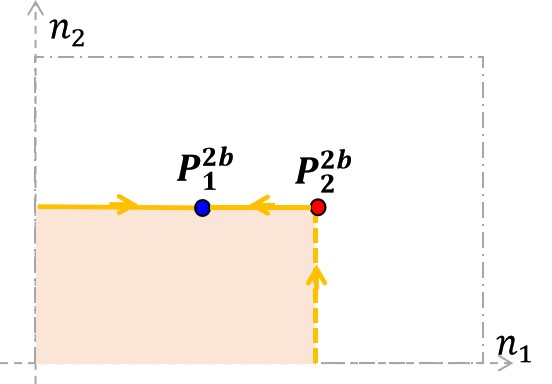}
		\caption{$\mathbb{K}^{2b} \wedge \mathbb{H}_4$}
		\label{fig06.8}
	\end{subfigure}
	\caption{Attraction region of the $P^{2b}_1$ for system \eqref{the model} under Condition $\mathbb{K}^{2b} \wedge \mathbb{H}_i$ ($i=1,2,\cdots,4$). Red circle: undesired saddle-node equilibrium. Blue circle: desired saddle-node equilibrium. The yellow curves: the separatrices, also serving as boundaries of attraction region.  Purple lines and bright blue lines denote the boundary of inner and outer estimations of attraction regions, respectively.}
	\label{fig06}
\end{figure}

	Similarly, for system \eqref{the model} under Condition $(\mathbb{K}^{2b})$, we can obtain the inner ($S^{2b}_{in}$) and outer ($S^{2b}_{out}$)  estimations of attraction regions for $P^{2b}_1$ under four different conditions, as shown in Fig.~\ref{fig06}.

\begin{figure}
		\centering
		\begin{subfigure}{.23\textwidth}
			\centering
			\includegraphics[width=4cm]{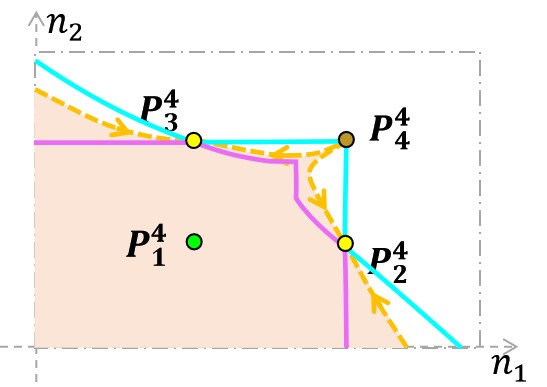}
			\caption{$\hat{\mathbb{K}}^{4}_1 \wedge \mathbb{H}_1$}
			\label{fig08.5}
		\end{subfigure}
		\begin{subfigure}{.23\textwidth}
			\centering
			\includegraphics[width=4cm]{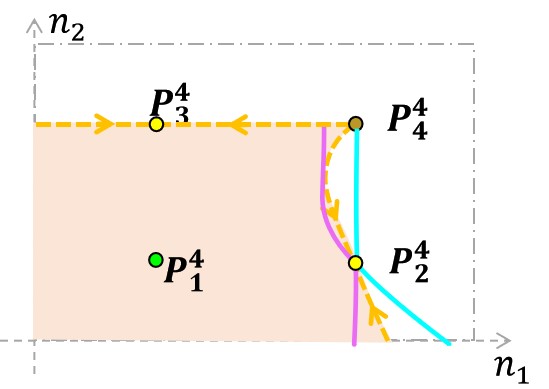}
			\caption{$\hat{\mathbb{K}}^{4}_1 \wedge \mathbb{H}_2$}
			\label{fig08.6}
		\end{subfigure}
		\begin{subfigure}{.23\textwidth}
			\centering
			\includegraphics[width=4cm]{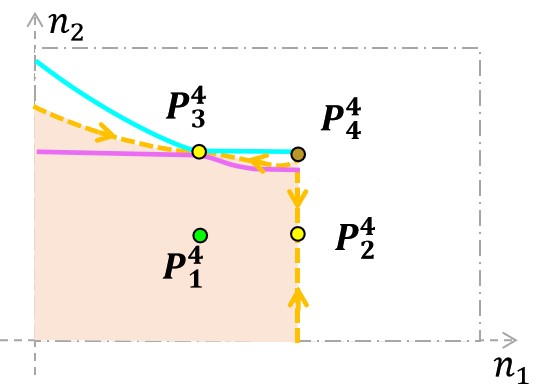}
			\caption{$\hat{\mathbb{K}}^{4}_1 \wedge \mathbb{H}_3$}
			\label{fig08.7}
		\end{subfigure}
		\begin{subfigure}{.23\textwidth}
			\centering
			\includegraphics[width=4cm]{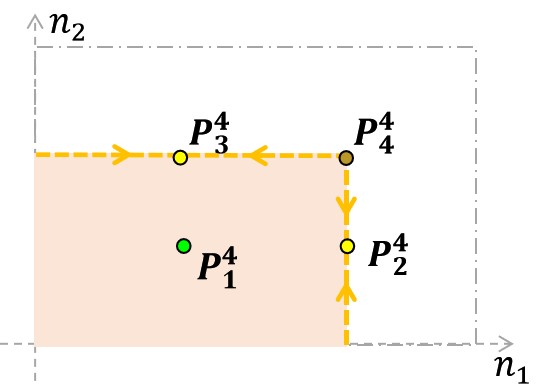}
			\caption{$\hat{\mathbb{K}}^{4}_1 \wedge \mathbb{H}_4$}
			\label{fig08.8}
		\end{subfigure}
		\begin{subfigure}{.23\textwidth}
			\centering
			\includegraphics[width=4cm]{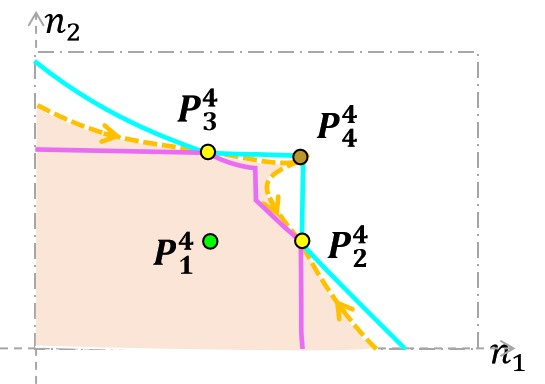}
			\caption{$\hat{\mathbb{K}}^{4}_2 \wedge \mathbb{H}_1$}
			\label{fig09.1}
		\end{subfigure}
		\begin{subfigure}{.23\textwidth}
			\centering
			\includegraphics[width=4cm]{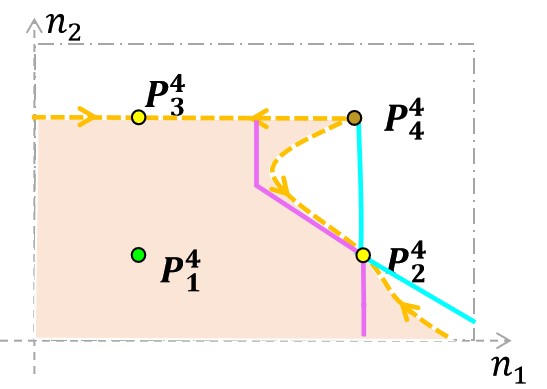}
			\caption{$\hat{\mathbb{K}}^{4}_2 \wedge \mathbb{H}_2$}
			\label{fig09.2}
		\end{subfigure}
		\begin{subfigure}{.23\textwidth}
			\centering
			\includegraphics[width=4cm]{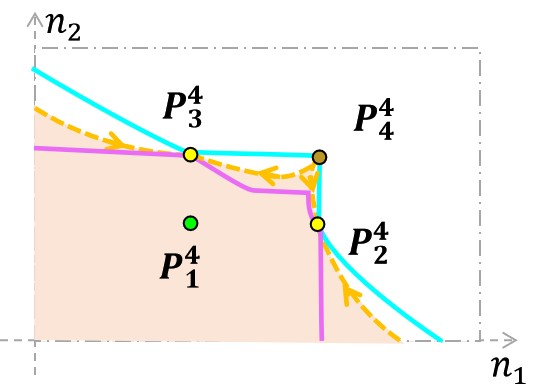}
			\caption{$\hat{\mathbb{K}}^{4}_3 \wedge \mathbb{H}_1$}
			\label{fig09.3}
		\end{subfigure}
		\begin{subfigure}{.23\textwidth}
			\centering
			\includegraphics[width=4cm]{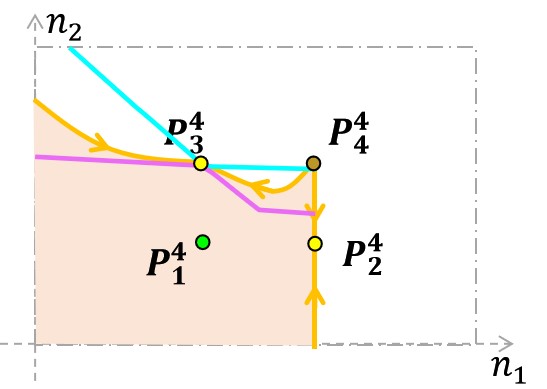}
			\caption{$\hat{\mathbb{K}}^{4}_3 \wedge \mathbb{H}_3$}
			\label{fig09.4}
		\end{subfigure}
		\begin{subfigure}{.23\textwidth}
			\centering
			\includegraphics[width=4cm]{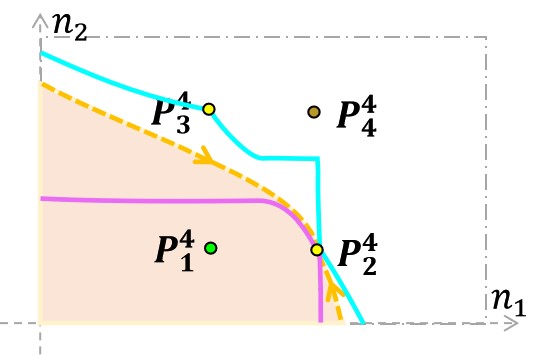}
			\caption{  $\hat{\mathbb{K}}^{4}_4 \wedge \mathbb{H}_1$, type 1}
			\label{fig09.5}
		\end{subfigure}
		\begin{subfigure}{.23\textwidth}
			\centering
			\includegraphics[width=4cm]{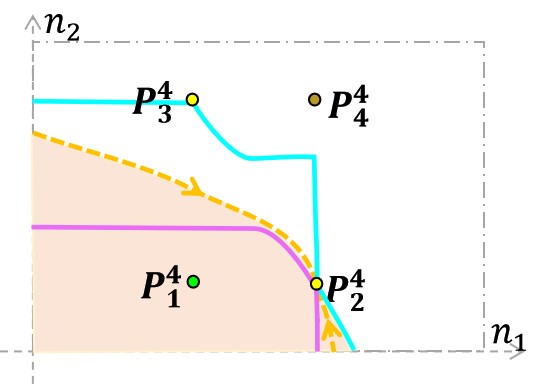}
			\caption{ $\hat{\mathbb{K}}^{4}_4 \wedge \mathbb{H}_2$, type 1}
			\label{fig09.6}
		\end{subfigure}
		\begin{subfigure}{.23\textwidth}
			\centering
			\includegraphics[width=4cm]{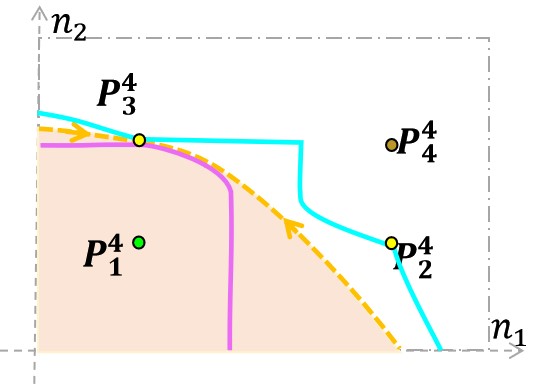}
			\caption{ $\hat{\mathbb{K}}^{4}_5 \wedge \mathbb{H}_1$, type 1}
			\label{fig09.7}
		\end{subfigure}
		\begin{subfigure}{.23\textwidth}
			\centering
			\includegraphics[width=4cm]{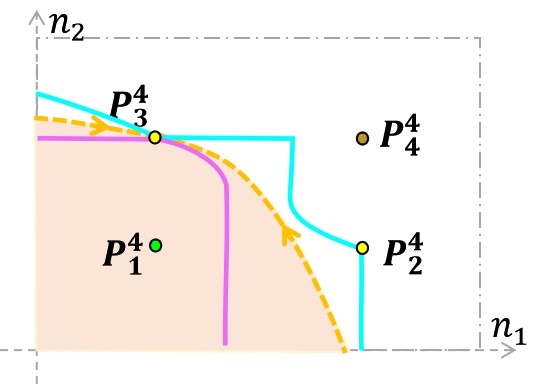}
			\caption{ $\hat{\mathbb{K}}^{4}_5 \wedge \mathbb{H}_3$, type 1}
			\label{fig09.8}
		\end{subfigure}
		
		\begin{subfigure}{.23\textwidth}
			\centering
			\includegraphics[width=4cm]{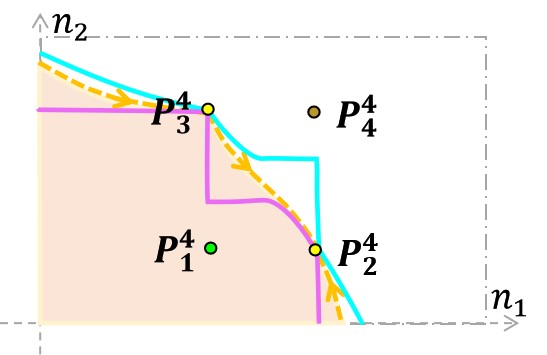}
			\caption{$\hat{\mathbb{K}}^{4}_4 \wedge \mathbb{H}_1$, type 2}
			\label{fig09.9}
		\end{subfigure}
		\begin{subfigure}{.23\textwidth}
			\centering
			\includegraphics[width=4cm]{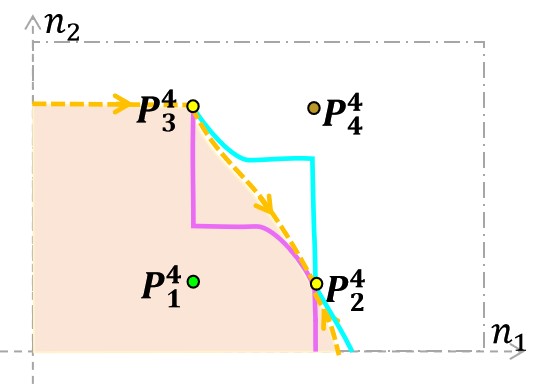}
			\caption{$\hat{\mathbb{K}}^{4}_4 \wedge \mathbb{H}_2$, type 2}
			\label{fig09.10}
		\end{subfigure}
		\begin{subfigure}{.23\textwidth}
			\centering
			\includegraphics[width=4cm]{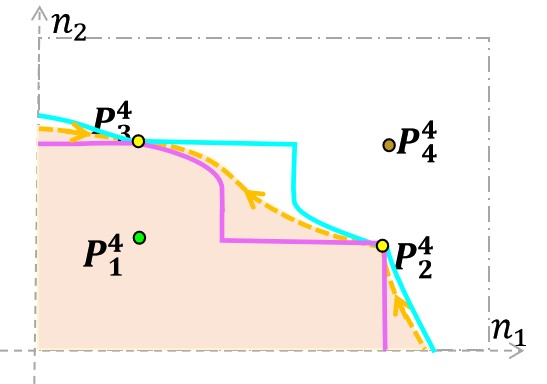}
			\caption{$\hat{\mathbb{K}}^{4}_5 \wedge \mathbb{H}_1$, type 2}
			\label{fig09.11}
		\end{subfigure}
		\begin{subfigure}{.23\textwidth}
			\centering
			\includegraphics[width=4cm]{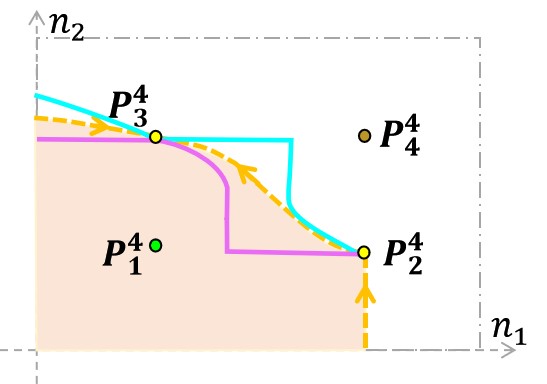}
			\caption{$\hat{\mathbb{K}}^{4}_5 \wedge \mathbb{H}_3$, type 2}
			\label{fig09.12}
		\end{subfigure}
		
		\begin{subfigure}{.23\textwidth}
			\centering
			\includegraphics[width=4cm]{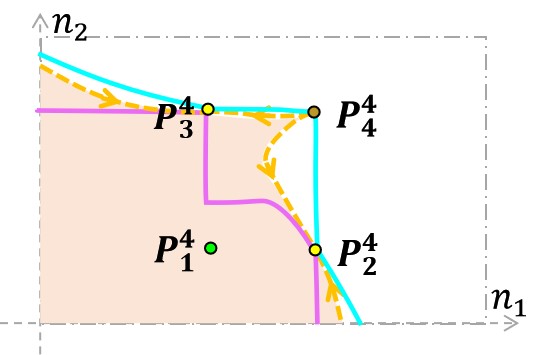}
			\caption{$\hat{\mathbb{K}}^{4}_4 \wedge \mathbb{H}_1$, type 3}
			\label{fig09.13}
		\end{subfigure}
		\begin{subfigure}{.23\textwidth}
			\centering
			\includegraphics[width=4cm]{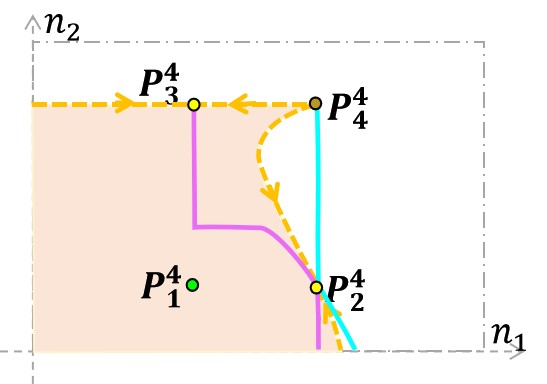}
			\caption{$\hat{\mathbb{K}}^{4}_4 \wedge \mathbb{H}_2$, type 3}
			\label{fig09.14}
		\end{subfigure}
		\begin{subfigure}{.23\textwidth}
			\centering
			\includegraphics[width=4cm]{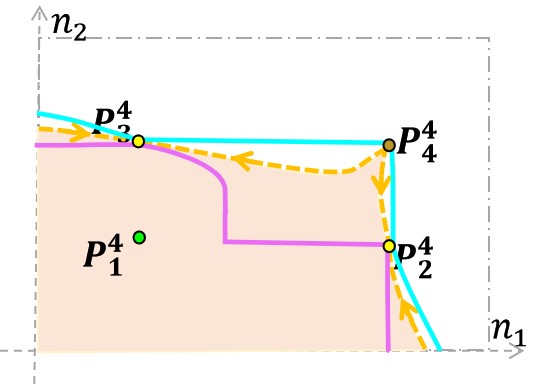}
			\caption{$\hat{\mathbb{K}}^{4}_5 \wedge \mathbb{H}_1$, type 3}
			\label{fig09.15}
		\end{subfigure}
		\begin{subfigure}{.23\textwidth}
			\centering
			\includegraphics[width=4cm]{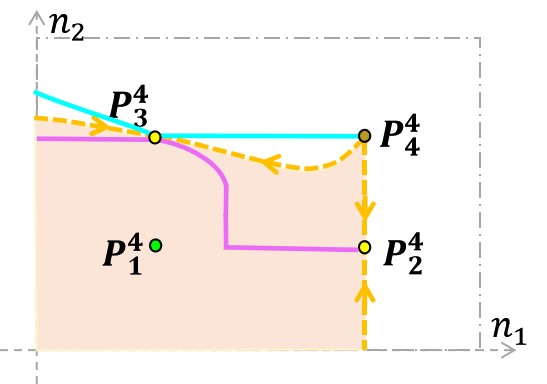}
			\caption{$\hat{\mathbb{K}}^{4}_5 \wedge \mathbb{H}_3$, type 3}
			\label{fig09.16}
		\end{subfigure}
		
		\caption{Attraction region of the $P^{4}_1$ of system \eqref{the model} under Condition $\hat{\mathbb{K}}^{4}_s\wedge \mathbb{H}_i$ ($s=1,2,\cdots,5$ and $i=1,2,\cdots,4$). Green circle: desired uncongested stable equilibrium. Yellow circle: saddle equilibrium. Yellow-brown circle: unstable equilibrium. Yellow curves with arrows: the separatrices, also serving as boundaries of attraction region.  Purple lines and bright blue lines denote the boundary of inner and outer estimations of attraction regions, respectively.}
		\label{fig09}
	\end{figure}

	For the four-equilibrium system, we also try to obtain the inner ($S^{4}_{in}$) and outer ($S^{4}_{out}$) estimations  of attraction regions, and the detailed analysis can be found in App. \ref{A.3}.
	Note that for the four-equilibrium system, the attraction regions become more intricate due to the increased number of equilibria. In App. \ref{A.3}, through a novel classification method, we obtained 20 distinct inner and outer estimations of  attraction regions, which can be summarized as Theorems \ref{inner and outer estimations of attraction regions in 4.1}, \ref{attraction region in 4.1}, and \ref{attraction region in 4.2} in App. \ref{A.3}. The geometries of the attraction region for the four-equilibrium system  are illustrated as the pale pink region shown in Fig.~\ref{fig09}.

	\section{Ecological resilience control design}\label{s4}
	In the previous two sections, we obtain the inner ($S_{in}$) and outer ($S_{out}$) estimations of attraction regions with explicit algebraic expressions under each condition for a multi-equilibria MFD-represented traffic dynamic system \eqref{the model}. 
	Note that for two-equilibria system, $S_{in}$ corresponds to $S^{2a}_{in}$ or $S^{2b}_{in}$, and $S_{out}$
	corresponds to $S^{2a}_{out}$ or $S^{2b}_{out}$;  for two-equilibria system, $S_{in}$ corresponds to $S^{4}_{in}$ and $S_{out}$
	corresponds to $S^{4}_{out}$. While 
	 In this section, we will design two distinct resilience control schemes, aiming to facilitate the recoverability of the system to its respective steady states while minimally altering the original landscape, utilizing the obtained $S_{in}$ and $S_{out}$. Sec. \ref{s4.1} illustrates the resilience control schemes utilizing $S_{in}$, denoted as RCS-1, and Sec. \ref{s4.2} provides the resilience control schemes utilizing $S_{out}$, denoted as RCS-2.
	
	The trajectories starting from inside attraction region (or inner estimations of attraction regions) can converge to the original equilibrium, while for those that starting from outside attraction region (or outer estimations of attraction regions) will move away from the equilibrium. Our primary task is to control the trajectories that move away from the equilibrium point and guide them towards an alternative steady state. For multi-equilibria systems \eqref{the model}, we set up an alternative steady state $\bm{n}^1$=$(ns_1,ns_2)$ in $\mathbb{D} \setminus S_{in}$ or $\mathbb{D} \setminus S_{out}$, respectively. 
	Then, the control schemes (RCS-1 and RCS-2)  shall steer the trajectories to the  $\varepsilon-Neighborhood$ of $\bm{n}^1$ within the recovery time $t_f$. 
	Note that the control does not necessarily bring the system  to  the equilibrium state $\bm{n}^0$, which allows the system to be adaptive (to a feasible steady state) and recoverable (to less congested state progressively).
	
	Generally speaking, there are three main differences between the RCS-1 and RCS-2 schemes. With respect to control “intensity”, RCS-1 is more aggressive than RCS-2. That is, the control region for RCS-1 is much larger than RCS-2, as control action $\bm{U}_1(\bm{n})$ ($\bm{U}_2(\bm{n})$) are imposed in $\mathbb{D} \setminus S_{in}$  ($\mathbb{D} \setminus S_{out}$) under RCS-1 (RCS-2) and we have  $S_{in}\subset S_{out}$. With respect to control mechanism, RCS-1 has a universal control algorithm  applied to  all alternative steady states, while RCS-2 applies three forms of control, depending on the location of the alternative steady state. Here, we choose three control methods to prevent the emergence of new equilibrium points on the boundary of outer estimation of attraction regions. For instance, trajectories from the boundary of $S_{out}$ tend to escape $S_{out}$, and one control scheme might lead to intersections between trajectories from $\mathbb{D} \setminus S_{out}$ and trajectories from $S_{out}$. This could potentially generate new equilibrium points on the boundary of the outer estimation of attraction regions, complicating theoretical derivations and causing control inefficiencies. With respect to control effect, the trajectories starting from $\mathbb{D} \setminus S_{out}$ will move towards $\bm{n}^1$, while trajectories starting from $\mathbb{D} \setminus S_{in}$ will move towards $\bm{n}^1$ (Fig.~\ref{fig10.2}), or  $\bm{n}^0$ (Fig.~\ref{fig10.3}).
	
	\subsection{Control formulation of RCS-1}\label{s4.1}
\begin{figure}
		\centering
		\begin{subfigure}{.32\textwidth}
			\centering
			\includegraphics[width=5.5cm]{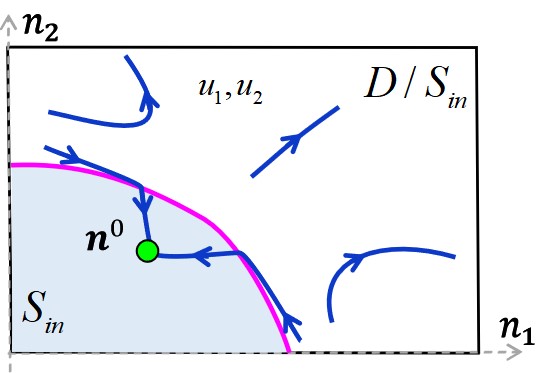}
			\caption{CPC}
			\label{fig10.1}
		\end{subfigure}
		\begin{subfigure}{.32\textwidth}
			\centering
			\includegraphics[width=5.5cm]{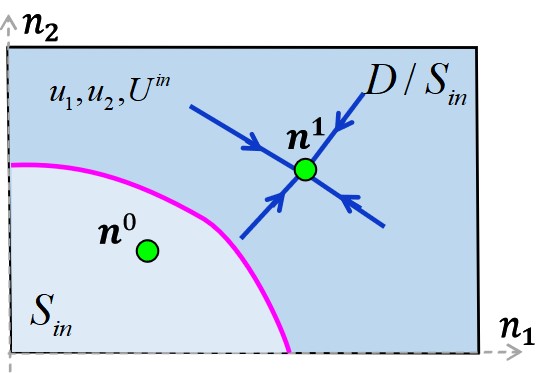}
			\caption{RCS-1}
			\label{fig10.2}
		\end{subfigure}
		\begin{subfigure}{.32\textwidth}
			\centering
			\includegraphics[width=5.5cm]{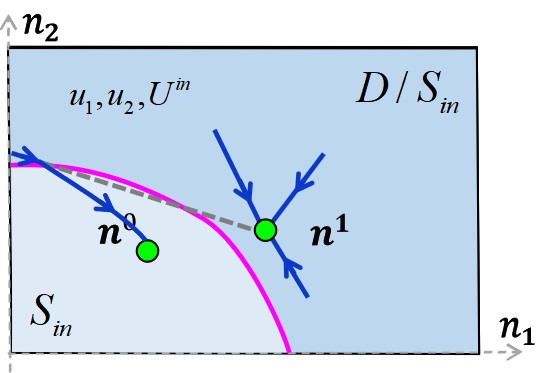}
			\caption{RCS-1}
			\label{fig10.3}
		\end{subfigure}
		\caption{Schematic representation of phase portraits under CPC (a)  and RCS-1 ((b) and (c)). Notice that in (c), the trajectories under RCS-1 intersect $ S_{in}$. Once intersecting $ S_{in}$, we rest the control $U^{in}= 0$ immediately, then trajectories spontaneously move towards the $\bm{n}^0$.}
		\label{fig10}
	\end{figure}

	The domain $\mathbb{D}$ can be divided into two subregions: $S_{in}$ and $\mathbb{D} \setminus S_{in}$, as shown in Fig.~\ref{fig10}.
	If the starting points are in $S_{in}$, 
	CPC $u_1$ and $u_2$  that satisfy Conditions $\mathbb{K}^{2a}$, $\mathbb{K}^{2b}$ or $\mathbb{K}^{4}$ are conducted. 
	According to Theorems \ref{attraction region in 3}, \ref{inner and outer estimations of attraction regions in 4.1} and \ref{attraction region in 4.2}, 
	trajectories from these starting points  will move towards the $\bm{n}^0$.
	Instead, 
	if the initial states are in $\mathbb{D} \setminus S_{in}$,
	new control action $\bm{U}^{in}(\bm{n})$ is  activated for recovering trajectories starting from $\mathbb{D} \setminus S_{in}$ to the  $\varepsilon-Neighborhood$ of the target point within the recovery time $t_f$, 
	The target point can be the uncongested equilibrium $\bm{n}^0$ or the alternative steady state $\bm{n}^1$ in $\mathbb{D} \setminus S_{in}$.
	Specifically, for system \eqref{the model}, 
	if the starting states are in $\mathbb{D} \setminus S_{in}$, we apply the following control $\bm{U}^{in}(n_1,n_2)=\left( U^{in}_1(n_1,n_2), U^{in}_2(n_1,n_2)\right)$, defined by:
	\begin{equation}\label{control rules}
		\begin{split}		
			U^{in}_1(n_1,n_2)= -\gamma_1\left( n_1-ns_1\right)-{u_2} {G_2}{(n_2 (t))}+{G_1}{(n_1 (t))}-d_1,\\
			U^{in}_2(n_1,n_2)= -\gamma_2\left( n_2-ns_2\right)-{u_1} {G_1}{(n_1 (t))}+{G_2}{(n_2 (t))}-d_2.
		\end{split}
	\end{equation}
	where $\gamma_i=\frac{2G_{i,max}}{p_i}$.
	By adding proposed controller \eqref{control rules} to system \eqref{the model}, 
	the trajectories starting from $\mathbb{D} \setminus S_{in}$ will move towards $\bm{n}^1$ directly (Fig.~\ref{fig10.2}), 
	or intersect $ S_{in}$ (Fig.~\ref{fig10.3}). 
	Once intersecting $ S_{in}$, 
	we rest the control, i.e., $U^{in}_1=U^{in}_2= 0$ immediately, as such that the trajectories will follow the initial system \eqref{the model}.  
	According to Theorems \ref{attraction region in 3}, \ref{inner and outer estimations of attraction regions in 4.1} and \ref{attraction region in 4.2}, 
	trajectories from these starting points will move towards the $\bm{n}^0$.
	
	By doing so, we have developed a switched controlled system ~(\cite{she2014discovering,skafidas1999stability,gao2022resilient,she2020design}) as:
	$$\dot{\bm{n}}=\bm{F}(\bm{n})+\bm{U}_1(\bm{n}),$$
	where 
	$\bm{U}_1(\bm{n})\equiv \bm{0}$ when $\bm{n}\in S_{in}$, 
	$\bm{U}_1(\bm{n})=\bm{U}^{in}(n_1,n_2)$ when $\bm{n}\in$ $\mathbb{D} \setminus S_{in}$.

	\begin{remark}
		For multi-equilibria systems \eqref{the model}, certain inner estimations of attraction regions do not contain its partial boundaries $l_{ex}$ (e.g. Figs. \ref{fig04.4}, \ref{fig06.8}, \ref{fig08.6} etc.). This is because the trajectories starting from the partial boundaries will deviate from the uncongested equilibrium point. In such cases, $\bm{U}^{in}(\bm{n})$ are imposed until the trajectories intersect with $\hat{l}_{ex}$, where $\hat{l}_{ex}$ is obtained by moving the line $l_{ex}$ down (or left) by $\varepsilon$.
	\end{remark}

	\subsection{Control formulation of RCS-2}\label{s4.2}
\begin{figure}
		\centering
		\begin{subfigure}{.33\textwidth}
			\centering
			\includegraphics[width=5.5cm]{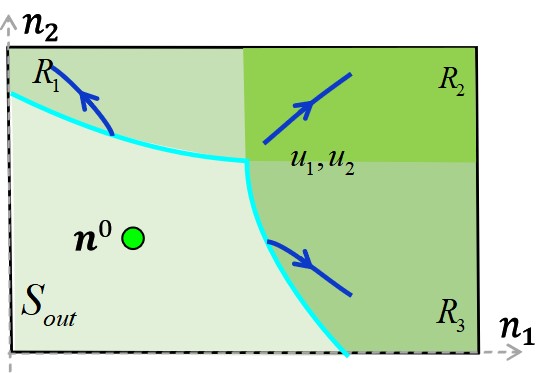}
			\caption{CPC}
			\label{fig24.1}
		\end{subfigure}
		\begin{subfigure}{.33\textwidth}
			\centering
			\includegraphics[width=5.5cm]{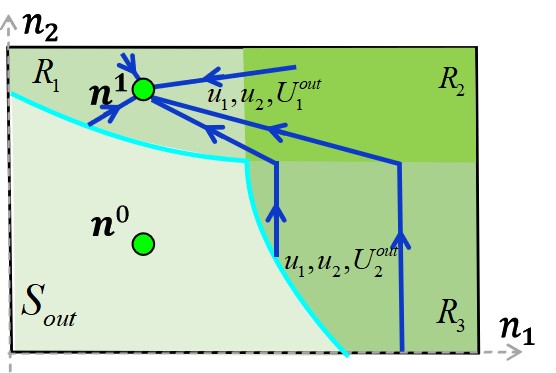}
			\caption{RCS-2, Case 1}
			\label{fig24.2}
		\end{subfigure}
		\begin{subfigure}{.33\textwidth}
			\centering
			\includegraphics[width=5.5cm]{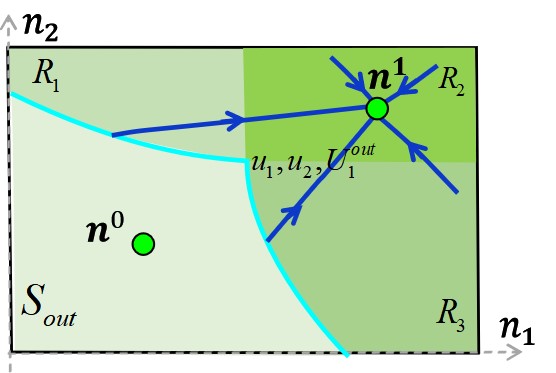}
			\caption{RCS-2, Case 2}
			\label{fig24.3}
		\end{subfigure}
		\begin{subfigure}{.33\textwidth}
			\centering
			\includegraphics[width=5.5cm]{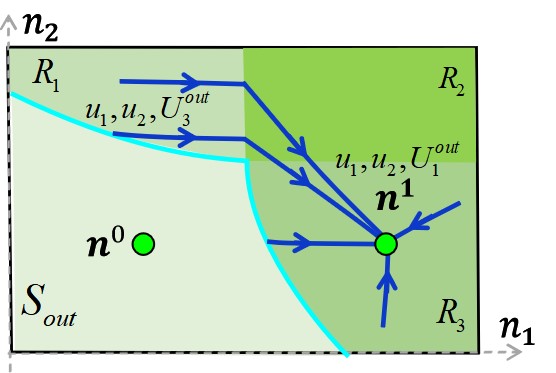}
			\caption{RCS-2, Case 3}
			\label{fig24.4}
		\end{subfigure}
		\caption{Schematic representation of phase portraits under CPC (a)  and RCS-2 ((b), (c) and (d)). Note that there are three cases of control action, depending on the location of $\bm{n}^1$. Three cases of control action are chosen to prevent the emergence of new equilibrium points on the boundary of $S_{out}$.}
		\label{fig24}
	\end{figure}

	RCS-2 is designed based on $S_{out}$. Likewise, 
	the domain $\mathbb{D}$ can be divided into $S_{out}$ and $\mathbb{D} \setminus S_{out}$, and $\mathbb{D} \setminus S_{out}$ can be further divided into $R_1$, $R_2$ and $R_3$, as illustrated  in Fig.~\ref{fig24}.
	If the starting states are in $S_{out}$, 
	CPC $u_1$ and $u_2$ that satisfy Conditions $\mathbb{K}^{2a}$, $\mathbb{K}^{2b}$ or $\mathbb{K}^{4}$ will be conducted.
	According to Theorems \ref{attraction region in 3}, \ref{inner and outer estimations of attraction regions in 4.1} and \ref{attraction region in 4.2}, trajectories from these starting points will move towards $\bm{n}^0$ or escape from the upper boundaries of $S_{out}$ and enter $\mathbb{D} \setminus S_{out}$.
	Once the states enter in $\mathbb{D} \setminus S_{out}$,
	control action $\bm{U}^{out}(\bm{n})$ will be activated for recovering trajectories to the  $\varepsilon-Neighborhood$ of $\bm{n}^1$ in $\mathbb{D} \setminus S_{out}$ within the recovery time $t_f$.
	
	Specifically, 
	for a given system \eqref{the model}, there are three cases of control action $\bm{U}^{out}(\bm{n})$, depending on the location of $\bm{n}^1$.  
	First consider Case 1, where $\bm{n}^1$ lies in $R_1$,  as depicted  in Fig.~\ref{fig24.2}.
	In this case, proposed RCS-2 is as follows:
	in addition to $u_1$ and $u_2$, 
	we will apply an additional control $\bm{U}^{out}_1(n_1,n_2)=\left( U^{out}_{1,1}(n_1,n_2), U^{out}_{1,2}(n_1,n_2)\right)$ in $R_1\vee R_2$, 
	and  $\bm{U}^{out}_2(n_1,n_2)=\left( U^{out}_{2,1}(n_1,n_2), U^{out}_{2,2}(n_1,n_2)\right)$ in $R_3$, 
	where $\bm{U}^{out}_1(n_1,n_2)$ are defined by:
	\begin{equation}\label{control rules 02-1}
		\begin{split}		
			U^{out}_{1,1}(n_1,n_2)= -\gamma_1\left( n_1-ns_1\right)-{u_2} {G_2}{(n_2 (t))}+{G_1}{(n_1 (t))}-d_1,\\
			U^{out}_{1,2}(n_1,n_2)= -\gamma_2\left( n_2-ns_2\right)-{u_1} {G_1}{(n_1 (t))}+{G_2}{(n_2 (t))}-d_2,
		\end{split}
	\end{equation}
	and  $\bm{U}^{out}_2(n_1,n_2)$ are defined by:
	\begin{eqnarray}\label{control rules 02-2}
		U^{out}_{2,1}(n_1,n_2)&=& -{u_2} {G_2}{(n_2 (t))}+{G_1}{(n_1 (t))}-d_1,\nonumber\\	
		U^{out}_{2,2}(n_1,n_2)&=& -\gamma_2 n_2-{u_1} {G_1}{(n_1 (t))}+{G_2}{(n_2 (t))}-d_2.
	\end{eqnarray}
	by adding proposed controller \eqref{control rules 02-2} to system \eqref{the model}, 
	the trajectories starting from $R_3$ will vertically escape from the upper boundary of $R_3$ and enter $R_2$.  
	Moreover, by adding proposed controller \eqref{control rules 02-1} to system \eqref{the model}, 
	the trajectories starting from $R_1\vee R_2$ will move towards $\bm{n}^1$ directly, as shown in Fig.~\ref{fig24.2}. 
	Thus, any trajectories entering $\mathbb{D} \setminus S_{out}$ will move towards  $\bm{n}^1$ under the controller.
	
	Then consider Case 2, where $\bm{n}^1$ lies in $R_2$. In this case, 
	$\bm{U}^{out}_1(n_1,n_2)=\left( U^{out}_{1,1}(n_1,n_2), U^{out}_{1,2}(n_1,n_2)\right)$ in $\mathbb{D} \setminus S_{out}$ will be applied. Subsequently, any trajectories going into $\mathbb{D} \setminus S_{out}$ will go to  $\bm{n}^1$ under our proposed controller, as shown in Fig.~\ref{fig24.3}.

	Finally consider Case 3,
	where $\bm{n}^1$ lies in $R_3$, as shown in Fig.~\ref{fig24.4}.
	similarly,  we will apply control action $\bm{U}^{out}_1(n_1,n_2)=\left( U^{out}_{1,1}(n_1,n_2), U^{out}_{1,2}(n_1,n_2)\right)$ in $R_3\vee R_2$, 
	and  $\bm{U}^{out}_3(n_1,n_2)=\left( U^{out}_{3,1}(n_1,n_2), U^{out}_{3,2}(n_1,n_2)\right)$ in $R_1$, 
	where $\bm{U}^{out}_3(n_1,n_2)$ are defined by:
	\begin{eqnarray}\label{control rules 02-3}
		U^{out}_{3,1}(n_1,n_2)&=& -\gamma_1n_1-{u_2} {G_2}{(n_2 (t))}+{G_1}{(n_1 (t))}-d_1,\nonumber\\	
		U^{out}_{3,2}(n_1,n_2)&=& -{u_1} {G_1}{(n_1 (t))}+{G_2}{(n_2 (t))}-d_2.
	\end{eqnarray}
	by adding proposed controller \eqref{control rules 02-3} to system \eqref{the model}, 
	the trajectories starting from $R_1$ will horizontally escape from the right boundary of $R_1$ and enter $R_2$.  
	Moreover, by adding proposed controller \eqref{control rules 02-1} to system \eqref{the model}, 
	the trajectories starting from $R_2\vee R_3$ will move towards $\bm{n}^1$ directly, as shown in Fig.~\ref{fig24.4}.
	Thus, any trajectories entering $\mathbb{D} \setminus S_{out}$ will move towards  $\bm{n}^1$ under our proposed controller.
	
	By doing so, we have developed another switched controlled system ~(\cite{she2014discovering,skafidas1999stability,gao2022resilient,she2020design}) as:
	$$\dot{\bm{n}}=\bm{F}(\bm{n})+\bm{U}_2(\bm{n}),$$
	where 
	$\bm{U}_2(\bm{n})\equiv \bm{0}$ when $\bm{n}\in S_{out}$, 
	and $\bm{U}_2(\bm{n})$ features different formulation in  different cases when $\bm{n}\in$ $\mathbb{D} \setminus S_{out}$. Specifically, 
	\begin{itemize}
		\item in case 1, $\bm{U}_2(\bm{n})=\bm{U}^{out}_1(n_1,n_2)$ when $\bm{n}\in$ $R_1\vee R_2$
		and  $\bm{U}_2(\bm{n})=\bm{U}^{out}_2(n_1,n_2)$ when $\bm{n}\in$ $R_3$;
		\item in case 2, $\bm{U}_2(\bm{n})=\bm{U}^{out}_1(n_1,n_2)$ when $\bm{n}\in$ $\mathbb{D} \setminus S_{out}$;
		\item in case 3, $\bm{U}_2(\bm{n})=\bm{U}^{out}_1(n_1,n_2)$ when $\bm{n}\in$ $R_2\vee R_3$
		and  $\bm{U}_2(\bm{n})=\bm{U}^{out}_3(n_1,n_2)$ when $\bm{n}\in$ $R_1$.
	\end{itemize}
	\begin{remark}\label{same}
		$\bm{U}^{in}(n_1,n_2)$
		and $\bm{U}^{out}_2(n_1,n_2)$ would have the same formulation if the alternative steady state $\bm{n}^1$ in both controllers are same. This implies  that there may exist initial states, such that the system may end up with identical recovery trajectories under RCS- 1 and RCS-2, respectively.
	\end{remark}
	
	%
	
	Furthermore, 
	for a given system \eqref{the model}, 
	if a stable equilibrium point exists (i.e., Condition $(\mathbb{K}^{4})$ holds), 
	a local Lyapunov function can be found (\cite{WANG2021100993,Willems1971,BLANCHINI1995451}) for the new controlled system with RCS-2.
	Specifically,
	denoting $\bm{n}=(n_1,n_2)^ \mathrm{T}$, we can prove that the function
	$V= \left( \bm{n}-\bm{n}^{0}\right)^ \mathrm{T} \bm{B} \left( \bm{n}-\bm{n}^{0}\right) \left( \bm{n}-\bm{n}^{1}\right)^ \mathrm{T} \left( \bm{n}-\bm{n}^{1}\right)$ is a local Lyapunov function for system \eqref{the model} with RCS-2, 
	where
	$\bm{B}=\left(\bm{P}^{-1} \right) ^ \mathrm{T}\bm{P}^{-1}$
	and $$\bm{P}=	\left(\begin{array}{cc} p_{11} & p_{12} \\ p_{21} & p_{22} \end{array}\right)=
	\left(\begin{array}{cc} \frac{-a_{1}\,s_{1}+a_{2}\,s_{2}+Q}{2\,a_{1}\,s_{1}\,u_{1}} & \frac{-a_{1}\,s_{1}+a_{2}\,s_{2}-Q}{2\,a_{1}\,s_{1}\,u_{1}}\\ 1 & 1 \end{array}\right)$$ 
	with $Q=\sqrt{{a_{1}}^2\,{s_{1}}^2+{a_{2}}^2\,{s_{2}}^2-2\,a_{1}\,a_{2}\,s_{1}\,s_{2}+4\,a_{1}\,a_{2}\,s_{1}\,s_{2}\,u_{1}\,u_{2}}$; $s_1=\frac{1}{2}\sqrt{{p_1}^2-\frac{4{d_1} + 4{u_2}{d_2}}{{a_1}(1-{u_1}{u_2})}}$; $s_2=\frac{1}{2}\sqrt{{p_2}^2-\frac{4{d_2} + 4{u_1}{d_1}}{{a_2}(1-{u_1}{u_2})}}$. 
	Obviously, $V>0$ holds for all $\bm{n}\in \mathbb{D}$ except for $\bm{n}=\bm{n}^{0}$ and $\bm{n}=\bm{n}^{1}$. Moreover, $V=0$ holds if and only if $\bm{n}=\bm{n}^{0}$ and $\bm{n}=\bm{n}^{1}$. 
	In addition, in the neighborhood of $\bm{n}^0$, we have  $	\frac{\mathrm{d}{V}(t)}{\mathrm{d}t}<0$ holds, and detailed proofs are as follows: 
	
	Letting $\bm{n}-\bm{n}^{0}=\bm{P}\bm{y}$, where $\bm{y}=(y_1,y_2)^ \mathrm{T}$,  we can transform the system \eqref{the model} into:
	\begin{equation}\label{the model_y}
		\begin{split}
			\frac{\mathrm{d}{y}_1(t)}{\mathrm{d}t}=\lambda_1 {y_1 (t)}+H_1({y}_1(t),{y}_2(t)),\\
			\frac{\mathrm{d}{y}_2(t)}{\mathrm{d}t}=\lambda_2 {y_2 (t)}+H_2({y}_1(t),{y}_2(t)),
		\end{split}
	\end{equation}
	where $\lambda_1=   -a_{1}\,s_{1}-a_{2}\,s_{2}+Q<0$; $\lambda_2=   -a_{1}\,s_{1}-a_{2}\,s_{2}-Q<0$; $H_1({y}_1(t),{y}_2(t))$ and $H_2({y}_1(t),{y}_2(t))$ represent second-order remained terms.
	
	By the above coordinate transformation, denoting $\bm{\delta}=\bm{n}^{1}-\bm{n}^{0}=(\delta_1,\delta_2)^\mathrm{T}$, we can obtain that:
	\begin{eqnarray}\label{Lyapunov-V}
		{V}(t)&=&{\left( \bm{n}-\bm{n}^{0}\right)^ \mathrm{T} \bm{B} \left( \bm{n}-\bm{n}^{0}\right)}\left( \bm{n}-\bm{n}^{1}\right)^ \mathrm{T} \left( \bm{n}-\bm{n}^{1}\right),\nonumber\\
		&=&{ (\bm{P}\bm{y})^ \mathrm{T} \bm{B} (\bm{P}\bm{y})}\left( \bm{P}\bm{y}-\bm{\delta}\right)^ \mathrm{T} \left( \bm{P}\bm{y}-\bm{\delta}\right),\nonumber\\
		&=&{\bm{y}^ \mathrm{T}\bm{y}} \left( \bm{P}\bm{y}-\bm{\delta}\right)^ \mathrm{T} \left( \bm{P}\bm{y}-\bm{\delta}\right),\nonumber\\
		&=&V_1\cdot V_2
	\end{eqnarray}
	where $V_1={y_1}^2+ {y_2}^2$ and $V_2=(p_{11} y_1+p_{12} y_2-\delta_1)^2+(p_{21} y_1+p_{22} y_2-\delta_2)^2$. Then we have:
	\begin{eqnarray}\label{Lyapunov-DV}
		\frac{\mathrm{d}{V}(t)}{\mathrm{d}t}
		&=&\frac{\partial {V}}{\partial {y}_1}\cdot \frac{\mathrm{d}{y}_1(t)}{\mathrm{d}{t}}
		+\frac{\partial {V}}{\partial {y}_2}\cdot \frac{\mathrm{d}{y}_2(t)}{\mathrm{d}{t}},\nonumber\\
		&=&2 (\delta_1^2+\delta_2^2)(\lambda_1  y_1(t)^2+ \lambda_2  y_2(t)^2)+H_v^1(y_1(t),y_2(t)).
	\end{eqnarray}
	where $H_v^1(y_1(t),y_2(t))$ represents the high-order remained terms, and the lowest power of it is cubic. Obviously, $	\frac{\mathrm{d}{V}(t)}{\mathrm{d}t}<0$ holds for arbitrary $(y_1,y_2)$  in the neighborhood of $(0,0)$ except $(0,0)$, as we have $\lambda_2<\lambda_1<0$. That is, $	\frac{\mathrm{d}{V}(t)}{\mathrm{d}t}<0$ holds for arbitrary $(n_1,n_2)$  in the neighborhood of $\bm{n}^0$, implying function ${V}$ serves as a local Lyapunov function for the system in the neighbor of $\bm{n}^0$. 
	
	Similarly, we can prove that $	\frac{\mathrm{d}{V}(t)}{\mathrm{d}t}<0$ holds in the neighborhood of $\bm{n}^0$. 
	Specifically, 
	letting $z_1=n_1-ns_1$ and $z_2=n_2-ns_2$, 
	we have 
	\begin{eqnarray}\label{Lyapunov-DV}
		\frac{\mathrm{d}{V}(t)}{\mathrm{d}t}
		=-2 (\delta_1\,\delta_2) B (\delta_1\,\delta_2)^ \mathrm{T} (z_1(t)^2+z_2(t)^2)+H_v^2(z_1(t),z_2(t)).
	\end{eqnarray}
	where $H_v^2(z_1(t),z_2(t))$ represents the high-order remained terms, and the lowest power of it is cubic. Obviously, $	\frac{\mathrm{d}{V}(t)}{\mathrm{d}t}<0$ holds for arbitrary $(z_1,z_2)$  in the neighborhood of $(0,0)$ except $(0,0)$, as $B$ is a symmetric positive definite matrix. In other words, $	\frac{\mathrm{d}{V}(t)}{\mathrm{d}t}<0$ holds for arbitrary $(n_1,n_2)$  in the neighborhood of $\bm{n}^1$, implying function ${V}$ is also local Lyapunov function for system in the neighbor of $\bm{n}^1$.

	\section{Case studies}\label{s5}
	In this section, the control performance of the proposed resilient control by comparative case studies are demonstrated. In particular,  we benchmark with  the classical perimeter control (CPC) and the traffic-resilience oriented control  proposed in  \cite{gao2022resilient}.
	
	\subsection{Numerical verification}\label{s5.1}
	A two-region system is set up for case studies, where the network and traffic settings follow the Downtown San Francisco road network from Aboudolas and Geroliminis’ work (\cite{aboudolas2013perimeter}). We adopt the function form of the MFD models from \cite{gao2022resilient} for comparison purposes, i.e., $G_1(n_1)={-\frac{28}{289}}{n_1}(n_1-850)$  for region 1 and $G_2(n_2)={-\frac{32}{961}}{n_2}(n_2-1550)$ for region 2, with  maximum capacity $G_{1,max}=7 \times 10^4$, $G_{2,max}=8 \times 10^4$ $[veh/h]$ and jam accumulations $p_1=1700$, $p_2=3100$ $[veh]$. 
	
\begin{figure}
		\centering
		\begin{subfigure}{.33\textwidth}
			\centering
			\includegraphics[width=5.5cm]{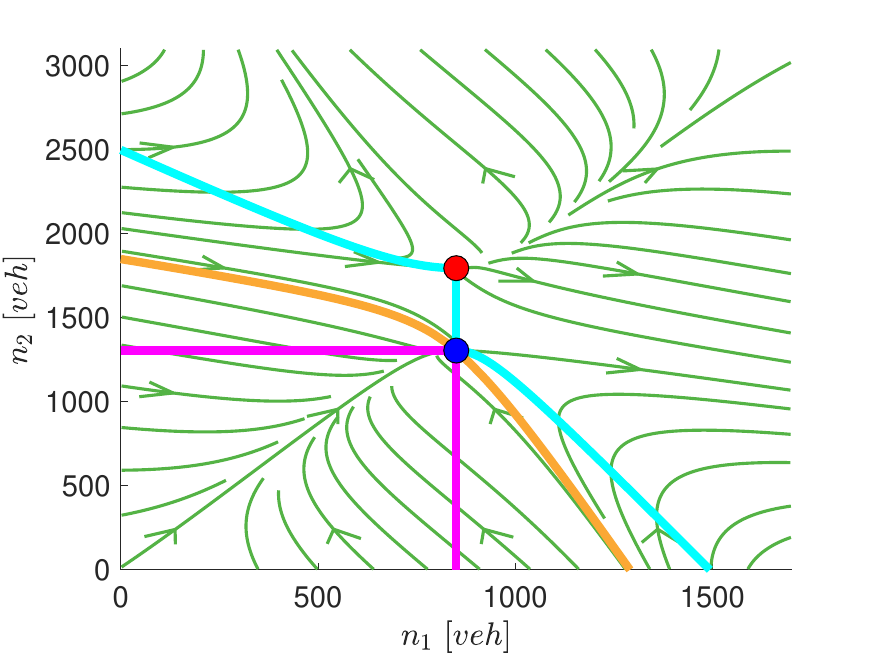}
			\caption{$\mathbb{K}^{2a}\wedge \mathbb{H}_1$ (scenario 1)}
			\label{fig32.1}
		\end{subfigure}
		\begin{subfigure}{.33\textwidth}
			\centering
			\includegraphics[width=5.5cm]{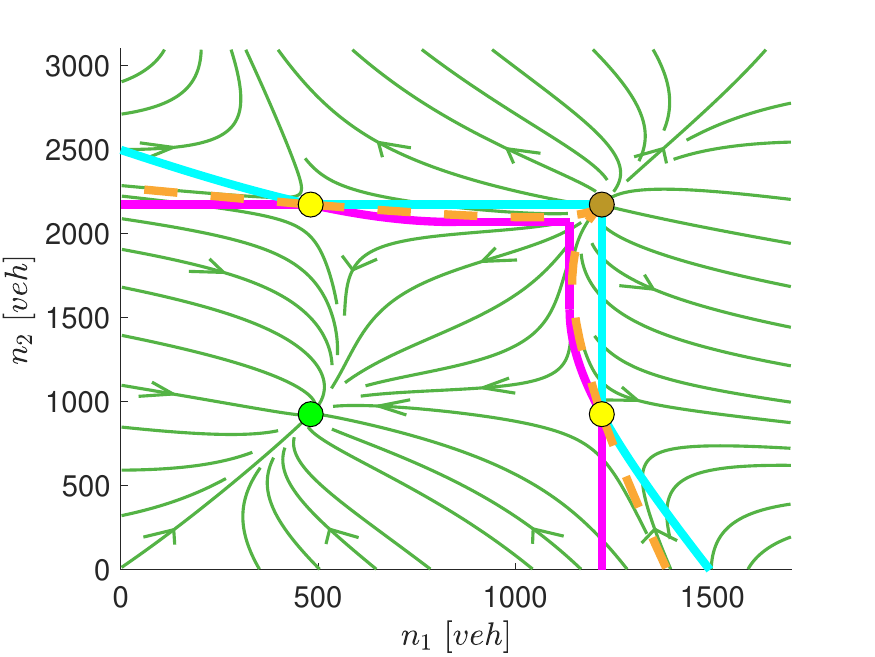}
			\caption{$\mathbb{K}^{4}_1 \wedge \mathbb{H}_1$ (scenario 9)}
			\label{fig32.2}
		\end{subfigure}
		\caption{
			Numerical simulation phase portraits for system \eqref{the model} under Conditions (a) $\mathbb{K}^{2a}\wedge \mathbb{H}_1$ (scenario 1) and (b) $\mathbb{K}^{4}_1 \wedge \mathbb{H}_2$ (scenario 9). 
			Red circle: undesired saddle-node equilibrium. Blue circle: desired uncongested saddle-node equilibrium. Green circle: desired uncongested stable equilibrium. Yellow circle: saddle equilibrium. Yellow-brown circle: unstable equilibrium. Green line with arrows: the trajectories. Purple line and  light blue line represent the inner and outer estimations of attraction regions, respectively. The yellow lines represent the numerical attraction region boundaries, solid yellow lines indicating the boundary line belongs to attraction region, while dashed yellow lines indicating the boundary line does not belong to attraction region.
			The agreement of this figure with Figs. \ref{fig05.1} and Fig. \ref{fig08.5} validate the theoretical analysis.}
		\label{fig32}
	\end{figure}

	Based on this MFD data, we consider traffic flow dynamics following Eq. \eqref{the model} and \eqref{boundary condition} as defined in Sec. \ref{s2.1}. We first validate the theoretical results of inner and outer estimation of attraction regions for both two and four equilibria derived in Sec. \ref{s3} (refer to Theorems \ref{attraction region in 3}, \ref{inner and outer estimations of attraction regions in 4.1} and \ref{attraction region in 4.2},  Figs. \ref{fig05}, \ref{fig06} and \ref{fig09}), accumulating to a total of 28 distinct phase portraits (8 for two-equilibria system and 20 for four-equilibria system). To achieve this, we consider 28 different scenarios, labeled as Scenario 1 to Scenario 28. Each scenario has distinct allowed flow rates $u_i$ and net fixed demand $d_i$ ($i=1,2$), corresponding to the 28 conditions in Sec. \ref{s3}. For example, Scenario 1 corresponds to Condition $\mathbb{K}^{2a}\wedge \mathbb{H}_1$. We showcase two scenarios (scenario 1 and scenario 9) in  Fig.~\ref{fig32}. Fig.~\ref{fig32.1} depicts the phase portrait for Scenario 1: $u_1=0.4,u_2=0.5128, d_1=3\times10^4, d_2=5\times10^4$ $[veh/h]$, satisfying Condition $\mathbb{K}^{2a}\wedge \mathbb{H}_1$, corresponding to the two-equilibrium case. It verifies Theorem 1 (or Fig. \ref{fig05.1}). Fig.~\ref{fig32.2} presents the phase portrait for Scenario 9: $u_1=0.3,u_2=0.4, d_1=3\times10^4, d_1=5\times10^4$ $[veh/h]$, satisfying Condition $\mathbb{K}^{4}_1 \wedge \mathbb{H}_1$, corresponding to the four-equilibrium case. It verifies Theorem 2 (or Fig. \ref{fig08.5}). Parameters and numerical results for other scenarios can be found in App. \ref{B}: Fig. \ref{fig11}, Fig. \ref{fig12} and Table \ref{tab5}. The results are also consistent with our theoretical findings.

\begin{figure}
		\centering
		\includegraphics[width=17cm]{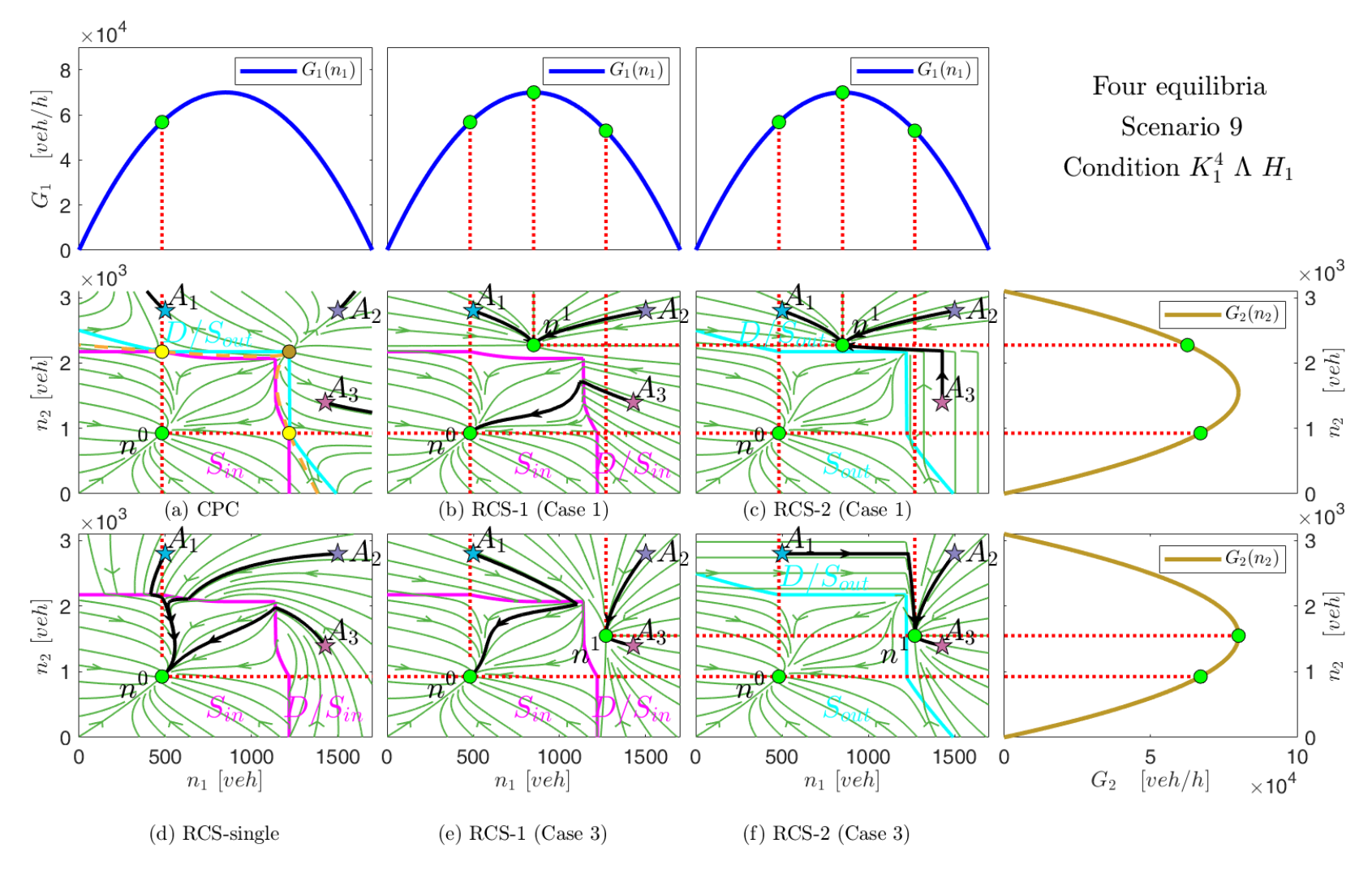}
		\caption{Numerical simulation phase portraits of system \eqref{the model} with (a) CPC, (b) RCS-1 (case 1), (c) RCS-2 (case 1), (d) RCS-single, (e) RCS-1 (case 3) and  (f) RCS-2 (case 3) under condition $\mathbb{K}^{4}_1 \wedge \mathbb{H}_1$ (scenario 9). Green circle: uncongested equilibrium $\bm{n}^0$ and alternative steady state $\bm{n}^1$. Pentagrams denote three 
			situations (i.e., three initial recovery states), black lines departing from the pentagrams are the corresponding recovery trajectories. CPC cannot recover the system; RCS-single recovers to the original equilibrium $\bm{n}^0=(481,926)$, while RCS-1 and RCS-2 can recover to alternative stable states $\bm{n}^1$. Here we choose $\bm{n}^1=(850,2274)$ for Case 1 and $\bm{n}^1=(1269,1550)$ for Case 3, both satisfying the selectable range of alternative stable states Eq. \ref{lastscope of n1}. Note that Case $i (i=1,2,3)$ denotes the case where the alternative steady state $\bm{n}^{1} \in R_i$. 
			The illustration of $R_i$ can be found in Fig. \ref{fig24}. 
		}
		\label{fig22}
	\end{figure}

	Next, 
	numerical phase portraits of the controlled systems under RCS-1 and RCS-2 are demonstrated,
	as depicted in Fig. \ref{fig22}(b), \ref{fig22}(c), and \ref{fig22}(e), \ref{fig22}(f). 
	Here we use Scenario 9 as a showcase, which corresponds to the complex four-equilibria case. For the simpler two-equilibria case, the control scheme can also be applicable.
	Note that Case $i (i=1,2,3)$ denotes the case where the alternative steady state $\bm{n}^{1} \in R_i$. 
	The illustration of $R_i$ can be found in Fig. \ref{fig24}. 
	Here, we take Case 1 ($\bm{n}^{1} \in R_1$) and Case 3 ($\bm{n}^{1} \in R_3$) as examples. 
	Fig. \ref{fig22}(a) represents the results under CPC control, 
	serving as a benchmark control. 
	Let us focus on the  part $\mathbb{D}\setminus S_{in}$ in Fig. \ref{fig22}(a), (b), and (e). 
	In Fig. \ref{fig22}(a), 
	a small portion of the trajectories in $\mathbb{D}\setminus S_{in}$ enters $S_{in}$, 
	while a larger portion of the trajectories moves towards the upper or right boundaries of the phase portrait, 
	approaching grid-lock, 
	i.e., high traffic density but low outflow. 
	In contrast, 
	Fig. \ref{fig22}(b) and \ref{fig22}(e) demonstrate that RCS-1 is capable of guiding trajectories starting from $\mathbb{D}\setminus S_{in}$ towards either the non-congested equilibrium point $\bm{n}^{0}$ 
	or the alternative steady state $\bm{n}^{1}$ in $\mathbb{D}\setminus S_{in}$, avoiding grid-lock. 
	Similarly, 
	in Fig. \ref{fig22}(a), trajectories in $\mathbb{D}\setminus S_{out}$ all move towards the upper and right boundaries of the phase portrait, while Fig. \ref{fig22}(c) and \ref{fig22}(f) confirm that RCS-2 directs trajectories in $\mathbb{D}\setminus S_{out}$ towards the alternative steady state $\bm{n}^{1}$ in $\mathbb{D}\setminus S_{out}$. 
	These results demonstrate the recoverability under RCS-1 and RCS-2. 
	Furthermore, the similar trajectories in $R1 \vee R2$ (upper region in $\mathbb{D}\setminus S_{out}$) in Fig. \ref{fig22}(b) and \ref{fig22}(c) confirm that there exist some initial states such that the recovery trajectories starting from them under RCS-1 and RCS-2 are identical, as illustrated in Remark 2. This is because, in $R1 \vee R2$, RCS-1 and RCS-2 have the same control formulation.

\begin{figure}
		\centering
		\begin{subfigure}{.33\textwidth}
			\centering
			\includegraphics[width=5.5cm]{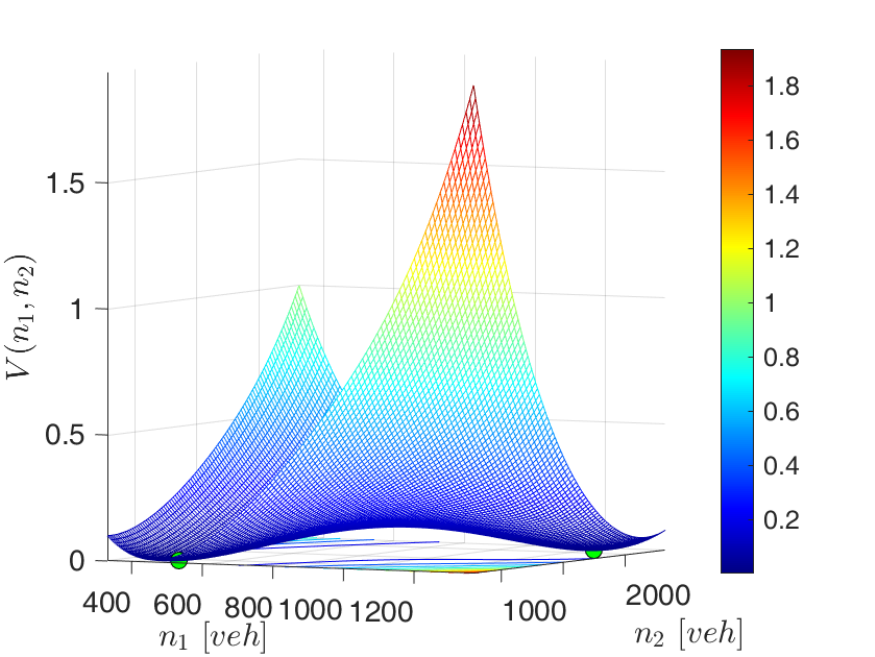}
			\caption{V function}
			\label{fig14.1}
		\end{subfigure}
		\begin{subfigure}{.33\textwidth}
			\centering
			\includegraphics[width=5.5cm]{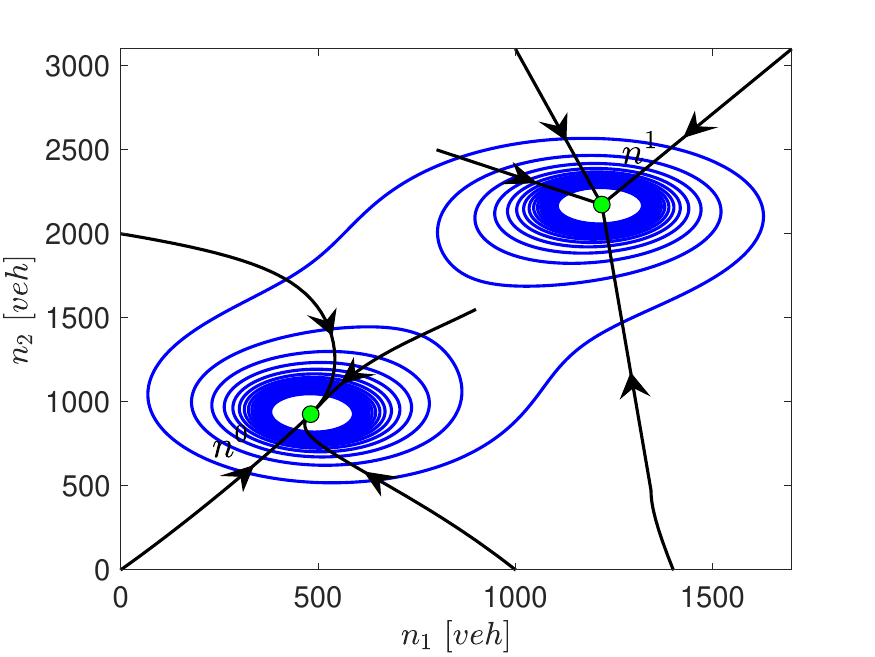}
			\caption{Projection of the V function}
			\label{fig14.2}
		\end{subfigure}
		\caption{(a) V function with respect to $n_1$ and $n_2$. (b) Projection of the V function on the $n_1-n_2$ plane. The black line with arrows represents several trajectories of system with RCS-2. Green circle: the equilibrium points of system with RCS-2 (case 2). Note that scenario 9 is considered. Here, $\bm{n}^{0}=(481,926)$ and $\bm{n}^1=(1219,2174)$.}
	\label{fig14}
\end{figure}

	Moreover, for the case where the system has a stable equilibrium (we here use scenario 9 as a showcase), we can find the function  $V(t)$ defined in Sec. \ref{s4.2}, as shown in Fig.~\ref{fig14.1}, is a local Lyapunov function for the controlled system with RCS-2. Moreover, the projection of $V(t)$ in the plane $n_1 - n_2$ is shown in the Fig.~\ref{fig14.2}. Note that the local Lyapunov is only applicable to four-equilibria system, and is not suitable for two-equilibria system, as the there exist no stable equilibria in two-equilibria system.
	
\subsection{Comparison on proposed two control schemes with CPC and RCS-single}\label{s5.2}

In this section, we compare our proposed control schemes (RCS-1 and RCS-2) with two classical perimeter control strategies (RCS-single in \cite{gao2022resilient} and CPC in \cite{haddad2012stability}) to show the resilience of our approaches. We maintain the same MFD setup as in the previous section, and use scenario 9 as an example. 
Fig. \ref{fig22}(d) depicts the numerical simulation of phase portrait for the control system with RCS-single, where trajectories within the whole region recover to the original equilibrium $\bm{n}^{0}$. The key distinction between RCS-single and proposed schemes lies in the trajectories within the part: our proposed control schemes can recover to the alternative steady state $\bm{n}^{1}$, while RCS-single can only recover to the original equilibrium $\bm{n}^{0}$.

\begin{figure}
	\centering
	\includegraphics[width=15cm]{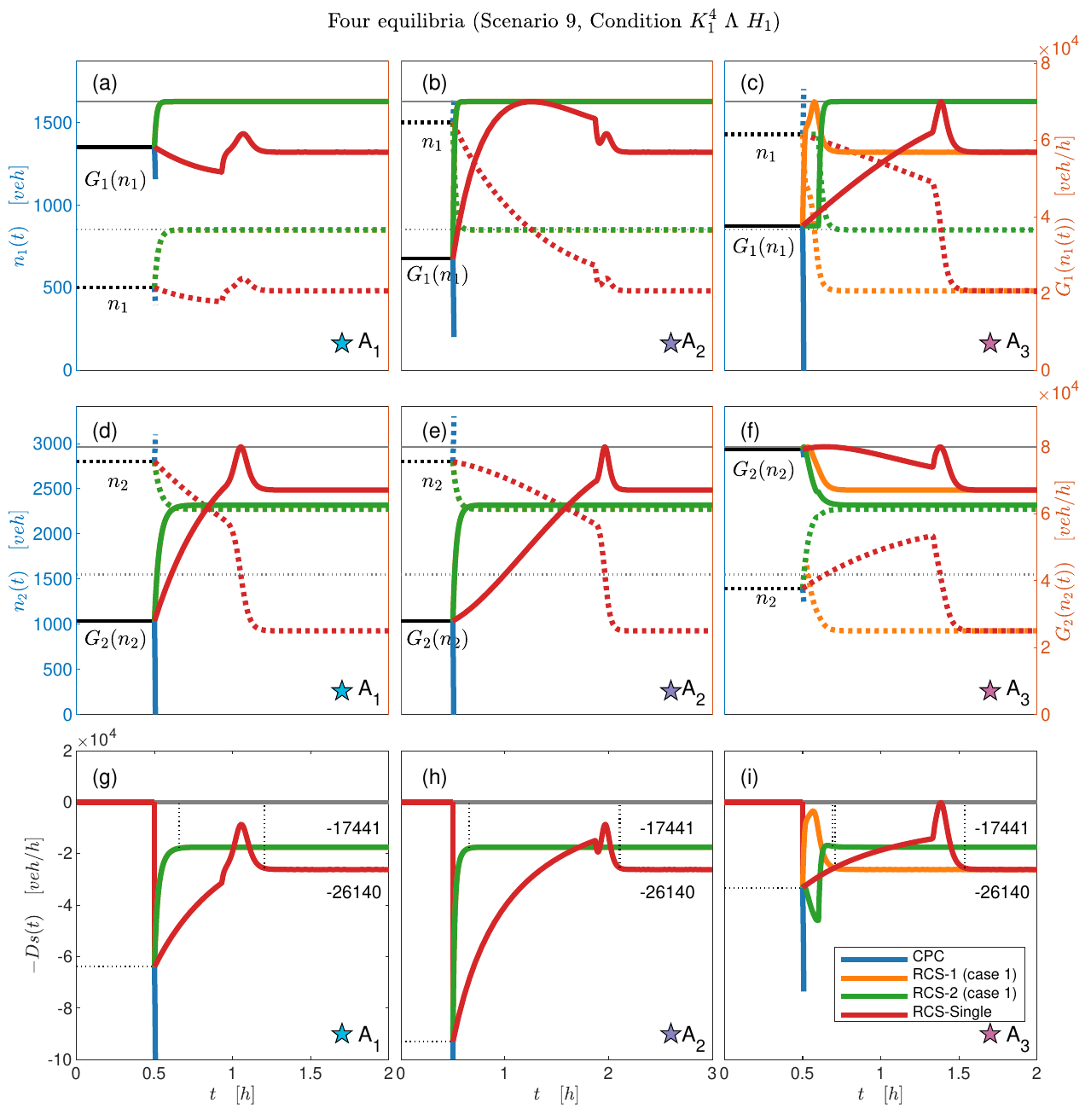}
	\caption{Vehicle accumulations (dotted line) and completion flow rates (solid line) of Region 1 (((a), (b) and (c))) and Region 2 ((d), (e) and (f)). 
	The numerical diagrams of deviation from maximum complete flow $-D_s$ ((g), (h) and (i)) evolving with time $t$ under CPC, RCS-1, RCS-2 and RCS-single. Scenario 9 is considered. 
	Note that for (a)-(f), we employ dual y-axes. The left y-axis represents vehicle accumulation (density), while the right y-axis represents completion flow.
	$A_i$ ($i=1,2,3$) are the same as the three pentagrams in Fig.~\ref{fig22}. The results in (a)(d)(g) ((b)(e)(h) and (c)(f)(i)) are obtained by following the recovery trajectories  (black lines in Fig.~\ref{fig22}) departing from the pentagram $A_1$ ($A_2$ and $A_3$). 
			Reaffirming that RCS-single, RCS-1, and RCS-2 can ensure recoverability, while CPC can not. The resilience triangles in (g), (h) and (i)  reveal that RCS-1 and RCS-2 outperform RCS-single, as RCS-1 and RCS-2 have smaller resilience triangles (loss) in most scenarios.}
	\label{fig15}
\end{figure}

Then, we calculate the resilience measure (defined in Eq. \eqref{tran resilience}) under different control schemes. Resilience measures are situation-dependent, where situations referring to states (vehicle accumulations) after disturbances, also known as initial recovery states. For instance, we assume state perturbations suddenly occur at moment $t = 0.5$ $[h]$. Situation $i$ denotes the disturbed state at $t = 0.5$ $[h]$ as $A_i$ ($i=1,2,3$). Here, we consider three situations, i.e., $A_1=(500,2800)$, $A_2=(1500,2800)$ and $A_3=(1428,1395)$, marked as three pentagrams in  Fig. \ref{fig22}. Following these trajectories (the black lines in  Fig. \ref{fig22}) departing from the pentagrams,
we plot the vehicle accumulation $n_1(t)$, 
completion flow rate $G_1(n_1(t))$,  
and the numerical diagrams of deviation from maximum completion flow ($G_{1,max}$ and $G_{2,max}$) evolving with time  under each control scheme (CPC, RCS-1, RCS-2, RCS-single) in Fig. \ref{fig15}. 
Here we use Case 1 (corresponding to Fig. \ref{fig22}(b) and \ref{fig22}(c)) as an example to illustrate performance the proposed control schemes, and the results for Case 2 and  Case 3 can be found in App. \ref{B}, Fig.~\ref{fig15.2} and Fig.~\ref{fig15.3}.

Clearly, as depicted in Fig. \ref{fig15}(g), (h), and (i), traffic system collapses rapidly under CPC (blue solid line) (within 0.006 $[h]$ for $A_1$, 0.01 $[h]$ for $A_2$ and 0.007 $[h]$ for $A_3$); under RCS-single (red solid line), system initially rises slowly approaching maximum completion flow, then rebounds and stabilizes at a deviation of 26140 $[veh/h]$ (within 0.70 $[h]$ for $A_1$, 1.61 $[h]$ time for $A_2$ and 1.04 $[h]$ for $A_3$); under the proposed RCS-1 (yellow solid line), the system swiftly recovers to a deviation of 17441 $[veh/h]$ (within 0.16 $[h]$ for $A_1$ and 0.16 $[h]$ for $A_2$, coinciding with the green solid line), or it first rises rapidly approaching the maximum completion flow, then rebounds and stabilizes at a deviation of 26140 $[veh/h]$ (within 0.21 $[h]$ for $A_3$); under the proposed RCS-2 (green solid line), the system swiftly recovers to a deviation of 17441 $[veh/h]$ (within 0.16 $[h]$ for $A_1$ and 0.16 $[h]$ for $A_2$), or it resists the continued deterioration of the system, guiding a swift recovery, and stabilizes at a deviation of 17441 $[veh/h]$ (within 0.19 $[h]$ for $A_3$).
Here, we take $A_1$ as an example to elucidate the reasons for the aforementioned phenomena. Under CPC, the vehicle accumulation $n_2(t)$ (Fig. \ref{fig15}(d), blue dotted line) develops from 2800 to a jam vehicle accumulation of 3100 $[veh]$ within 0.006 $[h]$, leading to the completion flow $G_2(n_2(t))$ (blue solid line) tending towards 0. This causes Region 2 to enter a gridlock state from which it cannot recover.
	Under RCS-single, the vehicle accumulation $n_1(t)$ (Fig. \ref{fig15}(a), red dotted line) initially decreases slowly from 500 $[veh]$, then slowly increases before decreasing again, finally stabilizing at the original equilibrium of 481 $[veh]$. Correspondingly, the completion flow decreases slowly, then increases slowly, before decreasing again and stabilizing at 56818 $[veh/h]$. Simultaneously, the vehicle accumulation $n_2(t)$ (Fig. \ref{fig15}(d), red dotted line) decreases slowly at first, then rapidly decreases and stabilizes at 926 $[veh]$. Consequently, the completion flow $G_2(n_2(t))$ (Fig. \ref{fig15}(d), red solid line) first rises slowly to maximum completion flow, then rapidly decreases and stabilizes at 67045 $[veh/h]$. Therefore, the recoverability of RCS-single is guaranteed, but with a longer duration.
	Under the proposed RCS-1 and RCS-2, due to high travel demand, the vehicle accumulation $n_1(t)$ (Fig. \ref{fig15}(a), green dotted line) rapidly increases from 500 $[veh]$ and stabilizes at 850, resulting in the completion flow (Fig. \ref{fig15}(a), green solid line) rapidly rising and stabilizing at the maximum completion flow of $7\times 10^4$ $[veh/h]$. Simultaneously, the vehicle accumulation $n_2(t)$ (Fig. \ref{fig15}(b), green dotted line) rapidly decreases from 2800 to 2274 $[veh]$, leading the completion flow $G_2(n_2(t))$ quickly stabilized at 62559 $[veh/h]$. Consequently, RCS-1 and RCS-2 can ensure rapid recoverability, outperforming RCS-single.
	It is noteworthy that in this scenario, RCS-1 and RCS-2 exhibit identical recovery trajectories. We also provide the cases of RCS-1 and RCS-2 employing different recovery trajectories, as shown in Fig. \ref{fig15}(c), (f), and (i).

For a more accurate comparison, 
based on Fig. \ref{fig15} and Eq. \eqref{tran resilience}, 
we integrate the curve of $-D_s$ versus $t$ to obtain the value of resilience measure, $R$ under RCS-1, RCS-2, and RCS-single. 
The results are outlined in Table \ref{tab4}. 
Noteworthy, the absolute value of $R$ corresponds to the area of the resilience triangle during the recovery period, and it signifies the loss of actual completed trips compared to the maximum number of trips that could have been completed during the recovery period. 
A larger resilience measure $R$ corresponds to a smaller area of the resilience triangle (less loss), 
indicating greater resilience. 
As presented in Table \ref{tab4}, 
for all three Situations ($A_1$, $A_2$ and $A_3$), 
resilience measures for RCS-1 and RCS-2 are significantly larger than RCS-single, demonstrating exceptional recovery capabilities. 

\begin{table}
	\caption{Resilience measure $[veh]$.}
	\label{tab4}
	\centering
	\fontsize{6.5}{8}\selectfont
	\setlength{\tabcolsep}{3mm}{
		\begin{threeparttable}
			\begin{tabular}{c|ccc|ccc|ccc}
				\toprule
				\multirow{2}{*}{resilience control}&
				\multicolumn{3}{c|}{$A_1$} &
				\multicolumn{3}{c|}{$A_2$} &
				\multicolumn{3}{c}{$A_3$} \cr
				&Case 1&Case 2&Case 3&Case 1&Case 2&Case 3&Case 1&Case 2&Case 3\cr		
				\midrule
				RCS-1     &-4031& -4384& -3110& -4284& -5646& -3681&  -4250&  -3302&  -1710\cr
				RCS-2     &-4031& -4384& -5800& -4284& -6558& -6615&  -3776&  -3302&  -1744\cr
				RCS-single&-24568& -56501& -20584&-24568& -56501& -20584&-24568& -56501& -20584\cr
				%
				\bottomrule
			\end{tabular}				
			\begin{tablenotes}
				\footnotesize
				\item Case $i (i=1,2,3)$ denotes the case where the alternative steady state $\bm{n}^{1} \in R_i$, as shown in Fig. \ref{fig24}.
				$A_i$ denotes three situations, corresponding to three pentagrams in Fig.~\ref{fig22} and Fig.~\ref{fig15}.  
			\end{tablenotes}
	\end{threeparttable}}
\end{table}

\section{Conclusion}
\label{s6}
In this paper, we define the ecological resilience of urban traffic, and further propose a resilience control methodology from the perspective of ecological resilience. Specifically, the ecological resilience of urban traffics is defined by the ability for a traffic system to resist uncertain perturbations by shifting to alternative states. The resilience control methodology comprises three aspects: portraying the recoverable scopes, designing alternative steady states, controlling system to shift to alternative steady states for adapting large disturbances. Among them, the recoverable scopes are portrayed by inner and outer estimations of attraction regions; the alternative steady states are set close to the optimal state and outside the recoverable scopes of the original equilibrium; the controller ensures the local stability of the alternative steady states, without changing the trajectories inside the recoverable scopes of the original equilibrium as much as possible. 
We implemented the proposed control framework in a two-region traffic network described by parabolic MFD dynamic, and designed proposed control schemes (RCS-1 and RCS-2). Comparisons with the classical urban traffic resilience control schemes (CPC and RCS-single, etc.) show that, proposed resilience control schemes (RCS-1 and RCS-2) has a better adaptability and generates a greater resilience measure. Noteworthy, RCS-1 differs with RCS-2 in control intensity, control formulation and control effect. Note that, for multi-region (more than two-region) MFD dynamics, our resilience control idea can still be applied. The difficulty in implementing the resilience control idea is that for N-dimensional system, the equilibrium point will be more complex, and the attraction region maybe more difficult to obtain. Moreover, the inner estimation of attraction region we can obtain may shrink and the outer estimation of attraction region may expand, potentially affecting the effectiveness of the regulation scheme.

There is still work to be done in the future. Given this paper considers mainly constant MFD dynamics, and it is worth to extend the methodology of this paper to more general time-varying MFD dynamics in the future. Moreover, a feasible resilience control is provided in this paper, and future work will be extended to find an optimal resilience control scheme.

\ACKNOWLEDGMENT{%
	This work was supported by the National Key R$\&$D Program of China (No.2022YFA1005103), the National Natural Science Foundation of China (12371452, 72225012, 72288101, 71822101, 12101030) and the Fundamental Research Funds for the Central Universities.}

\bibliographystyle{informs2014trsc} 
\bibliography{mybib} 


\newpage
\begin{APPENDICES}
	\section{Global phase portrait and attraction region for four-equilibria system} \label{A} 
	we will consider the four-equilibria cases,
	derive its qualitative characteristics under CPC ($u_1$ and $u_2$), 
	and further provide an inner and outer estimations of attraction regions.
	Specifically,
	local stability verification of the four-equilibria system will be shown in App. \ref{A.1};
	global phase portrait derivation
	is performed in App. \ref{A.2}; 
	and spontaneous attraction region estimation of four-equilibria system will be included in App. \ref{A.3}.
	
	\subsection{Local stability verification}\label{A.1}
	According to Condition $(\mathbb{K}^{4})$, we have:
	\begin{equation}\label{d1d2 in case3}
		\begin{split}
			{u_2}d_2+d_1-\left( 1-u_1u_2\right) G_{1,max}<0,\\
			{u_1}d_1+d_2-\left( 1-u_1u_2\right) G_{2,max}<0.
		\end{split}
	\end{equation}
	
	For system \eqref{the model} satisfying \eqref{d1d2 in case3}, the four equilibria are denoted as $P^4_m=  \left(p^4_{m,1},p^4_{m,2}\right)$, where $m=1,2,3,4$, $p^4_{m,1}=\frac{{p_1}+(-1)^m\sqrt{{p_1}^2-\frac{4{d_1} + 4{d_2}{u_2}}{{a_1}\left(1-{u_1}{u_2}\right)}}}{2}$; 
	$p^4_{m,2}=\frac{{p_2}-\sqrt{{p_2}^2-\frac{4{d_2} + 4{d_1}{u_1}}{{a_2}\left(1-{u_1}{u_2}\right)}}}{2}$ when $m=1,2$ and 
	$p^4_{m,2}=\frac{{p_2}+\sqrt{{p_2}^2-\frac{4{d_2} + 4{d_1}{u_1}}{{a_2}\left(1-{u_1}{u_2}\right)}}}{2}$ when $m=3,4$.

	We first verify the local stability of the four equilibria for system \eqref{the model} satisfying \eqref{d1d2 in case3}. Under Condition ($\mathbb{K}^{4} \wedge \mathbb{H}$), we have the following proposition for its four equilibria.
	
	\begin{proposition}~\label{stability in 4}
		Under Condition ($\mathbb{K}^{4} \wedge \mathbb{H}$), for system \eqref{the model},  $P^4_1$ is a locally stable node, $P^4_2$ and $P^4_3$ are saddle points, $P^4_4$ is an unstable node.
	\end{proposition}
	
	Proof:
		For system \eqref{the model}, we can get the derivative operator $A|_{P^4_m}$ ($m=1,2,3,4$) at the equilibrium $P^4_ m$ as follows:
		\begin{equation*}
			A|_{P^4_m}=Df|_{P^4_m}=\left(
			\begin{array}{ccccc}
				\frac{\partial F_1(n_1,n_2)}{\partial n_1} & \frac{\partial F_1(n_1,n_2)}{\partial n_2} \\
				\frac{\partial F_2(n_1,n_2)}{\partial n_1} & \frac{\partial F_2(n_1,n_2)}{\partial n_2}\\	\end{array}	\right)\Bigg|_{P^4_ m}=\left(
			\begin{array}{ccccc}
				2a_1(p^4_{m,1}-\frac{p_1}{2}) & -2 u_2 a_2 (p^4_{m,2}-\frac{p_2}{2}) \\
				-2 u_1 a_1 (p^4_{m,1}-\frac{p_1}{2}) & 2a_2(p^4_{m,2}-\frac{p_2}{2}) \\	\end{array}	\right).
		\end{equation*}
		Thus, the two eigenvalues $\lambda_{m,j}$ ($j=1,2$) of $A|_{P^4_m}$ satisfy:
		\begin{eqnarray}	
			\lambda^2_{m,j}-2\left[ a_1(p^4_{m,1}-\frac{p_1}{2})+  a_2(p^4_{m,2}-\frac{p_2}{2})\right] \lambda_{m,j}
			+4(1- u_1 u_2) a_1 a_2(p^4_{m,1}-\frac{p_1}{2})(p^4_{m,2}-\frac{p_2}{2})=0.
			\label{equ27}
		\end{eqnarray}
		Note that the discriminant $\Delta_m$ for Eq.~\eqref{equ27} is:
		\begin{eqnarray*}
			\Delta_m=4\left[ a_1(p^4_{m,1}-\frac{p_1}{2})+  a_2(p^4_{m,2}-\frac{p_2}{2})\right]^2
			-16(1- u_1 u_2) a_1 a_2(p^4_{m,1}-\frac{p_1}{2})(p^4_{m,2}-\frac{p_2}{2}).
		\end{eqnarray*}
		
		\begin{enumerate}
			\item For $P^4_1$=($p^4_{1,1}$,$p^4_{1,2}$), we can get $\Delta_1>0$, then Eq.~\eqref{equ27} has two real roots $\lambda_{1,1}$ and $\lambda_{1,2}$. Moreover, we have: $\lambda_{1,1}+\lambda_{1,2}=2\left[ a_1(p^4_{1,1}-\frac{p_1}{2})+  a_2(p^4_{1,2}-\frac{p_2}{2})\right]<0$, $\lambda_{1,1}\lambda_{1,2}=4(1- u_1 u_2) a_1 a_2(p^4_{1,1}-\frac{p_1}{2})(p^4_{1,2}-\frac{p_2}{2})>0$, which implies  $\lambda_{1,1}<0,\lambda_{1,2}<0$. Thus, we can conclude the equilibrium point $P^4_1$ is a locally stable node.

			\item For $P^4_2$=($p^4_{2,1}$,$p^4_{2,2}$), we can  get $\Delta_2>0$, then Eq.~\eqref{equ27} has two real roots $\lambda_{2,1}$ and $\lambda_{2,2}$, where $\lambda_{2,1}<\lambda_{2,2}$. Moreover, we have: $\lambda_{2,1}\lambda_{2,2}=4(1- u_1 u_2) a_1 a_2(P_{21}-\frac{p_1}{2})(p^4_{2,2}-\frac{p_2}{2})<0$, which implies  $\lambda_{2,1}<0<\lambda_{2,2}$. Thus, we have that the equilibrium point $P^4_2$ is a saddle point.
			
			\item For $P^4_3$=($p^4_{3,1}$,$p^4_{3,2}$), we also have $\Delta_3>0$, then Eq.~\eqref{equ27} has two real roots $\lambda_{3,1}$ and $\lambda_{3,2}$, where $\lambda_{3,1}<\lambda_{3,2}$. Moreover, we have: $\lambda_{3,1}\lambda_{3,2}<0$, which indicates  $\lambda_{3,1}<0<\lambda_{3,2}$. Thus, $P^4_3$ is a saddle point.
			
			\item For $P^4_4$=($p^4_{4,1}$,$p^4_{4,2}$), we  have $\Delta_4>0$, Moreover, the two real roots $\lambda_{41}$ and $\lambda_{4,2}$ of Eq.~\eqref{equ27} satisfy $\lambda_{4,1}+\lambda_{4,2}>0$, $\lambda_{4,1}\lambda_{4,2}>0$, which implies $\lambda_{4,1}>0,\lambda_{4,2}>0$. Thus, $P^4_4$ is an unstable node.
		\end{enumerate}
	$\hfill \square$

	\subsection{Global phase portrait derivation} \label{A.2}
	Subsequently, for the four-equilibria system \eqref{the model} under Condition  ($\mathbb{K}^{4} \wedge \mathbb{H}$),  the global phase portrait will be derived. First
	letting $x_1(t)= n_1(t)-p^4_{1,1}$ and $x_2(t)= n_2(t)-p^4_{1,2}$, we simplify system  \eqref{the model} as:
	\begin{equation}\label{Simplify system in case 3}
		\begin{split}
			&\frac{\mathrm{d}{x}_1(t)}{\mathrm{d}t}=
			-{u_2}{a_2}(x_2-s_2)^2+a_1(x_1-s_1)^2+M_1, \\
			&\frac{\mathrm{d}{x}_2(t)}{\mathrm{d}t}=
			-{u_1}{a_1}(x_1-s_1)^2+a_2(x_2-s_2)^2+M_2,
		\end{split}
	\end{equation}
	where $s_1=\frac{1}{2}\sqrt{{p_1}^2-\frac{4{d_1} + 4{u_2}{d_2}}{{a_1}(1-{u_1}{u_2})}}$, $s_2=\frac{1}{2}\sqrt{{p_2}^2-\frac{4{d_2} + 4{u_1}{d_1}}{{a_2}(1-{u_1}{u_2})}}$,
	$M_1=d_1 +u_2 \frac{{a_2 p_2}^2}{4}-\frac{{a_1 p_1}^2}{4}$, and $M_2=d_2 +u_1 \frac{{a_1 p_1}^2}{4}-\frac{{a_2 p_2}^2}{4}$.
	Obviously, the corresponding four equilibria of system  \eqref{Simplify system in case 3} are $\hat{P}^4_1=(0,0)$, $\hat{P}^4_2=(2s_1,0)$,
	$\hat{P}^4_3=(0,2s_2)$ and $\hat{P}^4_4=(2s_1,2s_2)$. Note that the four equilibria of the original system \eqref{the model} are ${P}^4_1=(\frac{p_1-2s_1}{2},\frac{p_2-2s_2}{2})$, ${P}^4_2=(\frac{p_1+2s_1}{2},\frac{p_2-2s_2}{2})$, ${P}^4_3=(\frac{p_1-2s_1}{2},\frac{p_2+2s_2}{2})$, and ${P}^4_4=(\frac{p_1+2s_1}{2},\frac{p_2+2s_2}{2})$. 
	Moreover, since $	M_1+u_2M_2= d_1+u_2d_2-\left( 1-u_1u_2\right) G_{1,max}$ and $
	M_2+u_1M_1= d_2+u_1d_1-\left( 1-u_1u_2\right) G_{2,max}$,  we have that Condition $\mathbb{K}^{4}$ (or inequality \eqref{d1d2 in case3}) holds if and only if Condition ($\hat{\mathbb{K}}^{4}$):
	\begin{equation}\label{condition 4}
		\begin{split}
			M_1+u_2M_2<0,\\
			M_2+u_1M_1<0,
		\end{split}
	\end{equation}
	holds. Further, Condition ($\hat{\mathbb{K}}^{4}$) can be divided into five sub-conditions: 
	\begin{enumerate}[(1)]
		\item Condition $(\hat{\mathbb{K}}^{4}_1)$: $M_1<0\wedge M_2<0$;
		\item Condition $(\hat{\mathbb{K}}^{4}_2)$: $M_1=0\wedge M_2<0$
		\item Condition $(\hat{\mathbb{K}}^{4}_3)$: $M_1<0\wedge M_2=0$
		\item Condition $(\hat{\mathbb{K}}^{4}_4)$: $M_1>0 \wedge M_2<\min\left\lbrace {-\frac{M_1}{u_2}}, {-u_1M_1}\right\rbrace$;
		\item Condition $(\hat{\mathbb{K}}^{4}_5)$: $M_1<\min\left\lbrace {-u_2M_2},{-\frac{M_2}{u_1}} \right\rbrace \wedge M_2>0$
		
	\end{enumerate}
	Thus in the following, we will discuss the global qualitative characteristics and attraction region for equivalence system \eqref{Simplify system in case 3} under Condition ($\hat{\mathbb{K}}^{4}_i \wedge \mathbb{H}$) respectively, where $i=1,2,\cdots ,5$. 
	
	To begin with, we consider the phase portrait under Condition ($\hat{\mathbb{K}}^{4}_1 \wedge \mathbb{H}$), which can be partitioned into four sub-Conditions:
	($\hat{\mathbb{K}}^{4}_1 \wedge \mathbb{H}_1$), ($\hat{\mathbb{K}}^{4}_1 \wedge \mathbb{H}_2$),
	($\hat{\mathbb{K}}^{4}_1 \wedge \mathbb{H}_3$) and ($\hat{\mathbb{K}}^{4}_1 \wedge \mathbb{H}_4$).

	First consider Condition  ($\hat{\mathbb{K}}^{4}_1 \wedge \mathbb{H}_1$). 
	Following the steps in Sec. \ref{s3.2}, we have that: 
	the  demarcation lines are $\frac{(x_1-s_1)^2}{-\frac{M_1}{a_1}}-\frac{(x_2-s_2)^2}{-\frac{ M_1}{u_2 a_2}}=1$ and $\frac{(x_2-s_2)^2}{-\frac{M_2}{a_2}}-\frac{(x_1-s_1)^2}{-\frac{ M_2}{u_1 a_1}}=1$, portrayed as the red and blue dotted curve respectively in Fig.~\ref{fig07.1}. In addition, we have $\frac{\mathrm{d}{x}_1(t)}{\mathrm{d}t}=0$ and $\frac{\mathrm{d}{x}_2(t)}{\mathrm{d}t}= \left(1-u_1 u_2\right) {a_2}x_2(t)\left( {x_2(t)-2s_2}\right)$ on these two red curves, indicating  $\frac{\mathrm{d}{x}_2(t)}{\mathrm{d}t} >0$ when $x_2(t)<0$ or $x_2(t)>2s_2$ holds and $\frac{\mathrm{d}{x}_2(t)}{\mathrm{d}t}<0$ when $0<x_2(t)<2s_2$ holds. Similarly, we have $\frac{\mathrm{d}{x}_2(t)}{\mathrm{d}t}=0$ and
	$\frac{\mathrm{d}{x}_1(t)}{\mathrm{d}t}= \left(1-u_1 u_2\right) {a_1}x_1(t)\left( {x_1(t)-2s_1}\right)$ on these two blue curves, implying  $\frac{\mathrm{d}{x}_1(t)}{\mathrm{d}t} >0$ when $x_1(t)<0$ or $x_1(t)>2s_1$ holds and $\frac{\mathrm{d}{x}_1(t)}{\mathrm{d}t} <0$ when $0<x_1(t)<2s_1$ holds. 
	Obviously, we can obtain nine unbounded regions in $\mathbb{R}^2$ divided by these four curves.
	Moreover, 
	Table \ref{tab3} summarizes the symbols of $\frac{\mathrm{d}{x}_i(t)}{\mathrm{d}t}$ ($i=1,2$) in these nine regions. 
	On the basis of Table \ref{tab3},
	the phase portrait of \eqref{Simplify system in case 3} under Condition ($\hat{\mathbb{K}}^{4}_1 \wedge \mathbb{H}_1$) can be obtained, as shown in Fig.~\ref{fig07.1}.
	Subsequently, based on the phase portrait \ref{fig03.1}, by utilizing proof by contradiction, we can verify that regions B, G and F are positive invariant sets. 
	Since the trajectories starting from B, G and F will not escape,
	there exist no close orbits for system \eqref{Simplify system in case 3} under Condition  ($\hat{\mathbb{K}}^{4}_1 \wedge \mathbb{H}_1$).
	
	\begin{figure}
		\centering
		\begin{subfigure}{.23\textwidth}
			\centering
			\includegraphics[width=4cm]{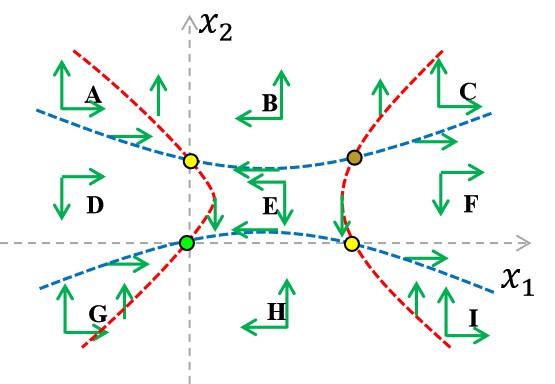}
			\caption{$\hat{\mathbb{K}}^{4}_1 \wedge \mathbb{H}_1$}
			\label{fig07.1}
		\end{subfigure}
		\begin{subfigure}{.23\textwidth}
			\centering
			\includegraphics[width=4cm]{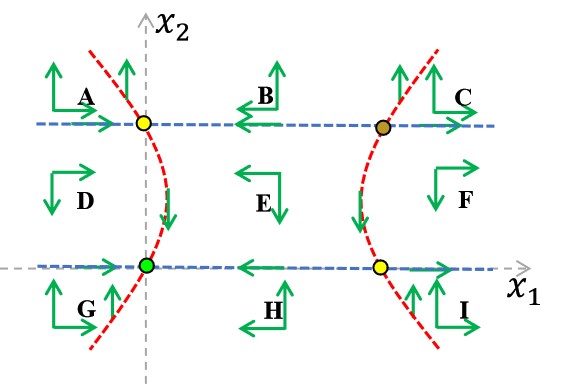}
			\caption{$\hat{\mathbb{K}}^{4}_1 \wedge \mathbb{H}_2$}
			\label{fig07.2}
		\end{subfigure}
		\begin{subfigure}{.23\textwidth}
			\centering
			\includegraphics[width=4cm]{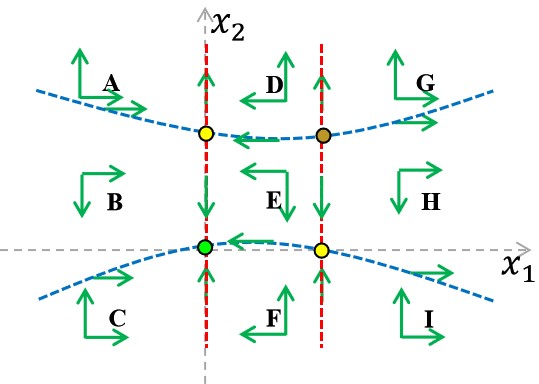}
			\caption{$\hat{\mathbb{K}}^{4}_1 \wedge \mathbb{H}_3$}
			\label{fig07.3}
		\end{subfigure}
		\begin{subfigure}{.23\textwidth}
			\centering
			\includegraphics[width=4cm]{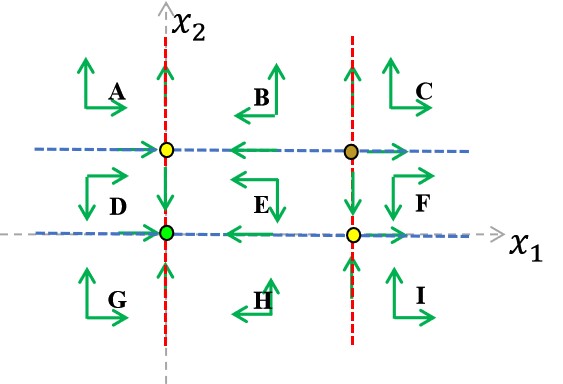}
			\caption{$\hat{\mathbb{K}}^{4}_1 \wedge \mathbb{H}_4$}
			\label{fig07.4}
		\end{subfigure}
		\caption{Theoretical schematic phase portraits of system \eqref{Simplify system in case 3} under Condition ($\hat{\mathbb{K}}^{4}_1 \wedge \mathbb{H}$). Green circle: desired uncongested stable equilibrium. Yellow circle: saddle equilibrium. Yellow-brown circle: unstable equilibrium. Red dotted curves: $\dot{x}_1=0$, blue dotted curves: $\dot{x}_2=0$, green arrows: trajectory directions. }
		\label{fig07}
	\end{figure}

	\begin{table}
		\caption{Symbols of the nine regions for system \eqref{Simplify system in case 3} under Condition $\hat{\mathbb{K}}^{4}_1 \wedge \mathbb{H}_1$.}
		\label{tab3}
		\centering
		\fontsize{6.5}{8}\selectfont
		\setlength{\tabcolsep}{4.5mm}{
			\begin{threeparttable}
				\begin{tabular}{cccccccccc}
					\toprule
					\multirow{2}{*}{Derivative}&
					\multicolumn{9}{c}{Symbol in the region}\cr
					\cmidrule(lr){2-10}
					&A&B&C&D&E&F&G&H&I\cr
					\midrule
					$\frac{\mathrm{d}{x}_1(t)}{\mathrm{d}t}$&$+$&$-$&$+$&$+$&$-$&$+$&$+$&$-$&$+$\cr
					$\frac{\mathrm{d}{x}_2(t)}{\mathrm{d}t}$&$+$&$+$&$+$&$-$&$-$&$-$&$+$&$+$&$+$\cr
					\bottomrule
				\end{tabular}
				
		\end{threeparttable}}
	\end{table}
	
	Then, for Conditions ($\hat{\mathbb{K}}^{4}_1 \wedge \mathbb{H}_2$), ($\hat{\mathbb{K}}^{4}_1 \wedge \mathbb{H}_3$) and ($\hat{\mathbb{K}}^{4}_1 \wedge \mathbb{H}_4$), we can also simplify the system \eqref{Simplify system in case 3} and then perform the above analysis.
	In a similar way, we can obtain four corresponding curves under each Condition, which separate $\mathbb{R}^2$ to nine unbounded regions.
	Then the corresponding phase portraits of \eqref{Simplify system in case 3} under these three Conditions can be acquired, as shown in Figs.~\ref{fig07.2}, \ref{fig07.3} and \ref{fig07.4}.
	Further, we can also obtain the positive invariant sets under each Condition, which verify that there is no close orbit for system \eqref{Simplify system in case 3} under Conditions ($\hat{\mathbb{K}}^{4}_1 \wedge \mathbb{H}_2$), ($\hat{\mathbb{K}}^{4}_1 \wedge \mathbb{H}_3$) and ($\hat{\mathbb{K}}^{4}_1 \wedge \mathbb{H}_4$).
	Thus, there exists no close orbit,  for the system \eqref{the model} under Condition $\hat{\mathbb{K}}^{4}_1 \wedge H$.
	
	\subsection{Spontaneous attraction region estimation} \label{A.3}
	Based on the nonexistence of close orbits and Fig. \ref{fig07}, we will next analyze the attraction region of the locally stable node $P^{4}_1$.
	For this purpose, we wish to find certain corresponding separatrices of system \eqref{Simplify system in case 3} under Conditions  ($\hat{\mathbb{K}}^{4}_1 \wedge \mathbb{H}_1$), ($\hat{\mathbb{K}}^{4}_1 \wedge \mathbb{H}_2$), ($\hat{\mathbb{K}}^{4}_1 \wedge \mathbb{H}_3$) and ($\hat{\mathbb{K}}^{4}_1 \wedge \mathbb{H}_4$). 
	However, 
	due to the increased number of equilibria, 
	it seems very difficult to find certain separatrices directly.
	Thus,
	we will first find inner and outer estimations of attraction regions, and then find certain corresponding separatrices under the above four conditions.

	To start with, we consider Condition  ($\hat{\mathbb{K}}^{4}_1 \wedge \mathbb{H}_1$). 
	The two red dotted demarcation lines  
	can be denoted as $x_1=\hat{r}^{4,1}_{i}(x_2)$ ($i=1,2$), where $\hat{r}^{4,1}_{i}(x_2)=(-1)^{i}\sqrt{\frac{u_2a_2}{a_1}\left( x_2-s_2\right) ^2-\frac{M_1}{a_1}}+s_1$. 
	Similarly,  
	the two blue dotted demarcation lines  
	can be denoted as $x_2=\hat{b}^{4}_{i}(x_1)$ ($i=1,2$), where $\hat{b}^{4}_{i}(x_1)=(-1)^{i}\sqrt{\frac{u_1a_1}{a_2}\left( x_1-s_1\right) ^2-\frac{M_2}{a_2}}+s_2$. 
	Obviously, the vertex of the line $x_1=\hat{r}^{4,1}_{i}(x_2)$ is $Ve_i=((-1)^{i}\sqrt{-\frac{M_1}{a_1}}+s_1,s_2)$ and the vertex of the line $x_2=\hat{b}^{4}_{i}(x_1)$ ($i=1,2$) is $Ve_{i+2}=(s_1,(-1)^{i+2}\sqrt{-\frac{M_2}{a_2}}+s_2)$.  
	Subsequently,
	in light of Fig.~\ref{fig07.1}, 
	we can similarly verify that the  region $\hat{U}^{4}$ (the light yellow region shown in Fig.~\ref{fig08.1}) is a positive invariant set, where
	$\hat{U}^{4}=\left\lbrace  (x_1,x_2)|x_1 \geq \hat{l}^{2}_{out}(x_2)\vee x_2 \geq \hat{l}^{1}_{out}(x_1) \right\rbrace$ with
	\begin{equation*}
		\hat{l}^{i}_{out}(x_i)=
		\begin{cases}
			\sqrt{\frac{u_ia_i}{a_j}\left( x_i-s_i\right) ^2-\frac{M_j}{a_j}}+s_j& \text{$x_i < 0$}\\
			2s_j& \text{$0 < x_i < 2s_i$}
		\end{cases}
	\end{equation*}
	Moreover, by similar analysis to $\hat{U}^{2a}$ in two-equilibria case, 
	we can arrive that 
	$\mathcal{R}(\hat{P}^{4}_1) \subset \mathbb{R}^2 \setminus \hat{U}^{4}$, i.e., $\hat{S}^{4}_{out}=\mathbb{R}^2 \setminus \hat{U}^{4}=\left\lbrace  (x_1,x_2)|x_1 < \hat{l}^{2}_{out}(x_2),~ x_2 < \hat{l}^{1}_{out}(x_1) \right\rbrace$ is an outer estimation of attraction region for $\hat{P}^{4}_1$.
	
	In addition, $\hat{S}^{4}_{in}$ (the light blue region shown in Fig.~\ref{fig08.1}) is also a positive invariant set, where 
	
	$$\hat{S}^{4}_{in}=\left\lbrace  (x_1,x_2)|x_1 \leq \hat{l}^{2}_{in}(x_2),~ x_2 \leq \hat{l}^{1}_{in}(x_1) \right\rbrace$$ with 
	\begin{equation*}
		\hat{l}^{i}_{in}(x_i)=
		\begin{cases}
			2s_j& \text{$ x_i < 0$}\\
			\sqrt{\frac{u_ia_i}{a_j}\left( x_i-s_i\right) ^2-\frac{M_j}{a_j}}+s_j& \text{$0 < x_i < s_i$}\\
			\sqrt{-\frac{M_j}{a_j}}+s_j& \text{$s_i \leq n_2 < 2s_i$}
		\end{cases}
	\end{equation*}
	Further, 
	we can prove that $\hat{S}^{4}_{in}$ is an inner estimation of attraction region for $\hat{P}^{4}_1$ by following two steps. 
	Firstly, 
	the trajectory starting from any point in $\hat{S}^{4}_{in}$ will entering the   bounded region $\hat{S}^{4}_{in}\wedge \mathbb{R}^2_1$ as $t\rightarrow +\infty$,
	where $$\mathbb{R}^2_1=\left\{ (x_1,x_2)|x_1 \geq -\frac{p_1-2s_1}{2},~x_2 \geq -\frac{p_2-2s_2}{2} \right\}, $$
	as we have $\dot{x}_1>0$ on the left boundary of $\hat{S}^{4}_{in}\wedge \mathbb{R}^2_1$ and $\dot{x}_2>0$ on the lower boundary of $\hat{S}^{4}_{in}\wedge \mathbb{R}^2_1$.
	Secondly,
	the trajectory starting from any point in bounded region $\hat{S}^{4}_{in}\wedge \mathbb{R}^2_1$ will go to $\hat{P}^{4}_1$ as $t\rightarrow +\infty$, since there is no closed orbit, and there exists only one stable node $\hat{P}^{4}_1$ in the bounded region~(\cite{zhang2006qualitative}).
	
	\begin{figure}
		\centering
		\begin{subfigure}{.23\textwidth}
			\centering
			\includegraphics[width=4cm]{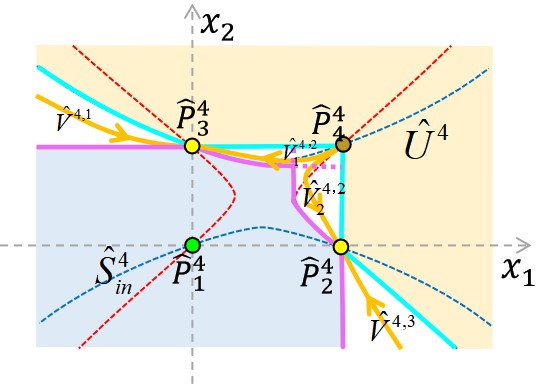}
			\caption{$\hat{\mathbb{K}}^{4}_1 \wedge \mathbb{H}_1$}
			\label{fig08.1}
		\end{subfigure}
		\begin{subfigure}{.23\textwidth}
			\centering
			\includegraphics[width=4cm]{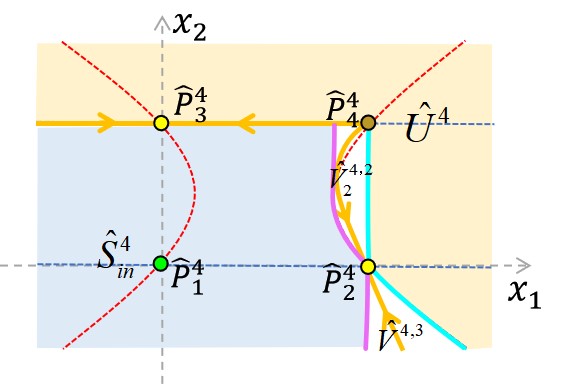}
			\caption{$\hat{\mathbb{K}}^{4}_1 \wedge \mathbb{H}_2$}
			\label{fig08.2}
		\end{subfigure}
		\begin{subfigure}{.23\textwidth}
			\centering
			\includegraphics[width=4cm]{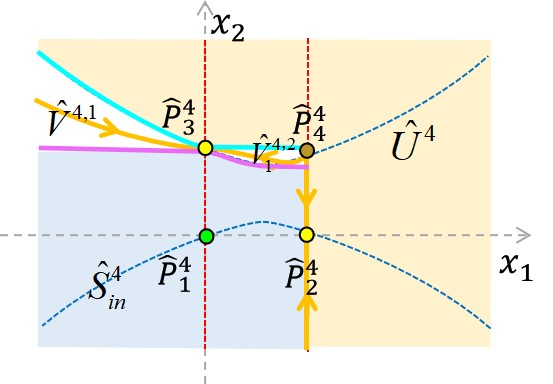}
			\caption{$\hat{\mathbb{K}}^{4}_1 \wedge \mathbb{H}_3$}
			\label{fig08.3}
		\end{subfigure}
		\begin{subfigure}{.23\textwidth}
			\centering
			\includegraphics[width=4cm]{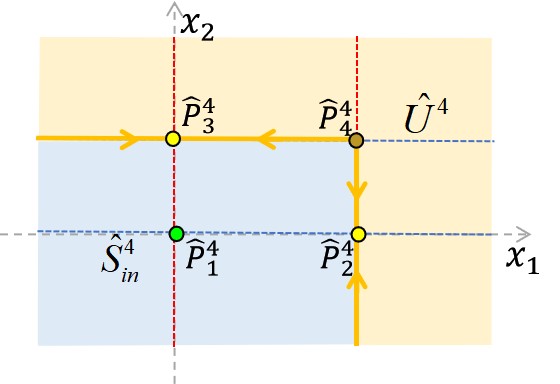}
			\caption{$\hat{\mathbb{K}}^{4}_1 \wedge \mathbb{H}_4$}
			\label{fig08.4}
		\end{subfigure}
		\caption{Qualitative characteristics of system \eqref{Simplify system in case 3} ((a)-(d)) and system \eqref{the model} ((e)-(h)) under Condition ($\hat{\mathbb{K}}^{4}_1 \wedge \mathbb{H}$). Green circle: desired uncongested stable equilibrium. Yellow circle: saddle equilibrium. Yellow-brown circle: unstable equilibrium. Red dotted curves: $\dot{x}_1=0$, blue dotted curves: $\dot{x}_2=0$, yellow curves: the separatrices, also serving as boundaries of attraction region. Light pink region: attraction region of $P^4_1$. }
		\label{fig08}
	\end{figure}

	As for Conditions ($\hat{\mathbb{K}}^{4}_1 \wedge \mathbb{H}_2$), 
	($\hat{\mathbb{K}}^{4}_1 \wedge \mathbb{H}_3$) and ($\hat{\mathbb{K}}^{4}_1 \wedge \mathbb{H}_4$), 
	through similar analysis, 
	we can obtain the inner and outer estimations of attraction regions for $\hat{P}^{4}_1$, 
	as shown in the Figs.~\ref{fig08.2}, \ref{fig08.3} and \ref{fig08.4}.
	Further, 
	letting  $n_1(t)= x_1(t)+\frac{p_1-2s_1}{2}$ and $n_2(t)= x_2(t)+\frac{p_2-2s_2}{2}$, 
	the inner and outer estimations of attraction regions for $P^{4}_1$ can be obtained. Denoting $l^{i}_{in}(n_i)=\hat{l}^{i}_{in}(n_i-\frac{p_i-2s_i}{2})+\frac{p_j-2s_j}{2}$ and $l^{i}_{out}(n_i)=\hat{l}^{i}_{out}(n_i-\frac{p_i-2s_i}{2})+\frac{p_j-2s_j}{2}$, where $i,j=1,2$ and $i \ne j$, we can arrive at the following theorem.
	\begin{theorem}\label{inner and outer estimations of attraction regions in 4.1}
		Under Condition ($\hat{\mathbb{K}}^{4}_1 \wedge \mathbb{H}$), for the local stable point $P^{4}_1$, we have $S^{4}_{in} \subset \mathcal{R}(P^{4}_1) \subset S^{4}_{out}$, 
		where the inner estimation of attraction region $S^{4}_{in}$ and the outer estimation of attraction region $S^{4}_{out}$ are defined as follows:
		\begin{enumerate}[(1)]
			\item Under Condition ($\hat{\mathbb{K}}^{4}_1 \wedge \mathbb{H}_1$), 
			$$S^{4}_{in}=\left\lbrace  (n_1,n_2)| n_1 \leq l^{2}_{in}(n_2),~  n_2 \leq l^{1}_{in}(n_1) \right\rbrace;
			S^{4}_{out}=\left\lbrace  (n_1,n_2)| n_1 \leq l^{2}_{out}(n_2),~  n_2 \leq l^{1}_{out}(n_1) \right\rbrace.$$
			
			\item Under Condition ($\hat{\mathbb{K}}^{4}_1 \wedge \mathbb{H}_2$), 
			$$S^{4}_{in}=\left\lbrace  (n_1,n_2)| n_1 \leq l^{2}_{in}(n_2),~  n_2 < \frac{n_2+2s_2}{2} \right\rbrace; ~
			S^{4}_{out}=\left\lbrace  (n_1,n_2)| n_1 \leq l^{2}_{out}(n_2),~  n_2 < \frac{n_2+2s_2}{2} \right\rbrace. $$
			
			\item Under Condition ($\hat{\mathbb{K}}^{4}_1 \wedge \mathbb{H}_3$), 
			$$S^{4}_{in}=\left\lbrace  (n_1,n_2)| n_1 < \frac{n_1+2s_1}{2},~  n_2 \leq l^{1}_{in}(n_1) \right\rbrace;~
			S^{4}_{out}=\left\lbrace  (n_1,n_2)| n_1 < \frac{n_1+2s_1}{2},~  n_2 \leq l^{1}_{out}(n_1) \right\rbrace.$$
			
			\item Under Condition ($\hat{\mathbb{K}}^{4}_1 \wedge \mathbb{H}_4$), 
			$S^{4}_{in}=S^{4}_{out}=\left\lbrace  (n_1,n_2)|n_1 < \frac{n_1+2s_1}{2},~  n_2 < \frac{n_2+2s_2}{2} \right\rbrace.$ 
		\end{enumerate}
		Note that $n_i\geq 0$ ($i=1,2$) holds all the time.
	\end{theorem}
	
	Clearly, according to Theorem \ref{inner and outer estimations of attraction regions in 4.1}, the inner and outer estimations of attraction regions for ${P}^{4}_1$ under ($\hat{\mathbb{K}}^{4}_1 \wedge \mathbb{H}_i$) with $i=1,2,3,4$ can be obtained, 
	as shown in the Figs.~\ref{fig08.5}–\ref{fig08.8}.
	
	Moreover,  
	we will derive the boundaries of the attraction region of ${P}^4_1$ under Condition ($\hat{\mathbb{K}}^{4}_1 \wedge \mathbb{H}_1$), by finding certain corresponding separatrices
	located in  $\bigcup\limits_{i=1}^3\hat{V}^{4,i}=\hat{S}^{4}_{out} \setminus \hat{S}^{4}_{in}$  of equivalent systems \eqref{Simplify system in case 3},  
	as shown in Fig.~\ref{fig08.1}, 
	where
	\begin{eqnarray*}
		\hat{V}^{4,1}&=&\left\{ (x_1,x_2)|x_1 < 0,~ \hat{l}^{1}_{in}(x_1) < x_2 < \hat{l}^{1}_{out}(x_1) \right\},\\
		\hat{V}^{4,3}&=&\left\{ (x_1,x_2)|\hat{l}^{2}_{in}(x_2)< x_1 < \hat{l}^{2}_{out}(x_2),~  x_2 < 0 \right\},
	\end{eqnarray*} 
	and $\hat{V}^{4,2}=\hat{V}^{4,2}_{1}\vee\hat{V}^{4,2}_{2}$, where 
	\begin{eqnarray*}
		\hat{V}^{4,2}_{1}&=&\left\{ (x_1,x_2)| 0<x_1<2s_1,~ \hat{l}^{1}_{in}(x_1) < x_2 < \hat{l}^{1}_{out}(x_1)   \right\},\\
		\hat{V}^{4,2}_{2}&=&\left\{ (x_1,x_2)|\hat{l}^{2}_{in}(x_2)< x_1 < \hat{l}^{2}_{out}(x_2) ,~ 0<x_2 < 2s_2\right\}.
	\end{eqnarray*}
	Obviously, we can obtain a stable separatrix $\hat{\Phi}^1(\hat{P}^{4}_3)\in\hat{V}^{4,1}$ for the saddle point $\hat{P}^{4}_3$  and a stable separatrix $\hat{\Phi}^1(\hat{P}^{4}_2)\in\hat{V}^{4,3}$ for the saddle point $\hat{P}^{4}_2$.
	Moreover,
	we can further obtain another stable separatrix  $\hat{\Phi}^2(\hat{P}^{4}_3)$ of $\hat{P}^{4}_3$, 
	which satisfy $\hat{\Phi}^2(\hat{P}^{4}_3) \in \hat{V}^{4,2}_{1}$ 
	and $\lim_{t \to -\infty}\hat{\Phi}^2(\hat{P}^{4}_3)=\hat{P}^{4}_4$,
	by following two steps. 
	Firstly, 
	we can verify that the region $\hat{V}^{4,2}_{1}$ is a negative invariant set, otherwise there will exist  contradiction with the trajectory direction at the boundary line of region $\hat{V}^{4,2}_{1}$.
	Secondly,
	the trajectory starting from any point in bounded region $\hat{V}^{4,2}_{1}$ will go to $\hat{P}^{4}_4$ as $t\rightarrow -\infty$, since there is no closed orbit, and there exists only one unstable node $\hat{P}^{4}_4$ in the bounded region~(\cite{zhang2006qualitative}).
	Similarly, we can obtain another stable separatrix  $\hat{\Phi}^2(\hat{P}^{4}_2)$ for $\hat{P}^{4}_2$, 
	which satisfies $\hat{\Phi}^2(\hat{P}^{4}_2) \in \hat{V}^{4,2}_{2}$ and $\lim_{t \to -\infty}\hat{\Phi}^2(\hat{P}^{4}_2)=\hat{P}^{4}_4$.  
	Clearly, 
	these four separatrices are exactly the boundaries of the attraction region for $\hat{P}^{4}_1$.
	
	Further, 
	letting  $n_1(t)= x_1(t)+\frac{p_1-2s_1}{2}$ and $n_2(t)= x_2(t)+\frac{p_2-2s_2}{2}$, 
	the  attraction region boundary for $P^{4}_1$ under Condition ($\hat{\mathbb{K}}^{4}_1 \wedge \mathbb{H}_1$) can be obtained. 
	Denoting $\Phi^{i}(P^4_2)$ ($i=1,2$) to be the two stable separatrices of $P^{4}_2$, 
	and $\Phi^{j}(P^4_3)$ ($j=1,2$) to be the two stable separatrices of $P^{4}_3$,  
	the below theorem can be arrived.
	\begin{theorem}\label{attraction region in 4.1}
		Under Condition ($\hat{\mathbb{K}}^{4}_1 \wedge \mathbb{H}_1$), for the local stable point $P^{4}_1$, we have $S^{4}_A=\mathcal{R}(P^{4}_1)$, where
		$$S^{4}_A:=\left\{ (n_1,n_2)| n_i < g^4_j(n_j),\ i,j=1,2,~j\neq i \right\}$$ satisfying	$l^{j}_{in}(n_j)<g^4_j(n_j)<l^{j}_{out}(n_j)$, and $g^4_j(n_j),\ j=1,2$ are defined as:
		\begin{equation*}
			g^4_1(n_1)=
			\begin{cases}
				\Phi^{1}(P^4_3)& \text{$ n_1 < \frac{p_1-2s_1}{2}$}\\
				\Phi^{2}(P^4_3)& \text{$\frac{p_1-2s_1}{2} < n_1 < \frac{p_1+2s_1}{2}$}
			\end{cases}
		\end{equation*}
		and
		\begin{equation*}
			g^4_2(n_2)=
			\begin{cases}
				\Phi^{1}(P^4_2)& \text{$ n_2 < \frac{p_2-2s_2}{2}$}\\
				\Phi^{2}(P^4_2)& \text{$\frac{p_2-2s_2}{2} < n_2 < \frac{p_2+2s_2}{2}$}
			\end{cases}
		\end{equation*}
		with $\lim_{t \to -\infty}\Phi^{2}(P^4_2)=P^4_4$ and  $\lim_{t \to -\infty}\Phi^{2}(P^4_3)=P^4_4$. 
		Note that $n_i\geq 0$ ($i=1,2$) holds all the time.
	\end{theorem}
	
	In summary, the attraction region of $P^{4}_1$ under corresponding conditions are plotted as light pink zones in the Figs.~\ref{fig08.5}-\ref{fig08.8}.

	Secondly, we consider Condition ($\hat{\mathbb{K}}^{4}_2 \wedge \mathbb{H}$), which can be partitioned into two sub-Conditions: ($\hat{\mathbb{K}}^{4}_2 \wedge \mathbb{H}_1$) and ($\hat{\mathbb{K}}^{4}_2 \wedge \mathbb{H}_2$). 
	Note that $\hat{\mathbb{K}}^{4}_2 \wedge \mathbb{H}_3=\varnothing$ and $\hat{\mathbb{K}}^{4}_2 \wedge \mathbb{H}_4=\varnothing$. 
	By repeating the analysis as above,
	the corresponding phase portraits of \eqref{Simplify system in case 3} under these two sub-conditions can be obtained;
	and further prove that there is no close orbit for system \eqref{Simplify system in case 3} under above two sub-conditions.
	In addition, 
	we can obtain the inner and outer estimations of attraction regions for $\hat{P}^{4}_1$, 
	and deduce the boundary of attraction region for $\hat{P}^{4}_1$, 
	as shown in the Figs.~\ref{fig16.1} and \ref{fig16.2}.
	Further,
	letting $n_1(t)= x_1(t)+\frac{p_1-2s_1}{2}$ and $n_2(t)= x_2(t)+\frac{p_2-2s_2}{2}$, 
	we can obtain the inner and outer estimations of attraction regions for ${P}^{4}_1$
	which are defined in 
	Theorem \ref{inner and outer estimations of attraction regions in 4.1} (1),
	and the attraction region boundary 
	for $P^4_1$ under Conditions ($\hat{\mathbb{K}}^{4}_2 \wedge \mathbb{H}_1$) and ($\hat{\mathbb{K}}^{4}_2 \wedge \mathbb{H}_2$),  
	as shown in Figs.~\ref{fig09.1} and \ref{fig09.2}.
	
	\begin{figure}
		\centering
		\begin{subfigure}{.23\textwidth}
			\centering
			\includegraphics[width=4cm]{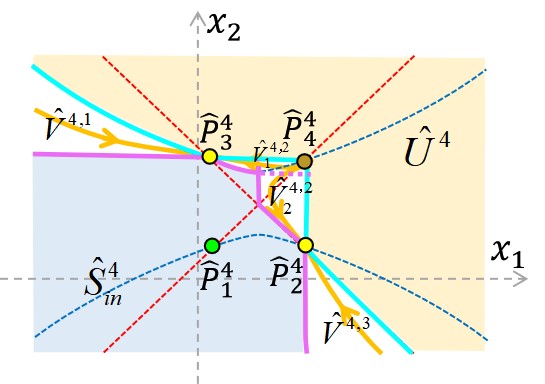}
			\caption{$\hat{\mathbb{K}}^{4}_2 \wedge \mathbb{H}_1$}
			\label{fig16.1}
		\end{subfigure}
		\begin{subfigure}{.23\textwidth}
			\centering
			\includegraphics[width=4cm]{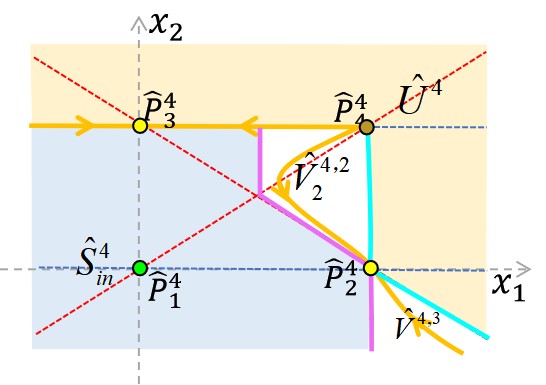}
			\caption{$\hat{\mathbb{K}}^{4}_2 \wedge \mathbb{H}_2$}
			\label{fig16.2}
		\end{subfigure}
		\begin{subfigure}{.23\textwidth}
			\centering
			\includegraphics[width=4cm]{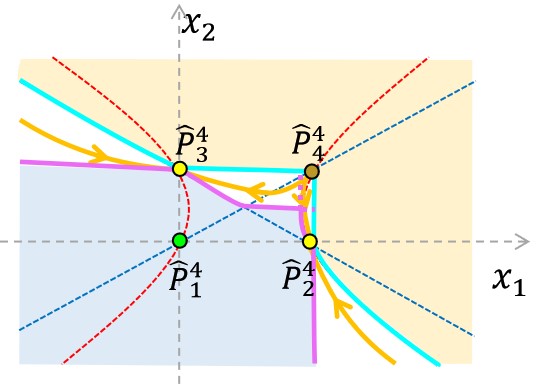}
			\caption{$\hat{\mathbb{K}}^{4}_3 \wedge \mathbb{H}_1$}
			\label{fig16.3}
		\end{subfigure}
		\begin{subfigure}{.23\textwidth}
			\centering
			\includegraphics[width=4cm]{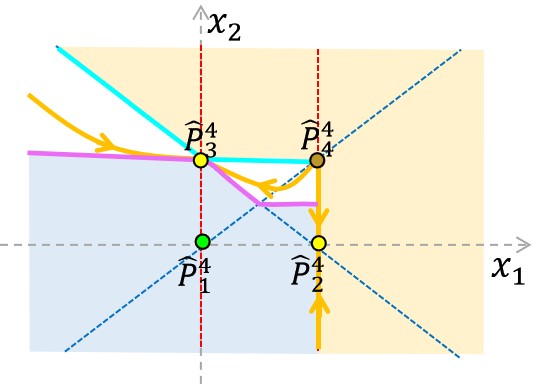}
			\caption{$\hat{\mathbb{K}}^{4}_3 \wedge \mathbb{H}_3$}
			\label{fig16.4}
		\end{subfigure}
		\caption{Qualitative characteristics of system \eqref{Simplify system in case 3} under corresponding conditions. }
		\label{fig16}
	\end{figure}
	
	%
		%

Thirdly, we analyze the phase portrait and the attraction region of $P^{4}_1$ under Condition ($\hat{\mathbb{K}}^{4}_3 \wedge \mathbb{H}$), which can be partitioned into two sub-Conditions: ($\hat{\mathbb{K}}^{4}_3 \wedge \mathbb{H}_1$) and ($\hat{\mathbb{K}}^{4}_3 \wedge \mathbb{H}_3$). 
Note that $\hat{\mathbb{K}}^{4}_3 \wedge \mathbb{H}_2=\varnothing$ and $\hat{\mathbb{K}}^{4}_3 \wedge \mathbb{H}_4=\varnothing$. Notice that Condition  ($\hat{\mathbb{K}}^{4}_3 \wedge \mathbb{H}$) is symmetric to Condition ($\hat{\mathbb{K}}^{4}_2 \wedge \mathbb{H}$), 
so the analysis process and results are similar. Thus, we can obtain the inner and outer estimations of attraction regions  for $\hat{P}^{4}_1$ in a similar way, as shown in Figs.~\ref{fig16.3} and \ref{fig16.4}.
Further, letting $n_1(t)= x_1(t)+\frac{p_1-2s_1}{2}$ and $n_2(t)= x_2(t)+\frac{p_2-2s_2}{2}$, 
the attraction region of $P^4_1$ under Conditions ($\hat{\mathbb{K}}^{4}_3 \wedge \mathbb{H}_1$) and ($\hat{\mathbb{K}}^{4}_3 \wedge \mathbb{H}_3$) can be obtained, as shown in Figs.~\ref{fig09.3} and \ref{fig09.4}.

Fourthly, we consider Condition ($\hat{\mathbb{K}}^{4}_4 \wedge \mathbb{H}$), which can be partitioned into two sub-Conditions: Condition ($\hat{\mathbb{K}}^{4}_4 \wedge \mathbb{H}_1$) and ($\hat{\mathbb{K}}^{4}_4 \wedge \mathbb{H}_2$).
Note that $\hat{\mathbb{K}}^{4}_4 \wedge \mathbb{H}_3=\varnothing$ and $\hat{\mathbb{K}}^{4}_4 \wedge \mathbb{H}_4=\varnothing$.

We first consider Condition ($\hat{\mathbb{K}}^{4}_4 \wedge \mathbb{H}_1$). 
For system \eqref{Simplify system in case 3} under Condition ($\hat{\mathbb{K}}^{4}_4 \wedge \mathbb{H}_1$), 
following the steps as Condition ($\hat{\mathbb{K}}^{4}_1 \wedge \mathbb{H}_1$), 
the global phase portrait can be obtained, as shown in Fig.~\ref{fig17.1}; 
further prove that there exist no close orbits; 
and subsequently derive the inner estimation of attraction region $\hat{S}^{4,1}_{in}=\left\lbrace  (x_1,x_2)|x_1 \leq 2s_1,~ x_2 \leq \hat{l}^{1,1}_{in}(x_1) \right\rbrace$
and the outer estimation of attraction region
$\hat{S}^{4,1}_{out}=\mathbb{R}^2 \setminus \hat{U}^{4,1}=\left\lbrace  (x_1,x_2)|x_1 \in \mathbb{R},~ x_2 < \hat{l}^{1,1}_{out}(x_1) \right\rbrace$,
where 
\begin{equation*}
	\hat{l}^{1,1}_{in}(x_1)=
	\begin{cases}
		-\sqrt{\frac{M_1}{u_2a_2}}+s_2& \text{$ x_1 < s_1$}\\
		-\sqrt{\frac{a_1}{u_2a_2}\left( x_1-s_1\right) ^2+\frac{M_1}{u_2a_2}}+s_2& \text{$s_1 < x_1 < 2s_1$}
	\end{cases}
\end{equation*}
and 
\begin{equation*}
	\hat{l}^{1,1}_{out}(x_1)=
	\begin{cases}
		\sqrt{\frac{u_1a_1}{a_2}\left( x_1-s_1\right) ^2-\frac{M_2}{a_2}}+s_2& \text{$ x_1 < 0$}\\
		2s_2& \text{$0 < x_1 < 2s_1$}\\
		-\sqrt{\frac{a_1}{u_2a_2}\left( x_1-s_1\right) ^2+\frac{M_1}{u_2a_2}}+s_2& \text{$ x_1 > 2s_1$}
	\end{cases}
\end{equation*}

\begin{figure}
	\centering
	\begin{subfigure}{.33\textwidth}
		\centering
		\includegraphics[width=7cm]{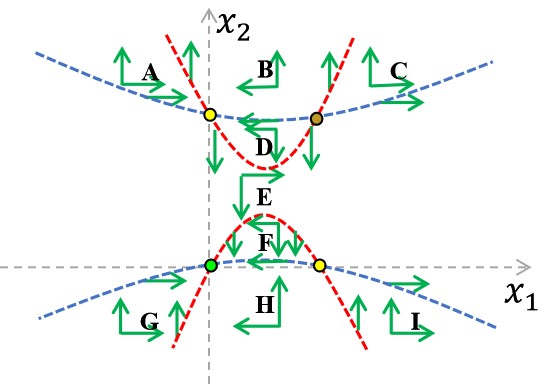}
		\caption{$\hat{\mathbb{K}}^{4}_3 \wedge \mathbb{H}_1$}
		\label{fig17.1}
	\end{subfigure}
	\begin{subfigure}{.48\textwidth}
		\centering
		\includegraphics[width=7cm]{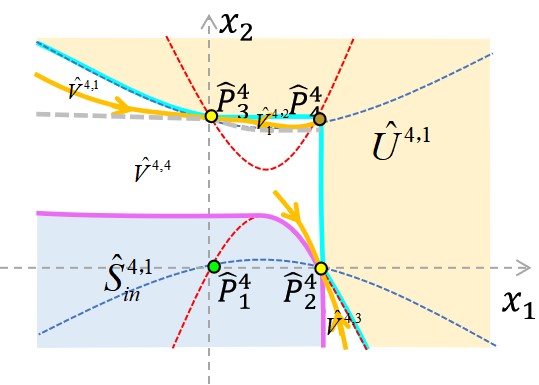}
		\caption{$\hat{\mathbb{K}}^{4}_3 \wedge \mathbb{H}_1$}
		\label{fig17.2}
	\end{subfigure}
	\caption{Phase diagram (a) and qualitative characteristics (b) of system \eqref{Simplify system in case 3} under Condition ($\hat{\mathbb{K}}^{4}_4 \wedge \mathbb{H}_1$).}
	\label{fig17}
\end{figure}

Thus,  we will next deduce the boundaries of the attraction region of ${P}^4_1$ by finding certain corresponding separatrices
located in  $\bigcup\limits_{i=3}^4\hat{V}^{4,i}=\hat{S}^{4,1}_{out} \setminus \hat{S}^{4,1}_{in}$  of equivalent systems \eqref{Simplify system in case 3},   
as shown in Fig.~\ref{fig17.2}, 
where 
$\hat{V}^{4,4}=\left\{ (x_1,x_2)|x_1 < 2s_1,~ \hat{l}^{1,1}_{in}(x_1) < x_2 < \hat{l}^{1,1}_{out}(x_1) \right\}$
and
$\hat{V}^{4,3}=\left\{ (x_1,x_2)|x_1 > 2s_1,~ x_2 < \hat{l}^{1,1}_{out}(x_1) \right\}$.
By similar analysis as above, we can obtain a stable separatrix $\hat{\Phi}^1(\hat{P}^{4}_3)\in\hat{V}^{4,1}$ for the saddle point $\hat{P}^{4}_3$;  
a stable separatrix $\hat{\Phi}^1(\hat{P}^{4}_2)\in\hat{V}^{4,3}$ for the saddle point $\hat{P}^{4}_2$;
and another stable separatrix  $\hat{\Phi}^2(\hat{P}^{4}_3)$ of $\hat{P}^{4}_3$, 
which satisfy $\hat{\Phi}^2(\hat{P}^{4}_3) \in \hat{V}^{4,2}_{1}$ 
and $\lim_{t \to -\infty}\hat{\Phi}^2(\hat{P}^{4}_3)=\hat{P}^{4}_4$.
Note that $\hat{V}^{4,1}$ and $\hat{V}^{4,2}_{1}$ are defined above.
As for another stable separatrix for $\hat{P}^{4}_2$ in $\hat{V}^{4,4}$, 
there are three types, 
where in type 1, 
$\lim_{t \to -\infty}\hat{\Phi}^2(\hat{P}^{4}_2)=-\infty$;
in type 2,
$\lim_{t \to -\infty}\hat{\Phi}^2(\hat{P}^{4}_2)=\hat{P}^{4}_3$;
and in type 3,
$\lim_{t \to -\infty}\hat{\Phi}^2(\hat{P}^{4}_2)=\hat{P}^{4}_4$;
as shown in Figs.~\ref{fig18.1}, \ref{fig18.2} and \ref{fig18.3}, respectively.

\begin{figure}
	\centering
	\begin{subfigure}{.32\textwidth}
		\centering
		\includegraphics[width=4cm]{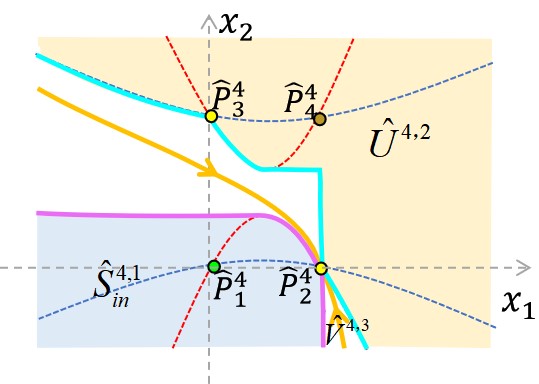}
		\caption{$\hat{\mathbb{K}}^{4}_4 \wedge \mathbb{H}_1$, type 1}
		\label{fig18.1}
	\end{subfigure}
	\begin{subfigure}{.32\textwidth}
		\centering
		\includegraphics[width=4cm]{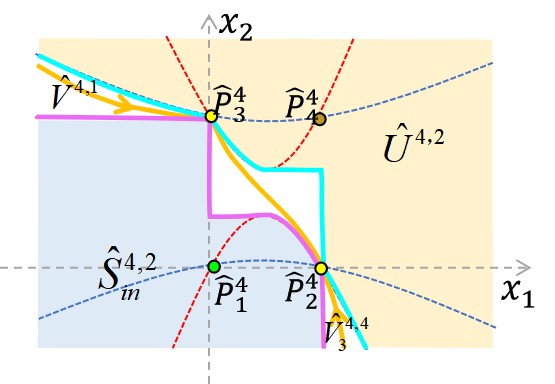}
		\caption{$\hat{\mathbb{K}}^{4}_4 \wedge \mathbb{H}_1$, type 2}
		\label{fig18.2}
	\end{subfigure}
	\begin{subfigure}{.32\textwidth}
		\centering
		\includegraphics[width=4cm]{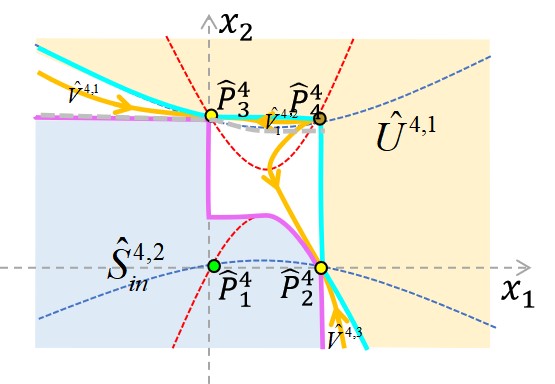}
		\caption{$\hat{\mathbb{K}}^{4}_4 \wedge \mathbb{H}_1$, type 3}
		\label{fig18.3}
	\end{subfigure}
	\caption{Three types of qualitative characteristics of system \eqref{Simplify system in case 3} ((a)-(c)) and system \eqref{the model} ((d)-(f)) under Condition ($\hat{\mathbb{K}}^{4}_4 \wedge \mathbb{H}_1$).}
	\label{fig18}
\end{figure}

Further, we can refine the inner and outer estimations of attraction regions under each type. 
Specifically,
In type 1,  
the outer estimation of attraction region can be refined as
$\hat{S}^{4,2}_{out} =\left\lbrace  (x_1,x_2)|x_1 \in \mathbb{R},~ x_2 < \hat{l}^{1,2}_{out}(x_1) \right\rbrace$, where
\begin{equation*}
	\hat{l}^{1,2}_{out}(x_1)=
	\begin{cases}
		\sqrt{\frac{u_1a_1}{a_2}\left( x_1-s_1\right) ^2-\frac{M_2}{a_2}}+s_2& \text{$ x_1 < 0$}\\
		\sqrt{\frac{a_1}{u_2a_2}\left( x_1-s_1\right) ^2+\frac{M_1}{u_2a_2}}+s_2& \text{$0 < x_1 < s_1$}\\
		\sqrt{\frac{M_1}{u_2a_2}}+s_2& \text{$s_1 < x_1 < 2s_1$}\\
		-\sqrt{\frac{a_1}{u_2a_2}\left( x_1-s_1\right) ^2+\frac{M_1}{u_2a_2}}+s_2& \text{$ x_1 > 2s_1$}
	\end{cases}
\end{equation*}
since the region $\hat{V}^{4,4}_1=\hat{S}^{4,2}_{out} \setminus \hat{S}^{4,1}_{out}$ is a negative invariant set 
and
the trajectory starting from any point in bounded region $\hat{V}^{4,4}_{1}$ will go to $\hat{P}^{4}_4$ as $t\rightarrow -\infty$.
In type 2, 
the outer estimation of attraction region can also be refined as  $\hat{S}^{4,2}_{out}$ for the same reason.
Moreover,
the inner estimation of attraction region can be refined as
$\hat{S}^{4,2}_{in} =\left\lbrace  (x_1,x_2)|x_1 \in \mathbb{R},~ x_2 < \hat{l}^{1,2}_{in}(x_1) \right\rbrace$, where
\begin{equation*}
	\hat{l}^{1,2}_{in}(x_1)=
	\begin{cases}
		2s_2& \text{$ x_1 < 0$}\\
		-\sqrt{\frac{M_1}{u_2a_2}}+s_2& \text{$ 0<x_1 < s_1$}\\
		-\sqrt{\frac{a_1}{u_2a_2}\left( x_1-s_1\right) ^2+\frac{M_1}{u_2a_2}}+s_2& \text{$s_1 < x_1 < 2s_1$}
	\end{cases}
\end{equation*}
since any trajectory starting from the region $\hat{V}^{4,4}_2=\hat{S}^{4,1}_{in} \setminus \hat{S}^{4,2}_{in}$ will also go to $\hat{P}^{4}_1$ as $t\rightarrow +\infty$.
In type 3, the inner estimation of attraction region can be refined as
$\hat{S}^{4,2}_{in}$ for the same reason.

In summary, the refined inner and outer estimations of attraction regions in each type are obtained,
as shown in Figs.~\ref{fig18.1}, \ref{fig18.2} and \ref{fig18.3}. Then, letting $n_1(t)= x_1(t)+\frac{p_1-2s_1}{2}$ and $n_2(t)= x_2(t)+\frac{p_2-2s_2}{2}$, 
we can obtain the attraction region, inner and outer estimations of attraction regions in each type of system \eqref{the model} under Conditions ($\hat{\mathbb{K}}^{4}_4 \wedge \mathbb{H}_1$), which are shown in Fig.~\ref{fig18}. 
Moreover, 
denoting  $l^{1,i}_{out}(n_1)=\hat{l}^{1,i}_{out}(n_1-\frac{p_1-2s_1}{2})+\frac{p_2-2s_2}{2}$ and $l^{1,i}_{in}(n_1)=\hat{l}^{1,i}_{in}(n_1-\frac{p_1-2s_1}{2})+\frac{p_2-2s_2}{2}$, where $i=1,2$, 
the below theorem can be arrived. 
\begin{theorem}\label{attraction region in 4.2}
	Under Condition ($\hat{\mathbb{K}}^{4}_4 \wedge \mathbb{H}_1$), for the local stable point $P^{4}_1$, we  have
	$S^{4}_{in} \subset S^{4}_A=\mathcal{R}(P^{4}_1) \subset S^{4}_{out}$,
	where $S^{4}_A$, the inner estimation of attraction region $S^{4}_{in}$ and outer estimation of attraction region $S^{4}_{out}$ have three types, which are defined as follows: 		
	\begin{enumerate}[(1)]
		\item In type 1: 
		$S^{4}_A=\left\{ (n_1,n_2)| n_1 \leq {p_1},\ ~ n_2 < g^4_3(n_1)\right\}$, where $g^4_3(n_1)$ represents $\Phi^{1}(P^{4}_2)$ when $ n_1 <\frac{p_1+2s_1}{2}$ and  $\Phi^{2}(P^{4}_2)$ when $\frac{p_1+2s_1}{2} < n_1 \leq {p_1}$;
		$${S}^{4}_{in} =\left\lbrace  (n_1,n_2)| n_1 \leq {p_1},\ ~ n_2 < {l}^{1,1}_{in}(n_1) \right\rbrace; 
		~{S}^{4}_{out} =\left\lbrace  (n_1,n_2)| n_1 \leq {p_1},\ ~ n_2 < {l}^{1,2}_{out}(n_1) \right\rbrace.$$
		
		\item In type 2: 
		$S^{4}_A=\left\{ (n_1,n_2)| n_1 \leq {p_1},\ ~ n_2 < g^4_4(n_1)\right\}$, 
		where $g^4_4(n_1)$ represents $\Phi^{1}(P^{4}_3)$ when $ n_1 <\frac{p_1-2s_1}{2}$, 
		represents $\Phi^{2}(P^{4}_2)$ when $\frac{p_1-2s_1}{2} < n_1 < \frac{p_1+2s_1}{2}$, 
		and represents $\Phi^{1}(P^{4}_2)$ when $\frac{p_1+2s_1}{2} < n_1 \leq {p_1}$;
		$${S}^{4}_{in} =\left\lbrace  (n_1,n_2)| n_1 \leq {p_1},\ ~ n_2 < {l}^{1,2}_{in}(n_1) \right\rbrace;~ 
		{S}^{4}_{out} =\left\lbrace  (n_1,n_2)| n_1 \leq {p_1},\ ~ n_2 < {l}^{1,2}_{out}(n_1) \right\rbrace.$$
		
		\item In type 3: 
		$S^{4}_A$ is defined in Theorem \ref{attraction region in 4.1};
		$${S}^{4}_{in} =\left\lbrace  (n_1,n_2)| n_1 \leq {p_1},\ ~ n_2 < {l}^{1,2}_{in}(n_1) \right\rbrace; ~
		{S}^{4}_{out} =\left\lbrace  (n_1,n_2)| n_1 \leq {p_1},\ ~ n_2 < {l}^{1,1}_{out}(n_1) \right\rbrace.$$
	\end{enumerate}
	Note that $n_i\geq 0$ ($i=1,2$) holds all the time.
\end{theorem}

Then, for Conditions ($\hat{\mathbb{K}}^{4}_4 \wedge \mathbb{H}_2$), 
we can also simplify the system \eqref{Simplify system in case 3} and then perform the above analysis. 
In a similar way, 
we can obtain the phase portrait for system \eqref{Simplify system in case 3} and further verify that there are no close orbits under Conditions ($\hat{\mathbb{K}}^{4}_4 \wedge \mathbb{H}_2$). 
Thus, we can obtain three types of attraction regions, and further obtain three types of refined the inner and outer estimations of attraction regions under each type, as shown in Figs.~\ref{fig19.1}, \ref{fig19.2} and \ref{fig19.3}. Then letting $n_1(t)= x_1(t)+\frac{p_1-2s_1}{2}$ and $n_2(t)= x_2(t)+\frac{p_2-2s_2}{2}$, we can obtain three types of attraction regions of \eqref{the model} under Conditions ($\hat{\mathbb{K}}^{4}_4 \wedge \mathbb{H}_2$), as shown in Figs.~\ref{fig09.6}, \ref{fig09.10} and \ref{fig09.14}.

\begin{figure}
	\centering
	\begin{subfigure}{.32\textwidth}
		\centering
		\includegraphics[width=4cm]{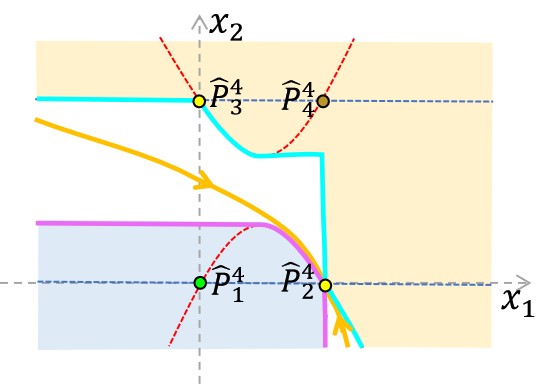}
		\caption{$\hat{\mathbb{K}}^{4}_4 \wedge \mathbb{H}_2$, type 1}
		\label{fig19.1}
	\end{subfigure}
	\begin{subfigure}{.32\textwidth}
		\centering
		\includegraphics[width=4cm]{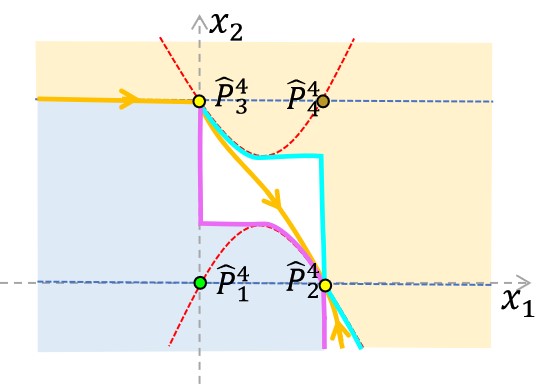}
		\caption{$\hat{\mathbb{K}}^{4}_4 \wedge \mathbb{H}_2$, type 2}
		\label{fig19.2}
	\end{subfigure}
	\begin{subfigure}{.32\textwidth}
		\centering
		\includegraphics[width=4cm]{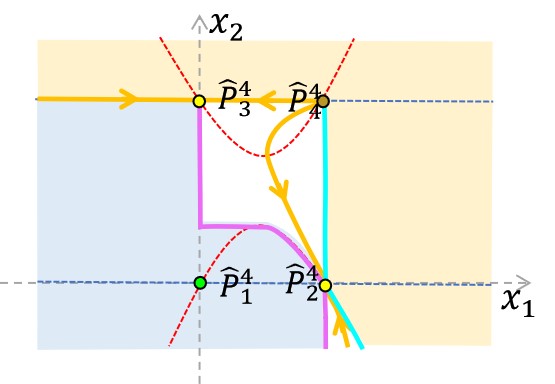}
		\caption{$\hat{\mathbb{K}}^{4}_4 \wedge \mathbb{H}_2$, type 3}
		\label{fig19.3}
	\end{subfigure}
	
	\begin{subfigure}{.32\textwidth}
		\centering
		\includegraphics[width=4cm]{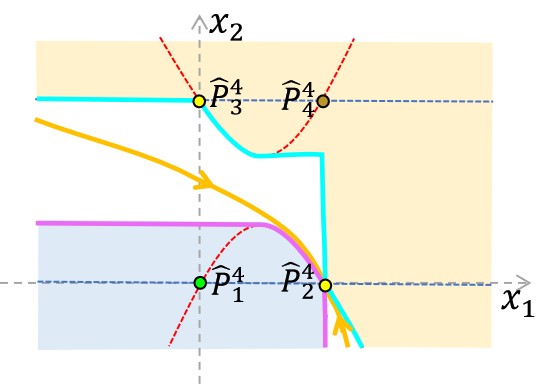}
		\caption{$\hat{\mathbb{K}}^{4}_5 \wedge \mathbb{H}_1$, type 1}
		\label{fig19.4}
	\end{subfigure}
	\begin{subfigure}{.32\textwidth}
		\centering
		\includegraphics[width=4cm]{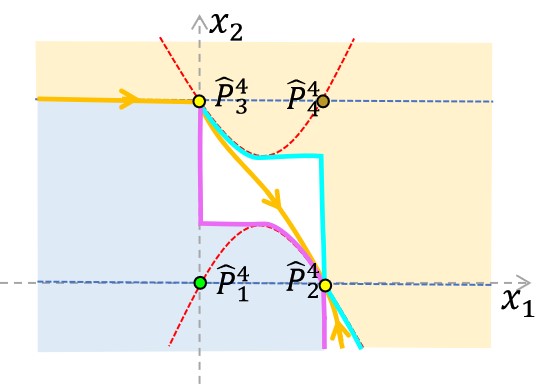}
		\caption{$\hat{\mathbb{K}}^{4}_5 \wedge \mathbb{H}_1$, type 2}
		\label{fig19.5}
	\end{subfigure}
	\begin{subfigure}{.32\textwidth}
		\centering
		\includegraphics[width=4cm]{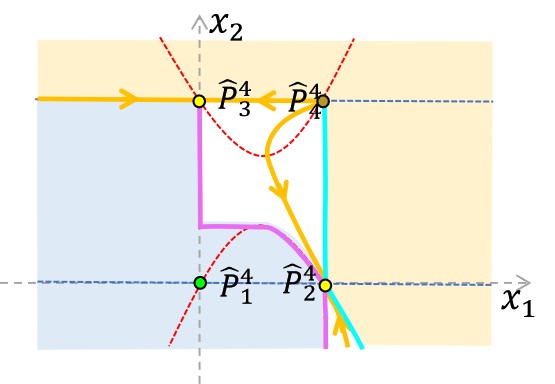}
		\caption{$\hat{\mathbb{K}}^{4}_5 \wedge \mathbb{H}_1$, type 3}
		\label{fig19.6}
	\end{subfigure}

	\begin{subfigure}{.32\textwidth}
		\centering
		\includegraphics[width=4cm]{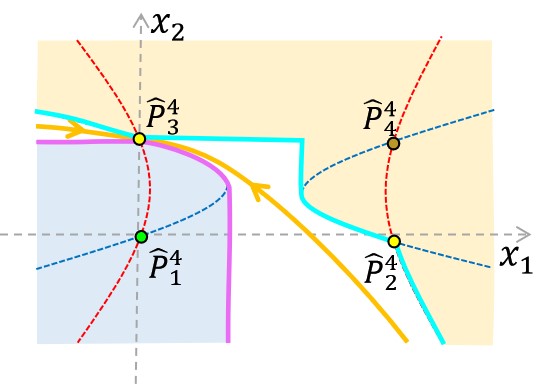}
		\caption{$\hat{\mathbb{K}}^{4}_5 \wedge \mathbb{H}_3$, type 1}
		\label{fig19.7}
	\end{subfigure}
	\begin{subfigure}{.32\textwidth}
		\centering
		\includegraphics[width=4cm]{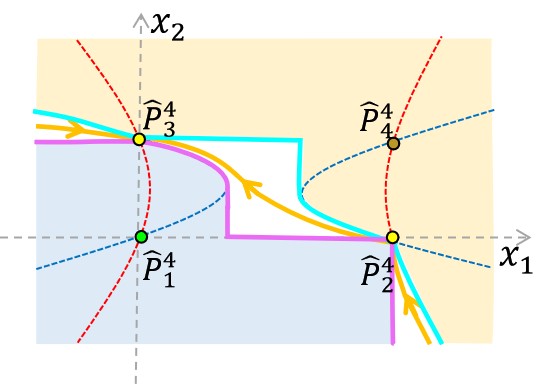}
		\caption{$\hat{\mathbb{K}}^{4}_5 \wedge \mathbb{H}_3$, type 2}
		\label{fig19.8}
	\end{subfigure}
	\begin{subfigure}{.32\textwidth}
		\centering
		\includegraphics[width=4cm]{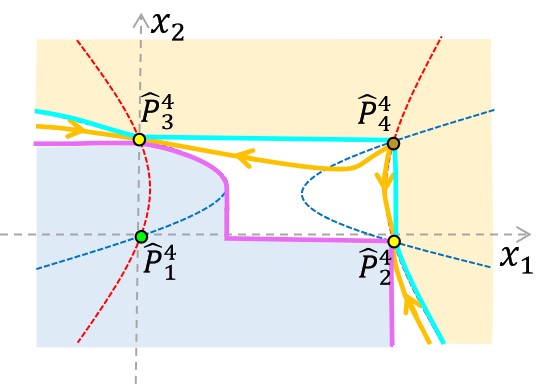}
		\caption{$\hat{\mathbb{K}}^{4}_5 \wedge \mathbb{H}_3$, type 3}
		\label{fig19.9}
	\end{subfigure}
	
	\caption{Three types of qualitative characteristics of system \eqref{Simplify system in case 3} under corresponding condition.}
	\label{fig19}
\end{figure}

Finally, we analyze the phase portrait and the attraction region of $P^{4}_1$ under Condition ($\hat{\mathbb{K}}^{4}_5 \wedge \mathbb{H}$), which can be partitioned into two sub-Conditions: ($\hat{\mathbb{K}}^{4}_5 \wedge \mathbb{H}_1$) and ($\hat{\mathbb{K}}^{4}_5 \wedge \mathbb{H}_3$). 
Note that $\hat{\mathbb{K}}^{4}_5 \wedge \mathbb{H}_2=\varnothing$ and $\hat{\mathbb{K}}^{4}_5 \wedge \mathbb{H}_4=\varnothing$. Notice that Condition  ($\hat{\mathbb{K}}^{4}_5 \wedge \mathbb{H}$) is symmetric to Condition ($\hat{\mathbb{K}}^{4}_4 \wedge \mathbb{H}$), 
so the analysis process and results are similar. 
By repeating the analysis as above, we can get the refined results under Conditions ($\hat{\mathbb{K}}^{4}_5 \wedge \mathbb{H}_1$) and ($\hat{\mathbb{K}}^{4}_5 \wedge \mathbb{H}_3$), as shown in Figs.~\ref{fig19} and \ref{fig09}.

\section{Numerical verification of the Global phase portrait and attraction region}\label{B}

\begin{figure}
	\centering
	\begin{subfigure}{.23\textwidth}
		\centering
		\includegraphics[width=4cm]{11_1.pdf}
		\caption{$\mathbb{K}^{2a}\wedge \mathbb{H}_1$}
		\label{fig11.1}
	\end{subfigure}
	\begin{subfigure}{.23\textwidth}
		\centering
		\includegraphics[width=4cm]{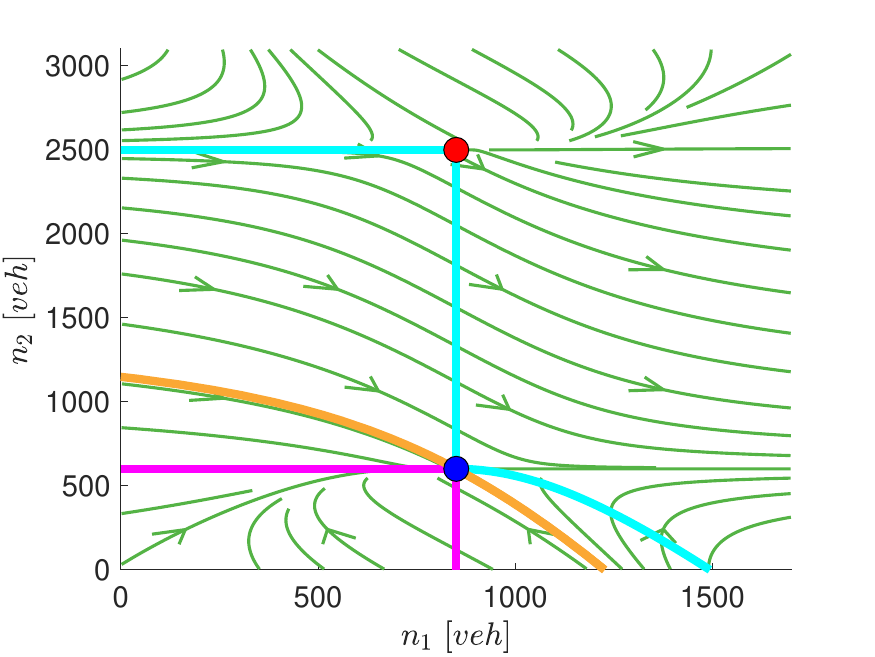}
		\caption{$\mathbb{K}^{2a} \wedge \mathbb{H}_2$}
		\label{fig11.2}
	\end{subfigure}
	\begin{subfigure}{.23\textwidth}
		\centering
		\includegraphics[width=4cm]{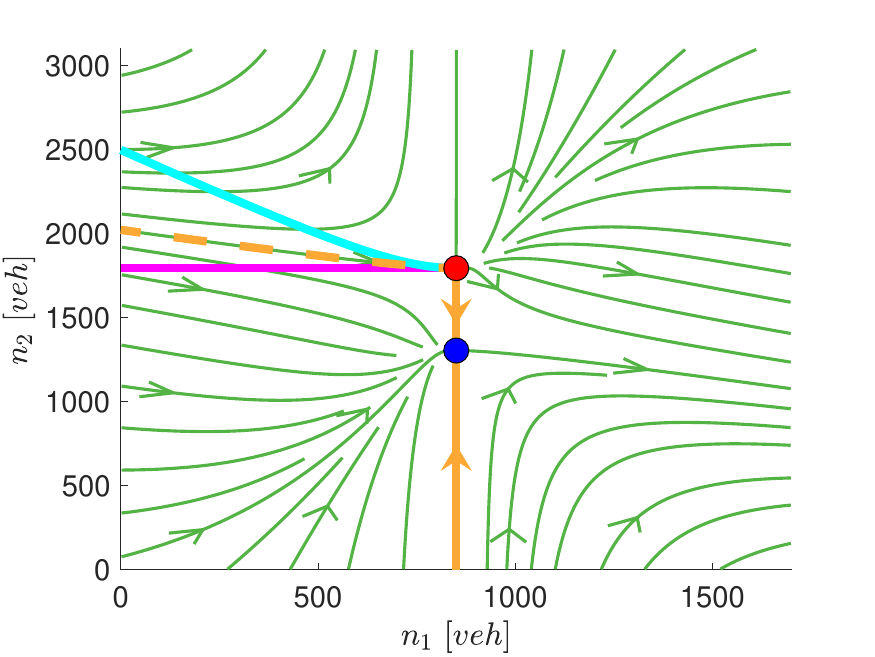}
		\caption{$\mathbb{K}^{2a} \wedge \mathbb{H}_3$}
		\label{fig11.3}
	\end{subfigure}
	\begin{subfigure}{.23\textwidth}
		\centering
		\includegraphics[width=4cm]{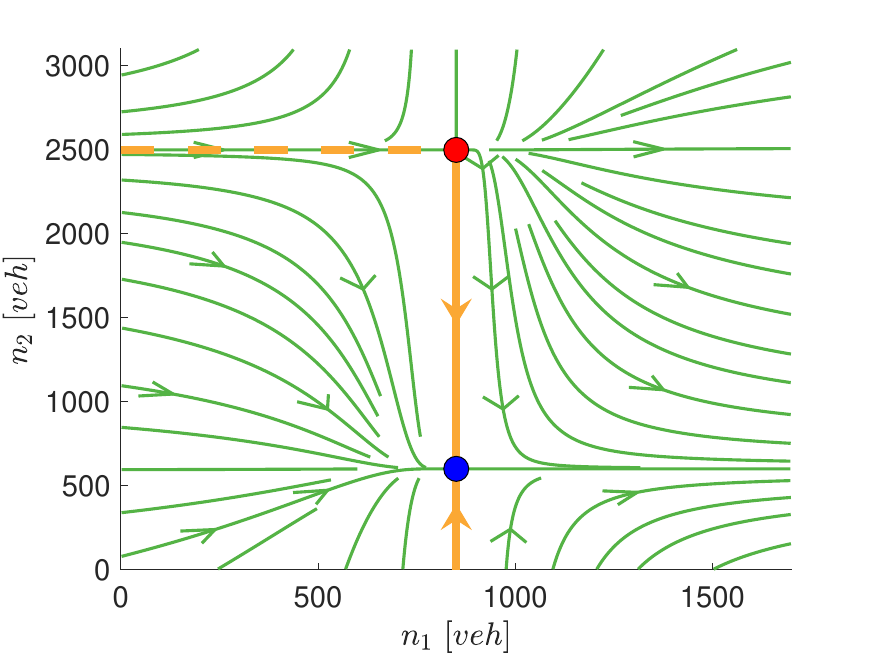}
		\caption{$\mathbb{K}^{2a} \wedge \mathbb{H}_4$}
		\label{fig11.4}
	\end{subfigure}
	\begin{subfigure}{.23\textwidth}
		\centering
		\includegraphics[width=4cm]{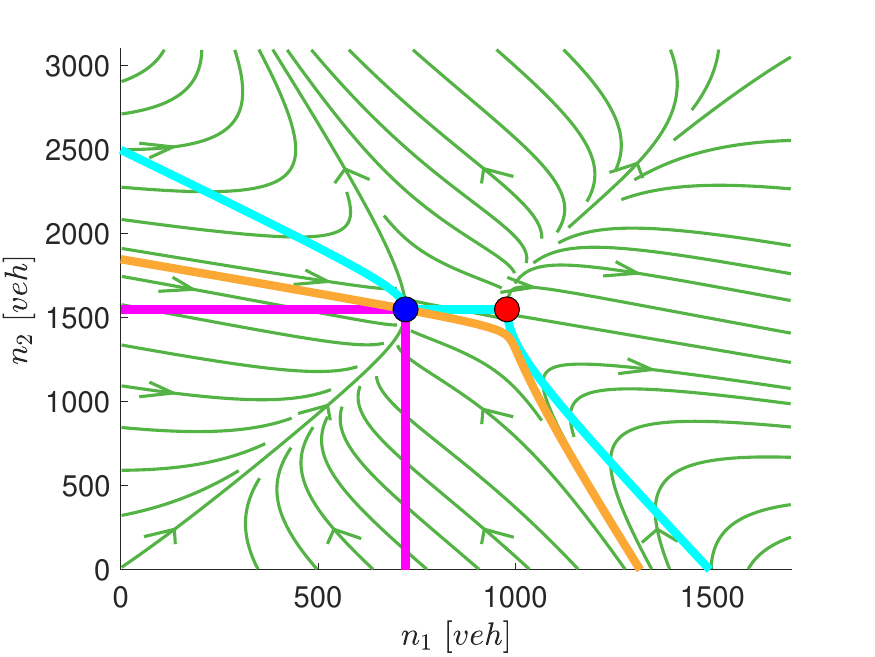}
		\caption{$\mathbb{K}^{2b} \wedge \mathbb{H}_1$}
		\label{fig11.5}
	\end{subfigure}
	\begin{subfigure}{.23\textwidth}
		\centering
		\includegraphics[width=4cm]{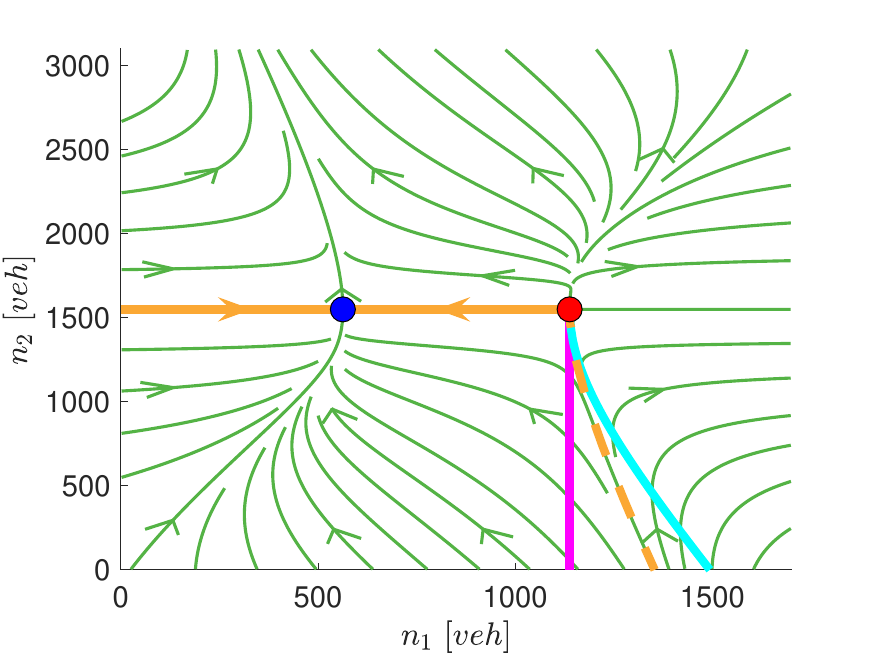}
		\caption{$\mathbb{K}^{2b} \wedge \mathbb{H}_2$}
		\label{fig11.6}
	\end{subfigure}
	\begin{subfigure}{.23\textwidth}
		\centering
		\includegraphics[width=4cm]{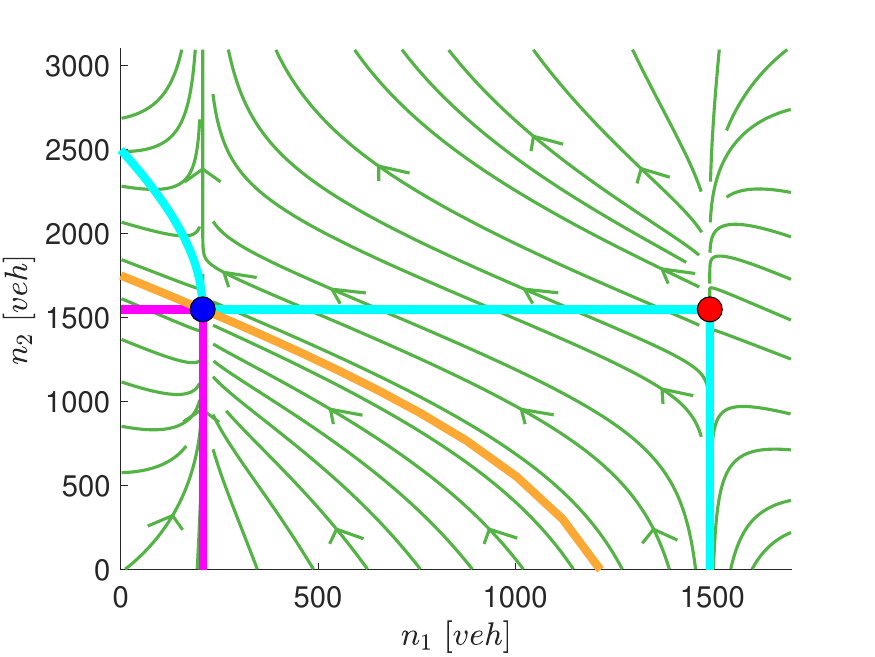}
		\caption{$\mathbb{K}^{2b} \wedge \mathbb{H}_3$}
		\label{fig11.7}
	\end{subfigure}
	\begin{subfigure}{.23\textwidth}
		\centering
		\includegraphics[width=4cm]{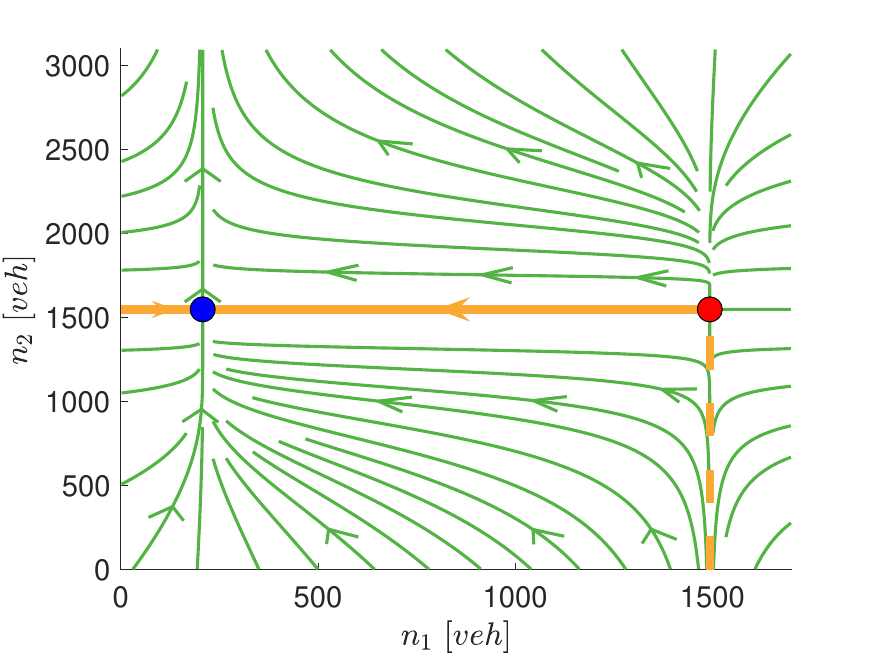}
		\caption{$\mathbb{K}^{2b} \wedge \mathbb{H}_4$}
		\label{fig11.8}
	\end{subfigure}
	\caption{
		Numerical simulation phase portraits for system \eqref{the model} under Condition under Conditions $(\mathbb{K}^{2a})$ and $(\mathbb{K}^{2b})$. Red circle: undesired saddle-node equilibrium. Blue circle: desired uncongested saddle-node equilibrium. Green line with arrows: the trajectories. Purple line and  light blue line represent the inner and outer estimations of attraction regions, respectively.  The yellow lines represent the numerical attraction region boundaries, solid yellow lines indicating the boundary line belongs to attraction region, while dashed yellow lines indicating the boundary line does not belong to attraction region. The agreement of this figure with Figs. \ref{fig05} and Fig. \ref{fig06} validate the theoretical analysis.}
	\label{fig11}
\end{figure}

\begin{figure}
	\centering
	\begin{subfigure}{.23\textwidth}
		\centering
		\includegraphics[width=4.5cm]{12_1.pdf}
		\caption{$\mathbb{K}^{4}_1\wedge \mathbb{H}_1$}
		\label{fig12.1}
	\end{subfigure}
	\begin{subfigure}{.23\textwidth}
		\centering
		\includegraphics[width=4.5cm]{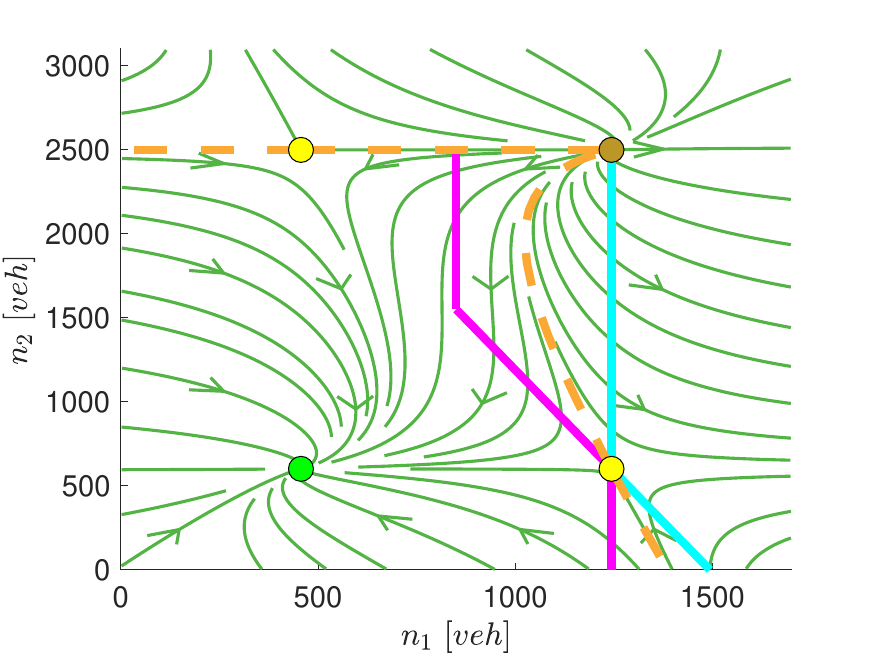}
		\caption{$\mathbb{K}^{4}_1 \wedge \mathbb{H}_2$}
		\label{fig12.2}
	\end{subfigure}
	\begin{subfigure}{.23\textwidth}
		\centering
		\includegraphics[width=4.5cm]{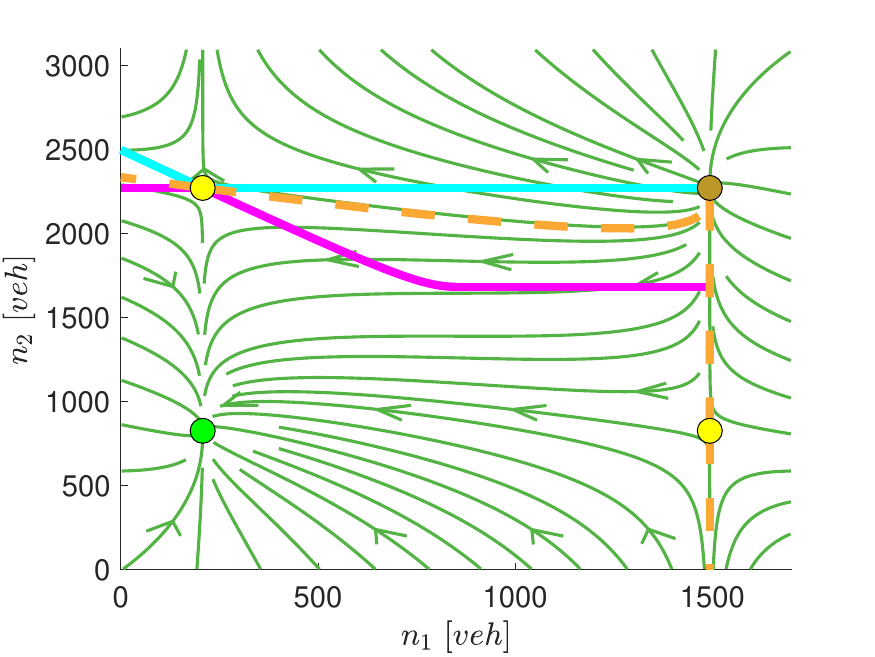}
		\caption{$\mathbb{K}^{4}_1 \wedge \mathbb{H}_3$}
		\label{fig12.3}
	\end{subfigure}
	\begin{subfigure}{.23\textwidth}
		\centering
		\includegraphics[width=4.5cm]{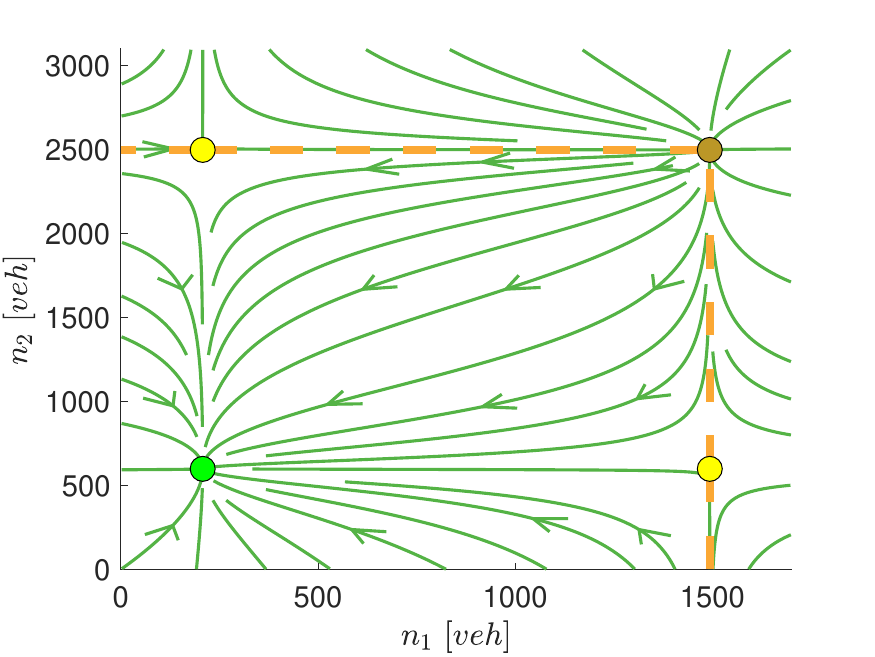}
		\caption{$\mathbb{K}^{4}_1 \wedge \mathbb{H}_4$}
		\label{fig12.4}
	\end{subfigure}
	\begin{subfigure}{.23\textwidth}
		\centering
		\includegraphics[width=4.5cm]{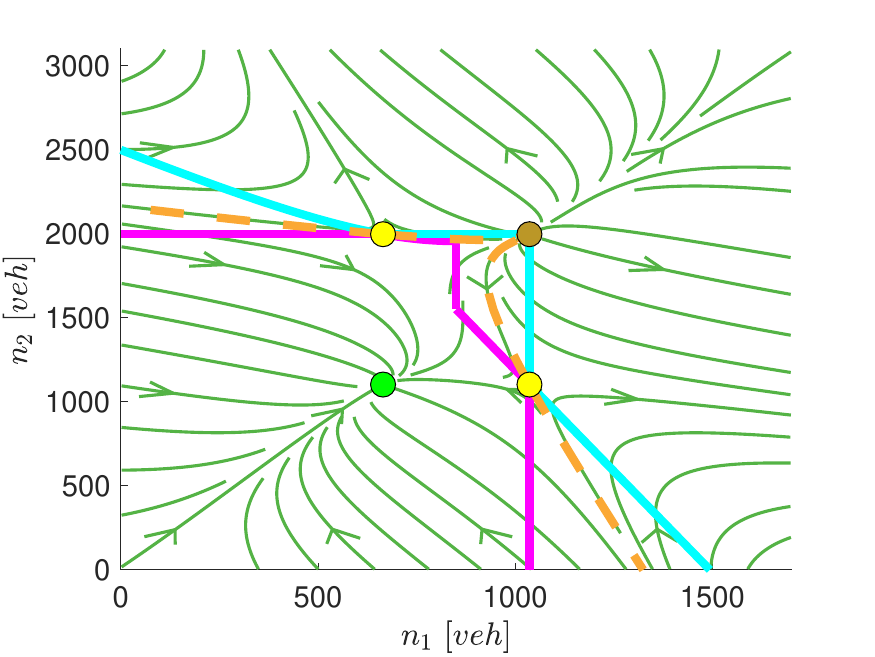}
		\caption{$\mathbb{K}^{4}_2 \wedge \mathbb{H}_1$}
		\label{fig12.5}
	\end{subfigure}
	\begin{subfigure}{.23\textwidth}
		\centering
		\includegraphics[width=4.5cm]{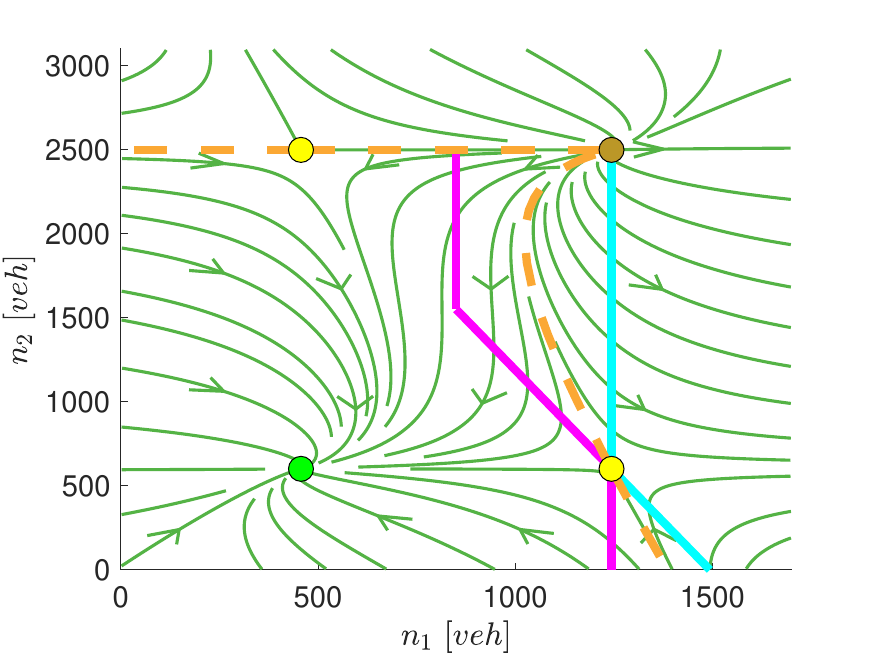}
		\caption{$\mathbb{K}^{4}_2 \wedge \mathbb{H}_2$}
		\label{fig12.6}
	\end{subfigure}
	\begin{subfigure}{.23\textwidth}
		\centering
		\includegraphics[width=4.5cm]{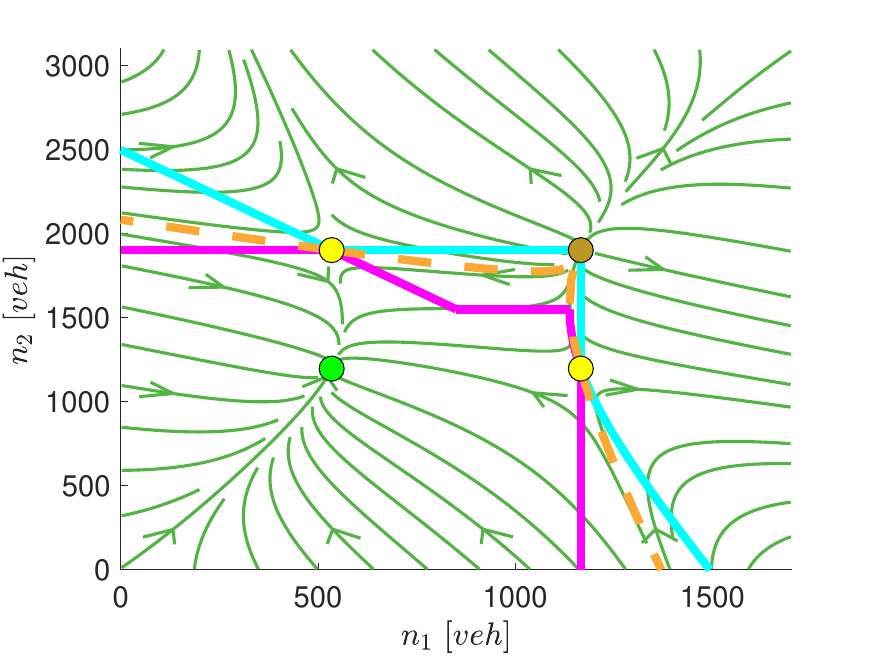}
		\caption{$\mathbb{K}^{4}_3 \wedge \mathbb{H}_1$}
		\label{fig12.7}
	\end{subfigure}
	\begin{subfigure}{.23\textwidth}
		\centering
		\includegraphics[width=4.5cm]{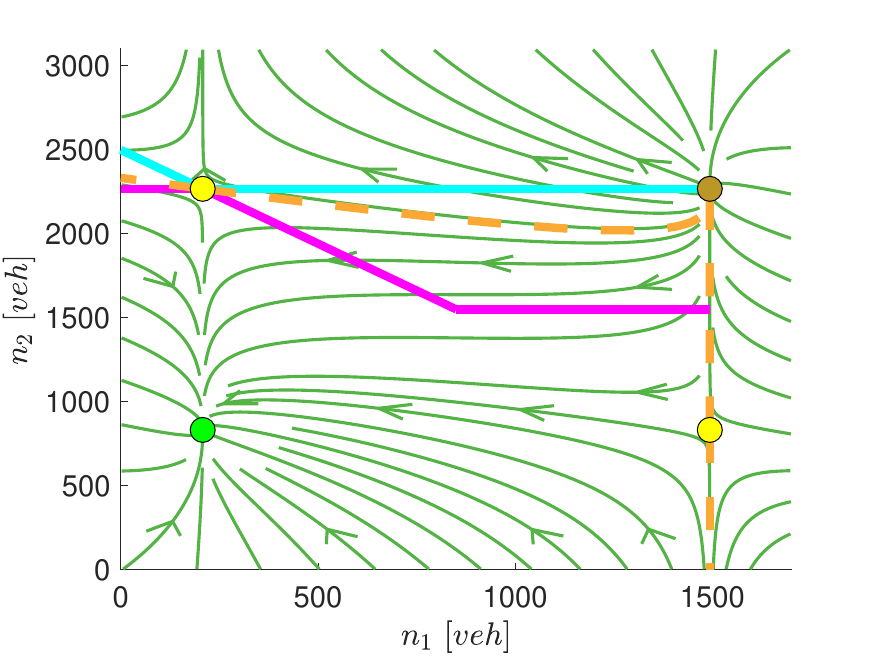}
		\caption{$\mathbb{K}^{4}_3 \wedge \mathbb{H}_3$}
		\label{fig12.8}
	\end{subfigure}
	\begin{subfigure}{.23\textwidth}
		\centering
		\includegraphics[width=4.5cm]{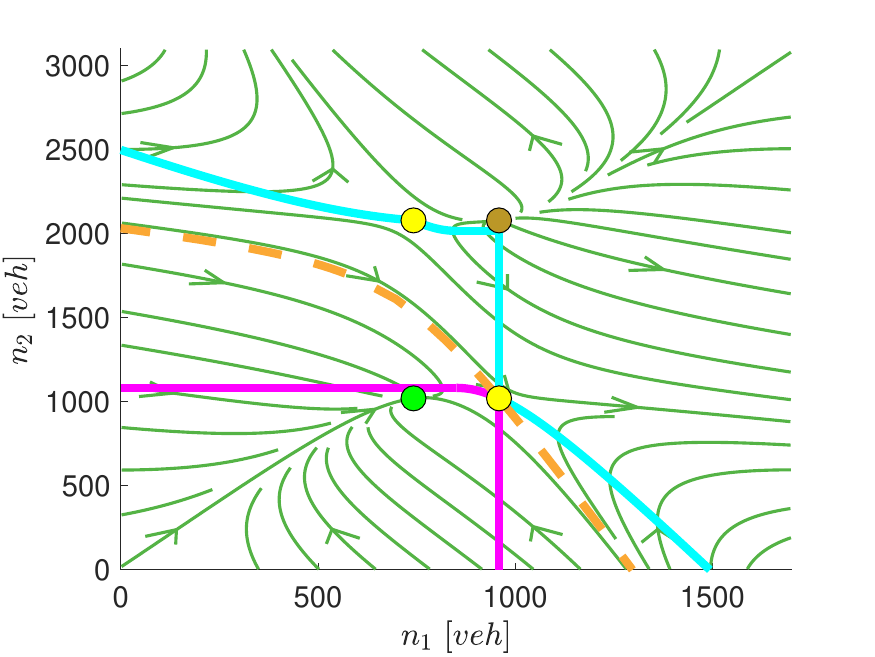}
		\caption{$\mathbb{K}^{4}_4 \wedge \mathbb{H}_1$, type 1}
		\label{fig12.9}
	\end{subfigure}
	\begin{subfigure}{.23\textwidth}
		\centering
		\includegraphics[width=4.5cm]{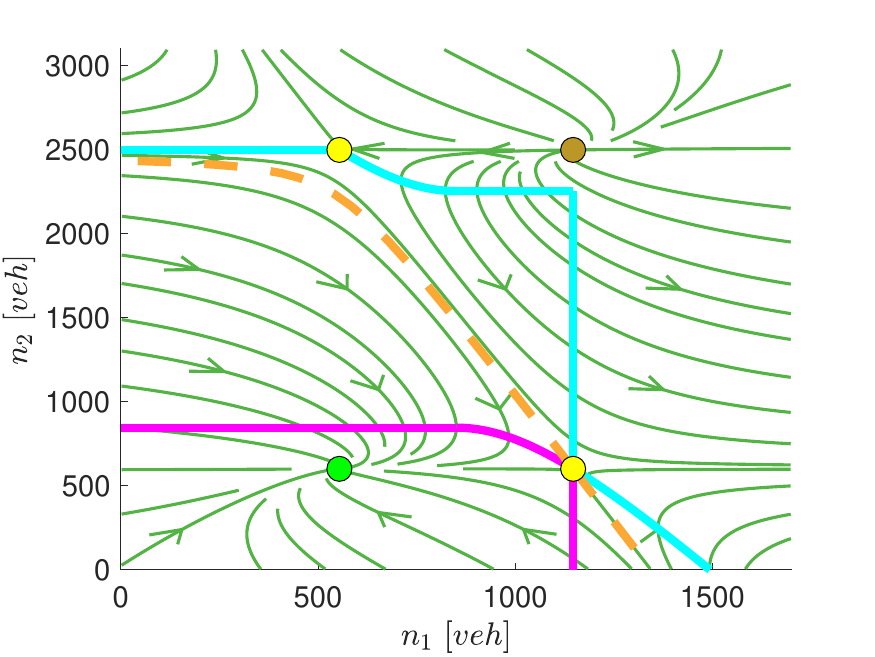}
		\caption{$\mathbb{K}^{4}_4 \wedge \mathbb{H}_2$, type 1}
		\label{fig12.10}
	\end{subfigure}
	\begin{subfigure}{.23\textwidth}
		\centering
		\includegraphics[width=4.5cm]{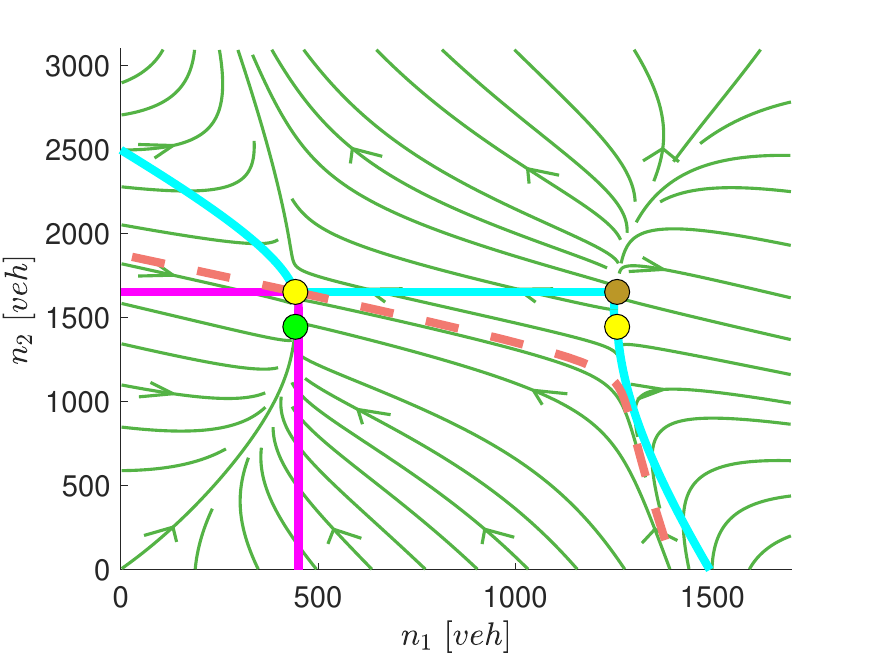}
		\caption{$\mathbb{K}^{4}_5 \wedge \mathbb{H}_1$, type 1}
		\label{fig12.11}
	\end{subfigure}
	\begin{subfigure}{.23\textwidth}
		\centering
		\includegraphics[width=4.5cm]{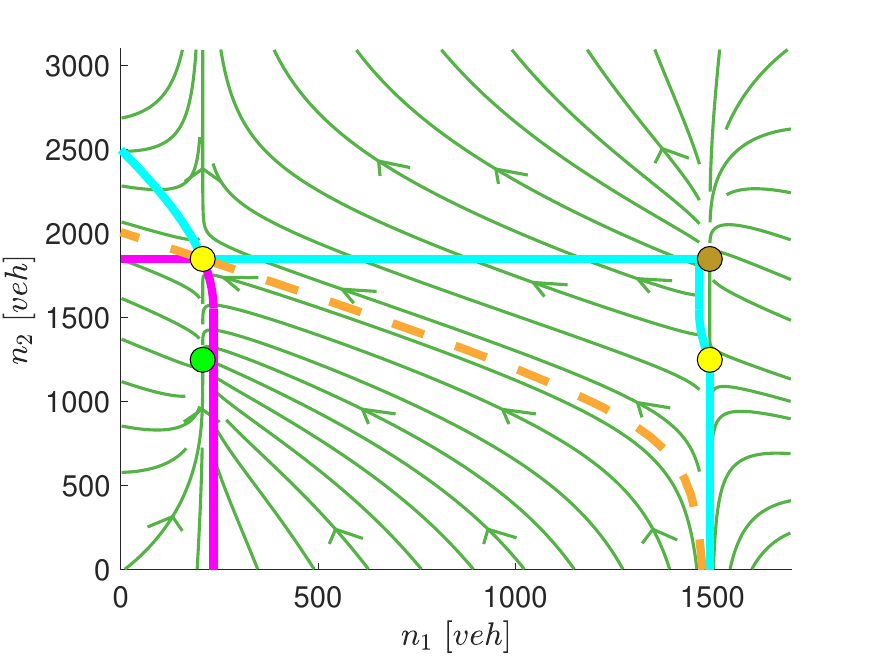}
		\caption{$\mathbb{K}^{4}_5 \wedge \mathbb{H}_3$, type 1}
		\label{fig12.12}
	\end{subfigure}
	\begin{subfigure}{.23\textwidth}
		\centering
		\includegraphics[width=4.5cm]{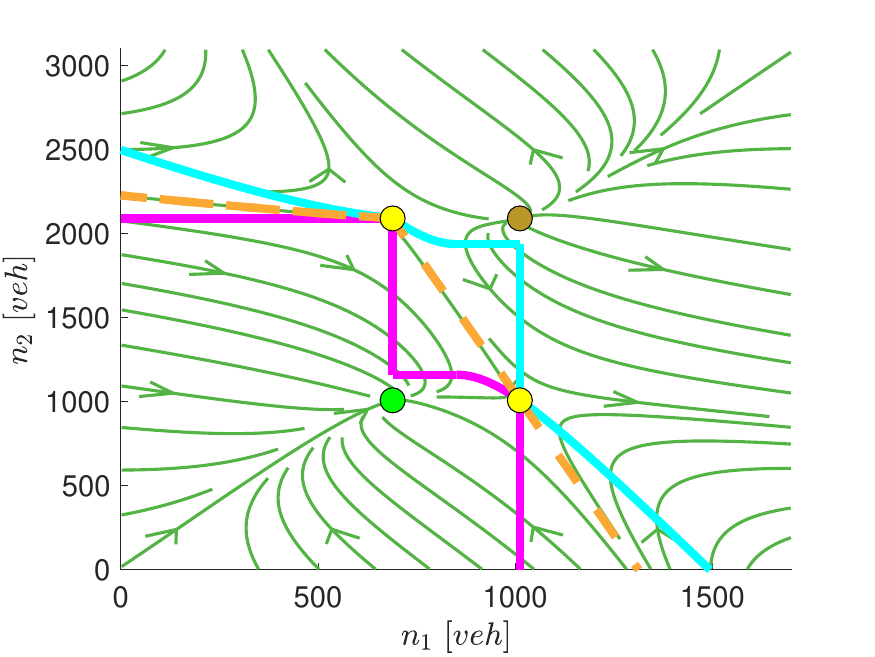}
		\caption{$\mathbb{K}^{4}_4 \wedge \mathbb{H}_1$, type 2}
		\label{fig12.13}
	\end{subfigure}
	\begin{subfigure}{.23\textwidth}
		\centering
		\includegraphics[width=4.5cm]{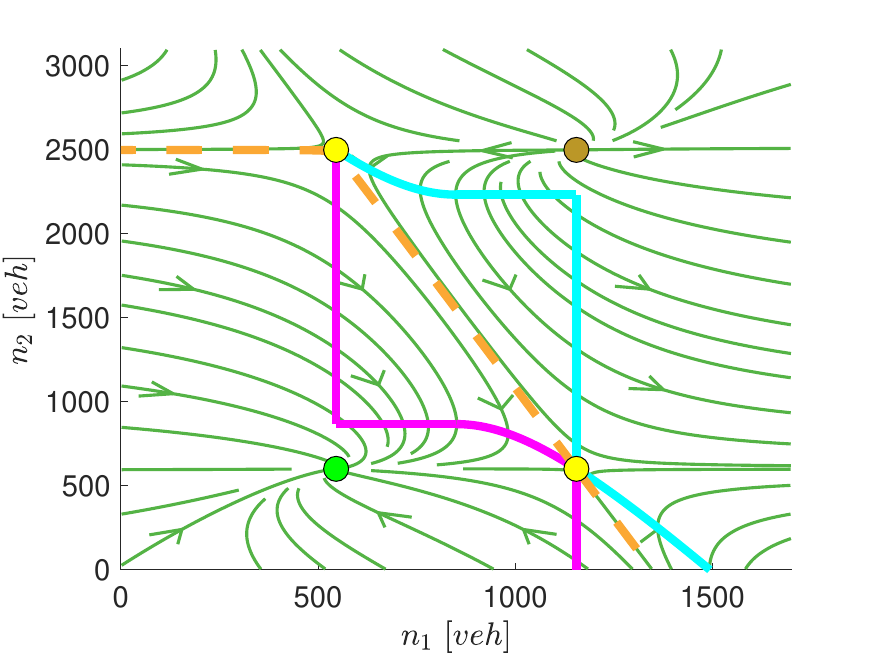}
		\caption{$\mathbb{K}^{4}_4 \wedge \mathbb{H}_2$, type 2}
		\label{fig12.14}
	\end{subfigure}
	\begin{subfigure}{.23\textwidth}
		\centering
		\includegraphics[width=4.5cm]{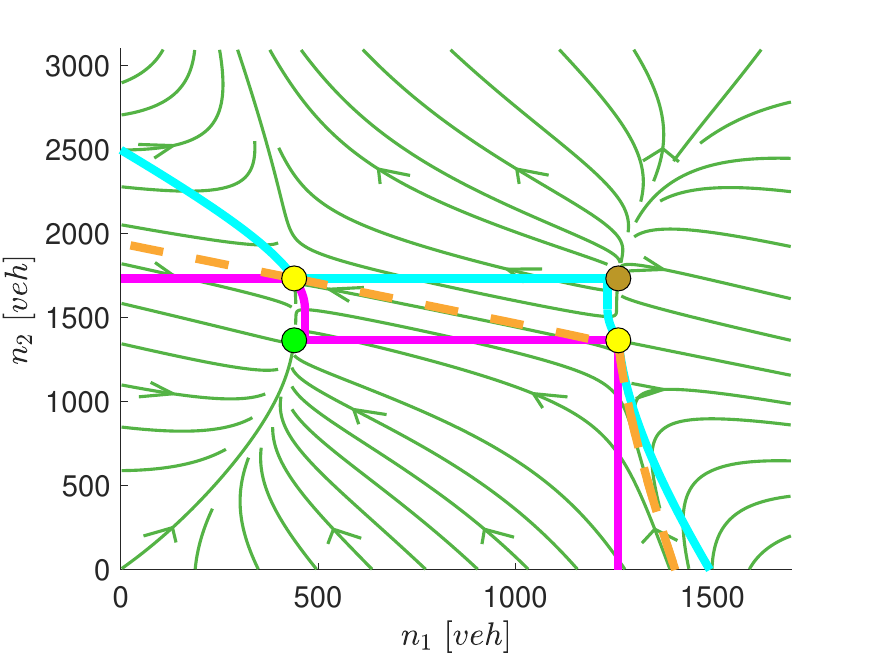}
		\caption{$\mathbb{K}^{4}_5 \wedge \mathbb{H}_1$, type 2}
		\label{fig12.15}
	\end{subfigure}
	\begin{subfigure}{.23\textwidth}
		\centering
		\includegraphics[width=4.5cm]{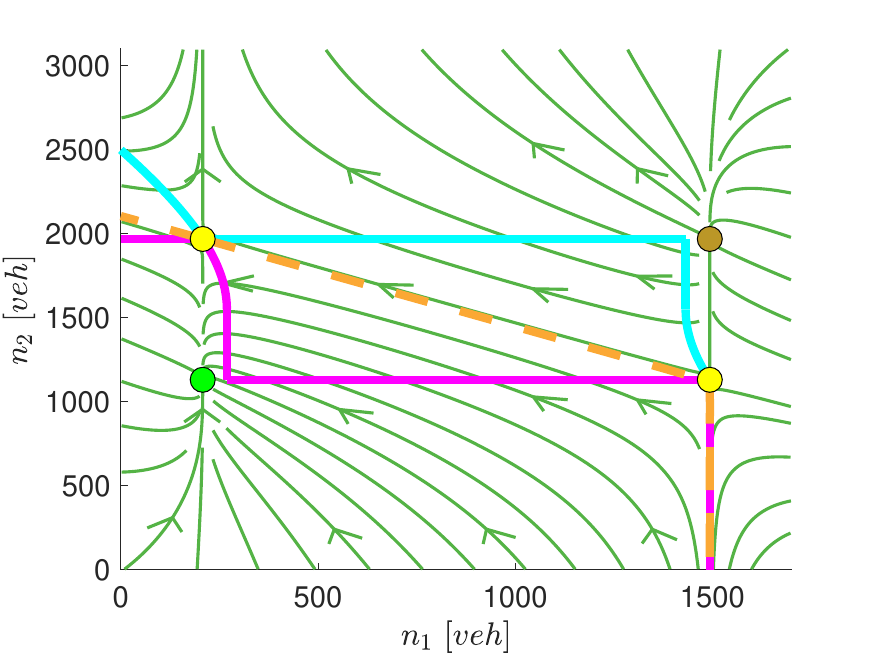}
		\caption{$\mathbb{K}^{4}_5 \wedge \mathbb{H}_3$, type 2}
		\label{fig12.16}
	\end{subfigure}
	\begin{subfigure}{.23\textwidth}
		\centering
		\includegraphics[width=4.5cm]{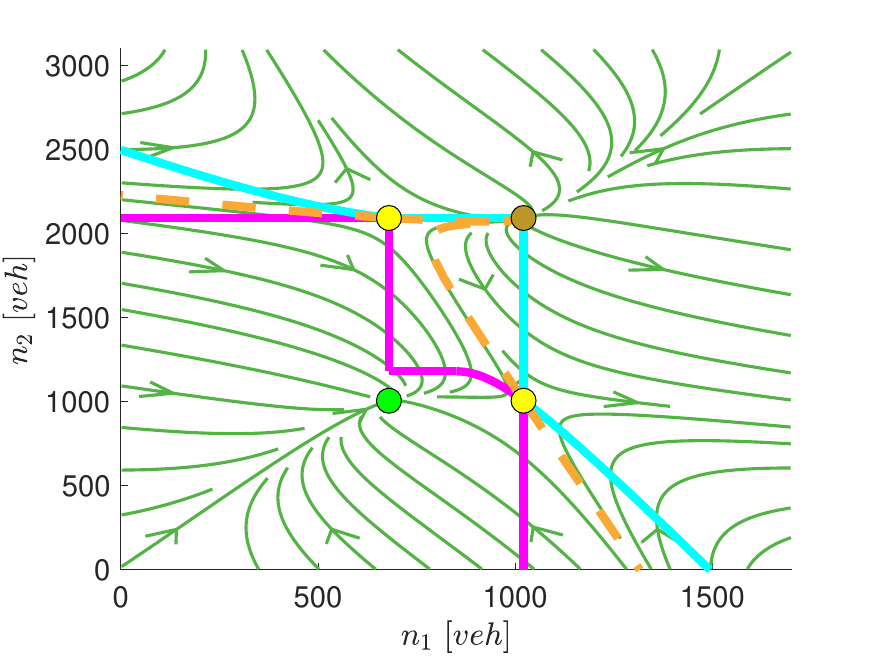}
		\caption{$\mathbb{K}^{4}_4 \wedge \mathbb{H}_1$, type 3}
		\label{fig12.17}
	\end{subfigure}
	\begin{subfigure}{.23\textwidth}
		\centering
		\includegraphics[width=4.5cm]{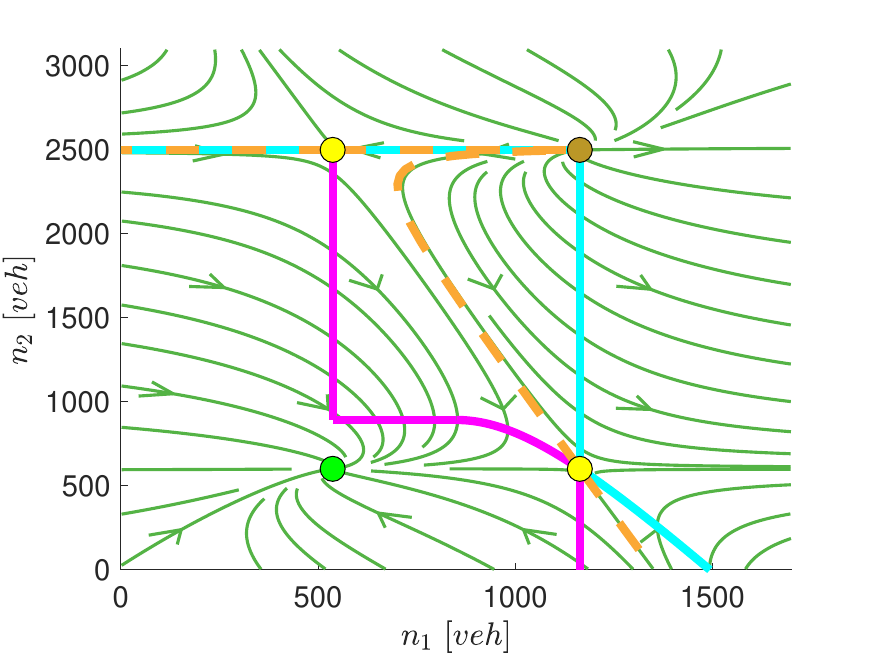}
		\caption{$\mathbb{K}^{4}_4 \wedge \mathbb{H}_2$, type 3}
		\label{fig12.18}
	\end{subfigure}
	\begin{subfigure}{.23\textwidth}
		\centering
		\includegraphics[width=4.5cm]{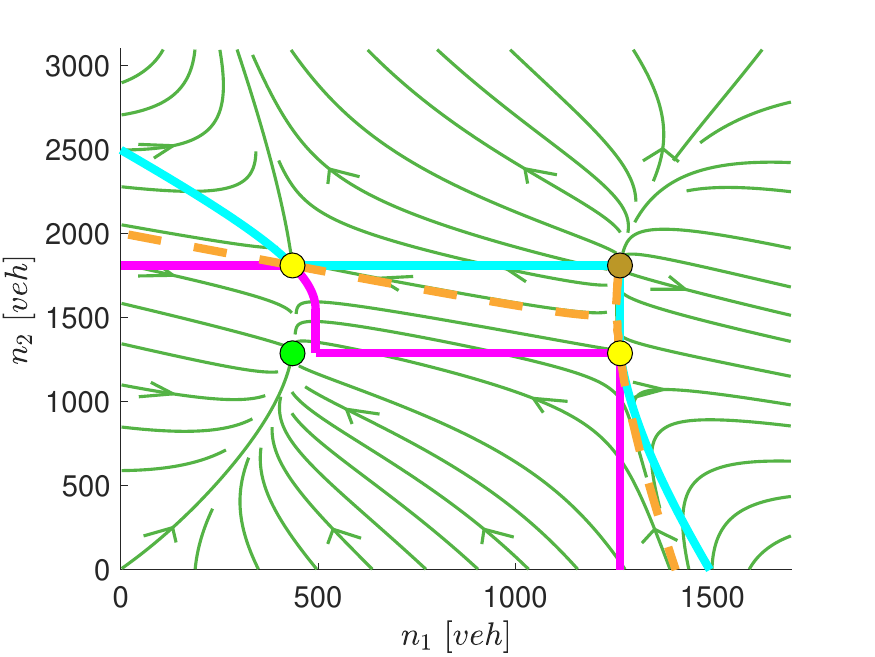}
		\caption{$\mathbb{K}^{4}_5 \wedge \mathbb{H}_1$, type 3}
		\label{fig12.19}
	\end{subfigure}
	\begin{subfigure}{.23\textwidth}
		\centering
		\includegraphics[width=4.5cm]{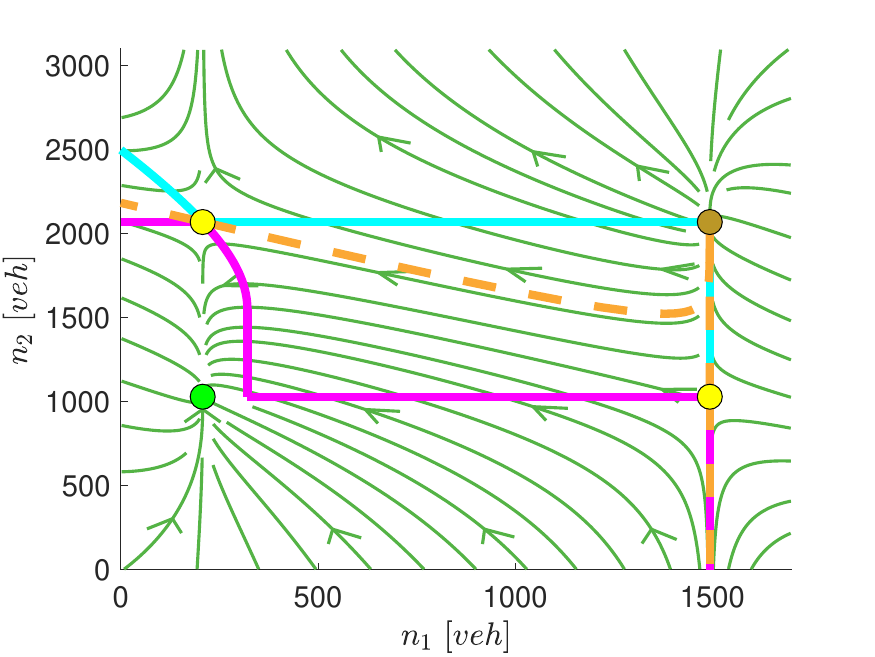}
		\caption{$\mathbb{K}^{4}_5 \wedge \mathbb{H}_3$, type 3}
		\label{fig12.20}
	\end{subfigure}
	\caption{Numerical simulation phase portraits for system \eqref{the model} under Condition $(\mathbb{K}^{4})$. Green circle: desired uncongested stable equilibrium. Yellow circle: saddle equilibrium. Yellow-brown circle: unstable equilibrium. Green line with arrows: the trajectories. Purple line and  light blue line represent the inner and outer estimations of attraction regions, respectively.  The yellow lines represent the numerical attraction region boundaries, solid yellow lines indicating the boundary line belongs to attraction region, while dashed yellow lines indicating the boundary line does not belong to attraction region. The agreement of this figure with Fig. \ref{fig09} validate the theoretical analysis.}
	\label{fig12}
\end{figure}

\begin{table}
	\caption{Parameter values for different scenarios}
	\label{tab5}
	\centering
	\fontsize{6.5}{8}\selectfont
	\setlength{\tabcolsep}{3mm}{
		\begin{threeparttable}
			\begin{tabular}{c|cc|cc|c|c}
				\toprule
				\multirow{1}{*}{Scenarios}&
				\multicolumn{2}{c|}{Flow rates} &
				\multicolumn{2}{c|}{Network-outside demands $[veh/h]$} &
				\multicolumn{1}{c|}{Satisfying Condition} &
				\multicolumn{1}{c}{Corresponding phase portrait}\cr
				\midrule
				Scenario 1&$u_1=0.3$&$u_2=0.4$& $d_1=3\times10^4$& $d_2=5\times10^4$&Condition $(\mathbb{K}^{2a} \wedge \mathbb{H}_1)$& Fig. \ref{fig11.1}\cr
				
				Scenario 2&$u_1=0$&$u_2=0.8$& $d_1=3\times10^4$& $d_2=5\times10^4$&Condition $(\mathbb{K}^{2a} \wedge \mathbb{H}_2)$& Fig. \ref{fig11.2}\cr
				
				Scenario 3&$u_1=0.4$&$u_2=0$& $d_1=7\times10^4$& $d_2=5\times10^4$&Condition $(\mathbb{K}^{2a} \wedge \mathbb{H}_3)$& Fig. \ref{fig11.3}\cr
				
				Scenario 4&$u_1=0$&$u_2=0$& $d_1=7\times10^4$& $d_2=5\times10^4$&Condition $(\mathbb{K}^{2a} \wedge \mathbb{H}_4)$& Fig. \ref{fig11.4}\cr
				
				Scenario 5&$u_1=0.4386$&$u_2=0.48$& $d_1=3\times10^4$& $d_2=5\times10^4$&Condition $(\mathbb{K}^{2b} \wedge \mathbb{H}_1)$& Fig. \ref{fig11.5}\cr
				
				Scenario 6&$u_1=0$&$u_2=0.4$& $d_1=3\times10^4$& $d_2=8\times10^4$&Condition $(\mathbb{K}^{2b} \wedge \mathbb{H}_2)$& Fig. \ref{fig11.6}\cr
				
				Scenario 7&$u_1=0.8$&$u_2=0$& $d_1=0.5\times10^4$& $d_2=8\times10^4$&Condition $(\mathbb{K}^{2b} \wedge \mathbb{H}_3)$& Fig. \ref{fig11.7}\cr
				
				Scenario 8&$u_1=0$&$u_2=0.8$& $d_1=6\times10^4$& $d_2=8\times10^4$&Condition $(\mathbb{K}^{2b} \wedge \mathbb{H}_4)$& Fig. \ref{fig11.8}\cr
				
				Scenario 9&$u_1=0.4$&$u_2=0.4$& $d_1=3\times10^4$& $d_2=5\times10^4$&Condition $(\hat{\mathbb{K}}^{4}_1 \wedge \mathbb{H}_1)$& Fig. \ref{fig12.1}\cr
				
				Scenario 10&$u_1=0$&$u_2=0.8$& $d_1=5\times10^4$& $d_2=1\times10^4$&Condition $(\hat{\mathbb{K}}^{4}_1 \wedge \mathbb{H}_2)$& Fig. \ref{fig12.2}\cr
				
				Scenario 11&$u_1=0.8$&$u_2=0$& $d_1=0.2\times10^4$& $d_2=5\times10^4$&Condition $(\hat{\mathbb{K}}^{4}_1 \wedge \mathbb{H}_3)$& Fig. \ref{fig12.3}\cr
				
				Scenario 12&$u_1=0$&$u_2=0$& $d_1=4\times10^4$& $d_2=3\times10^4$&Condition $(\hat{\mathbb{K}}^{4}_1 \wedge \mathbb{H}_4)$& Fig. \ref{fig12.4}\cr
				
				Scenario 13&$u_1=0.8$&$u_2=0.8$& $d_1=0.6\times10^4$& $d_2=2\times10^4$&Condition $(\hat{\mathbb{K}}^{4}_2 \wedge \mathbb{H}_1)$& Fig. \ref{fig12.5}\cr
				
				Scenario 14&$u_1=0$&$u_2=0.8$& $d_1=0.6\times10^4$& $d_2=3\times10^4$&Condition $(\hat{\mathbb{K}}^{4}_2 \wedge \mathbb{H}_2)$& Fig. \ref{fig12.6}\cr
				
				Scenario 15&$u_1=0.8$&$u_2=0.8$& $d_1=0.3\times10^4$& $d_2=2.4\times10^4$&Condition $(\hat{\mathbb{K}}^{4}_3 \wedge \mathbb{H}_1)$& Fig. \ref{fig12.7}\cr
				
				Scenario 16&$u_1=0.8$&$u_2=0$& $d_1=4\times10^4$& $d_2=2.4\times10^4$&Condition $(\hat{\mathbb{K}}^{4}_1 \wedge \mathbb{H}_3)$& Fig. \ref{fig12.8}\cr
				
				Scenario 17&$u_1=0.8$&$u_2=0.8$& $d_1=2\times10^4$& $d_2=0.5\times10^4$&Condition $(\hat{\mathbb{K}}^{4}_4 \wedge \mathbb{H}_1)$& Fig. \ref{fig12.9}\cr
				
				Scenario 18&$u_1=0.8$&$u_2=0.8$& $d_1=0.78\times10^4$& $d_2=1.99\times10^4$&Condition $(\hat{\mathbb{K}}^{4}_4 \wedge \mathbb{H}_1)$& Fig. \ref{fig12.13}\cr
				
				Scenario 19&$u_1=0.8$&$u_2=0.8$& $d_1=0.72\times10^4$& $d_2=2\times10^4$&Condition $(\hat{\mathbb{K}}^{4}_4 \wedge \mathbb{H}_1)$& Fig. \ref{fig12.17}\cr	
				
				Scenario 20&$u_1=0$&$u_2=0.8$& $d_1=1.5\times10^4$& $d_2=6\times10^4$&Condition $(\hat{\mathbb{K}}^{4}_4 \wedge \mathbb{H}_2)$& Fig. \ref{fig12.10}\cr
				
				Scenario 21&$u_1=0$&$u_2=0.8$& $d_1=1.5\times10^4$& $d_2=5.6\times10^4$&Condition $(\hat{\mathbb{K}}^{4}_4 \wedge \mathbb{H}_2)$& Fig. \ref{fig12.14}\cr
				
				Scenario 22&$u_1=0$&$u_2=0.8$& $d_1=1.5\times10^4$& $d_2=5\times10^4$&Condition $(\hat{\mathbb{K}}^{4}_4 \wedge \mathbb{H}_2)$& Fig. \ref{fig12.18}\cr
				
				Scenario 23&$u_1=0.8$&$u_2=0.8$& $d_1=0.02\times10^4$& $d_2=2.85\times10^4$&Condition $(\hat{\mathbb{K}}^{4}_5 \wedge \mathbb{H}_1)$& Fig. \ref{fig12.11}\cr
				
				Scenario 24&$u_1=0.8$&$u_2=0.8$& $d_1=0.02\times10^4$& $d_2=2.82\times10^4$&Condition $(\hat{\mathbb{K}}^{4}_5 \wedge \mathbb{H}_1)$& Fig. \ref{fig12.15}\cr
				
				Scenario 25&$u_1=0$&$u_2=0.8$& $d_1=0.02\times10^4$& $d_2=2.6\times10^4$&Condition $(\hat{\mathbb{K}}^{4}_5 \wedge \mathbb{H}_1)$& Fig. \ref{fig12.19}\cr
				
				Scenario 26&$u_1=0.8$&$u_2=0$& $d_1=5\times10^4$& $d_2=3.9\times10^4$&Condition $(\hat{\mathbb{K}}^{4}_5 \wedge \mathbb{H}_3)$& Fig. \ref{fig12.12}\cr
				
				Scenario 27&$u_1=0.8$&$u_2=0$& $d_1=5\times10^4$& $d_2=3.78\times10^4$&Condition $(\hat{\mathbb{K}}^{4}_5 \wedge \mathbb{H}_3)$& Fig. \ref{fig12.16}\cr
				
				Scenario 28&$u_1=0.8$&$u_2=0$& $d_1=5\times10^4$& $d_2=3\times10^4$&Condition $(\hat{\mathbb{K}}^{4}_5 \wedge \mathbb{H}_3)$& Fig. \ref{fig12.20}\cr				
				\bottomrule
			\end{tabular}	
			\begin{tablenotes}
				\footnotesize
				\item Note that these are all the scenarios used in this paper. In each scenario, the different formula of the inner and outer estimations of attraction regions are derived (see in Sec. \ref{s3}), and we have plotted the phase portraits in each scenario, as shown in Figs. \ref{fig11} and \ref{fig12}. 
			\end{tablenotes}		
	\end{threeparttable}}
\end{table}

\begin{figure}
	\centering
	\includegraphics[width=15cm]{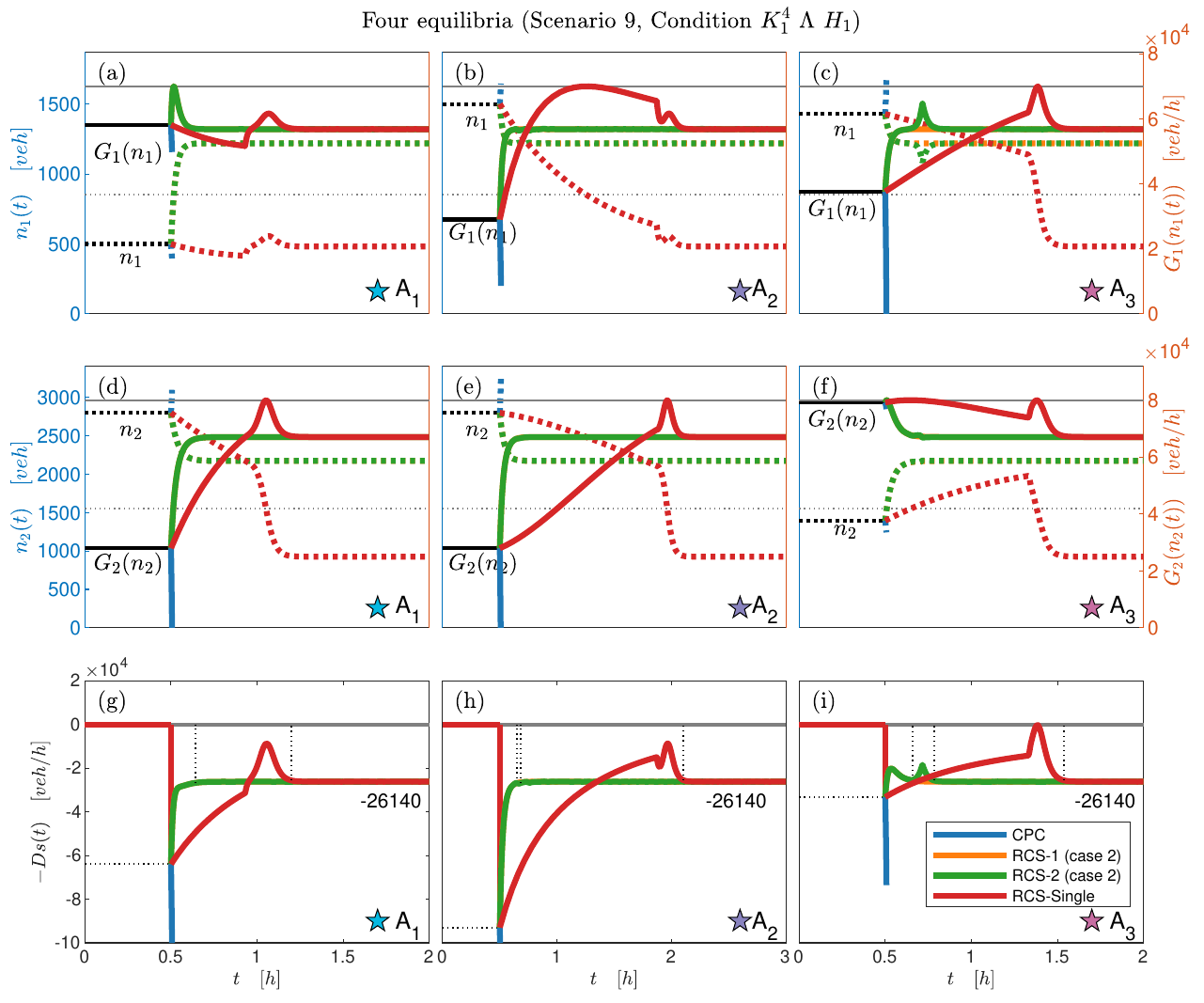}
	\caption{Vehicle accumulations (dotted line) and completion flow rates (solid line) of Region 1 (((a), (b) and (c))) and Region 2 ((d), (e) and (f)). 
	The numerical diagrams of deviation from maximum complete flow $-D_s$ ((g), (h) and (i)) evolving with time $t$ under CPC, RCS-1, RCS-2 and RCS-single. Scenario 9 is considered. 
			Note that for (a)-(f), we employ dual y-axes. The left y-axis represents vehicle accumulation (density), while the right y-axis represents completion flow. Scenario 9 is considered. Here we consider case 2 (in Fig. \ref{fig15} we consider case 1). Other data settings are the same as Fig. \ref{fig15}.}
	\label{fig15.2}
\end{figure}

\begin{figure}
	\centering
	\includegraphics[width=15cm]{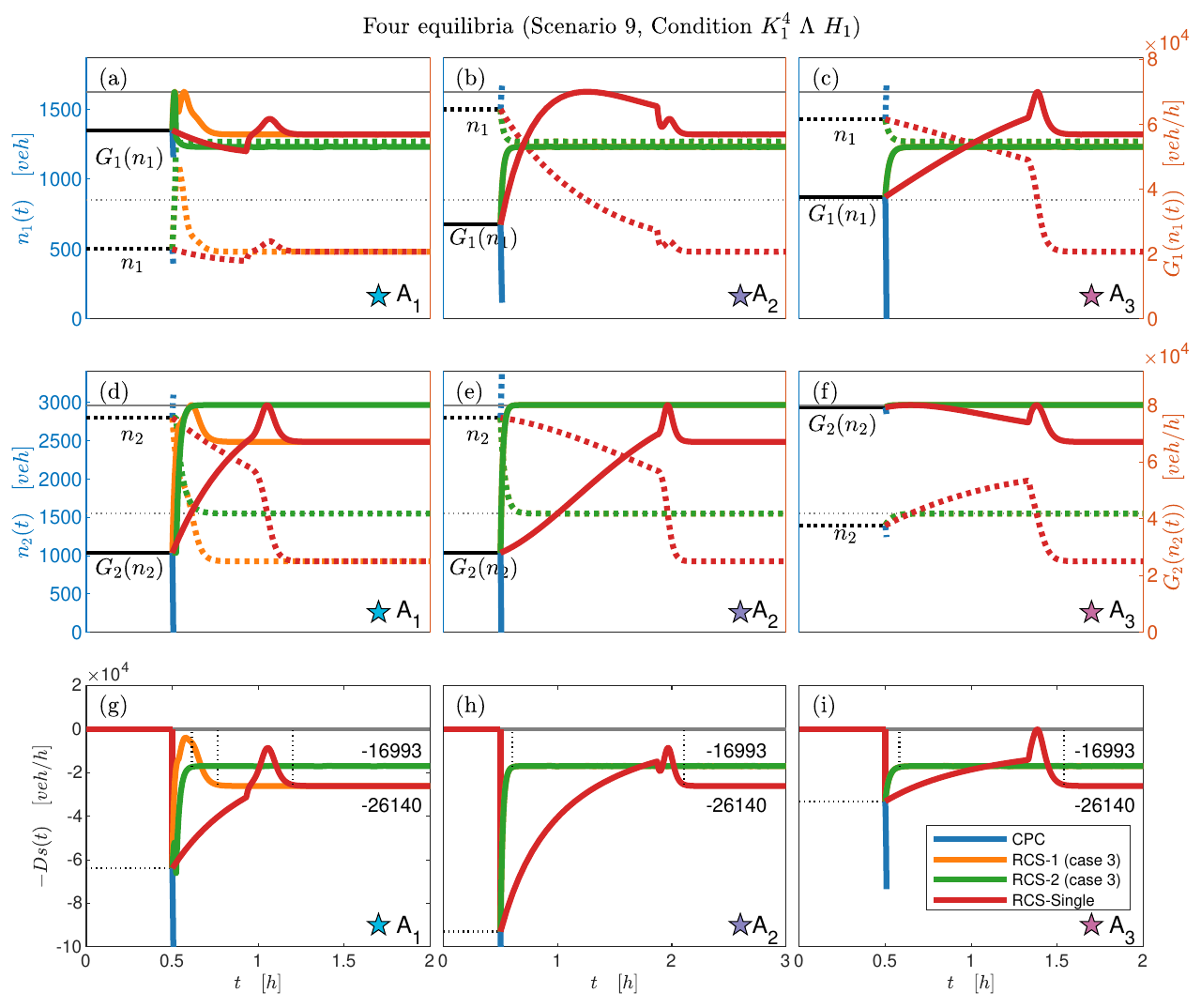}
	\caption{Vehicle accumulations (dotted line) and completion flow rates (solid line) of Region 1 (((a), (b) and (c))) and Region 2 ((d), (e) and (f)). 
			The numerical diagrams of deviation from maximum complete flow $-D_s$ ((g), (h) and (i)) evolving with time $t$ under CPC, RCS-1, RCS-2 and RCS-single. Scenario 9 is considered. 
			Note that for (a)-(f), we employ dual y-axes. The left y-axis represents vehicle accumulation (density), while the right y-axis represents completion flow. Scenario 9 is considered. Here we consider case 3 (in Fig. \ref{fig15} we consider case 1). Other data settings are the same as Fig. \ref{fig15}.}
	\label{fig15.3}
\end{figure}

\end{APPENDICES}

\end{document}